\documentclass{JHEP3}
\pdfoutput=1

\usepackage{graphicx}
\usepackage{mathrsfs}
\usepackage{amsmath, amsthm, amssymb}
\usepackage{multirow} 
\usepackage{color}
\let\normalcolor\relax

\newcommand{\be}{\begin{equation}}
\newcommand{\ee}{\end{equation}}
\newcommand{\beq}{\begin{equation}}
\newcommand{\eeq}{\end{equation}}
\newcommand{\bea}{\begin{eqnarray}}
\newcommand{\eea}{\end{eqnarray}}

\newcommand{\gsim}{\lower.7ex\hbox{$\;\stackrel{\textstyle>}{\sim}\;$}}
\newcommand{\lsim}{\lower.7ex\hbox{$\;\stackrel{\textstyle<}{\sim}\;$}}

\newcommand{\tev}{\mbox{ TeV}}

\newcommand{\cl}{\text{CL}}

\newcommand{\sineff}{\sin^2 \theta_{\rm{eff}}}

\newcommand{\BR}{BR}

\newcommand{\brbsgamma}{\BR(\overline{B}\rightarrow X_s\gamma)}

\newcommand\brbsmumu{\BR(\overline{B}_s\to\mu^+\mu^-)}

\newcommand\RBtaunu{\frac{\BR(B_u \to \tau \nu)}{\BR(B_u \to \tau \nu)_{SM}}}
\newcommand\DeltaO{\Delta_{0-}}
\newcommand\RBDtaunuBDenu{\frac{\BR(B \to D \tau \nu)}{\BR(B \to D e \nu)}}
\newcommand\Rl{R_{l23}}

\newcommand\Dstaunu{\BR(D_s \to \tau \nu)}
\newcommand\Dsmunu{\BR(D_s\to \mu \nu)}

\newcommand{\mhl}{m_h}

\newcommand\Dmunu{\BR(D \to \mu \nu)} 
\newcommand{\sigsip}{\sigma_p^\text{SI}}
\newcommand{\sigmaSI}{\sigma_{\chi-p}^{\mathrm{SI}}}
\newcommand{\neut}{{\tilde{\chi}^0_1}}
\newcommand{\like}{\mathcal{L}}
\newcommand{\OhLSP}{\Omega_\text{LSP} h^2}
\newcommand{\siW}{\sigma_{\text{WMAP}}}
\newcommand{\muW}{\mu_\text{WMAP}}

\title{The health of SUSY after the Higgs discovery and the XENON100 data}

\author{Maria Eugenia Cabrera\\
        Institute of Theoretical Physics, GRAPPA, \\
        University of Amsterdam, \\
        Amsterdam, The Netherlands}

\author{J. Alberto Casas\\
        Instituto de F\'isica Te\'orica, IFT-UAM/CSIC, \\
        U.A.M., Cantoblanco, \\
        28049 Madrid, Spain} 

\author{Roberto Ruiz de Austri\\
        Instituto de F\'isica Corpuscular, IFIC-UV/CSIC, \\
        Valencia, Spain}

\abstract{ \small

We analyze the implications for the status and prospects of  
supersymmetry of the Higgs discovery and the last XENON data. We focus  
mainly, but not only, on the CMSSM and NUHM models. Using a Bayesian  
approach we determine the distribution of probability in the parameter  
space of these scenarios. This shows that, most probably, they are now  
beyond the LHC reach . This negative chances increase further (at more  
than 95\% c.l.) if one includes dark matter constraints in the  
analysis, in particular the last XENON100 data. However, the models  
would be probed completely by XENON1T. The mass of the LSP neutralino  
gets essentially fixed around 1 TeV. We do not incorporate ad hoc  
measures of the fine-tuning to penalize unnatural possibilities: such  
penalization arises automatically from the careful Bayesian analysis  
itself, and allows to scan the whole parameter space. In this way, we  
can explain and resolve the apparent discrepancies between the  
previous results in the literature. Although SUSY has become hard to  
detect at LHC, this does not necessarily mean that is very fine-tuned.  
We use Bayesian techniques to show the experimental Higgs mass is at  
$\sim 2\ \sigma$ off the  CMSSM or NUHM expectation. This is  
substantial but not dramatic. Although the CMSSM or the NUHM are  
unlikely to show up at the LHC, they are still interesting and  
plausible models after the Higgs observation; and, if they are true,  
the chances of discovering them in future dark matter experiments are  
quite high.}

\preprint{IFT-UAM/CSIC-09-59}

\begin{document}

\section{Introduction}

The ATLAS and CMS collaborations have reported \cite{:2012gk, :2012gu} the strong evidence evidence of a Standard Model (SM) -like Higgs boson with $m_h\sim 126$ GeV. According to the most recent analysis presented in the Kyoto conference \cite{ATLAS-CONF-2012-162,CMS-PAS-HIG-12-045}, which combines the $\sqrt{s}=7 \tev$ (L=5.0 ${\rm fb}^{-1}$) and $\sqrt{s}=8$ TeV (L=12.1 ${\rm fb}^{-1}$) data, the ``Higgs-boson" mass is
\bea
\label{mhATLASCMS}
m_h\ &=&\ 126.0\pm 0.4\ {\rm (stat)}\ \pm 0.4\ {\rm (syst)}\ {\rm GeV}\hspace{0.5cm}{\rm (ATLAS)} 
\nonumber\\
m_h\ &=&\ 125.7\pm 0.4\ {\rm (stat)}\ \pm 0.4\ {\rm (syst)}\ {\rm GeV}\hspace{0.5cm}{\rm (CMS)}.
\eea
Adding the statistical and systematic uncertainties in quadrature, and combining further the ATLAS and CMS results, using the principle of maximum likelihood, one finds the following central value  and uncertainty
\bea
\label{mhATLASCMScomb}
m_h\ &=&\ 125.85\pm 0.4.
\eea

Beyond the crucial importance of this discovery by itself, this result has far-reaching consequences for well-motivated candidates for physics beyond the SM, such as Supersymmetry (SUSY), and in particular the Minimal Supersymmetric Standard Model (MSSM). 

Many works have explored these implications from different points of view, often relying on popular versions of the MSSM, such as the constrained MSSM (CMSSM) or the (slightly less restrictive) non-universal Higgs masses model (NUHM). In addition, some analyses incorporate as well the recent XENON100 constraints on fermionic dark matter \cite{Farina:2011bh,Balazs:2012qc,Fowlie:2012im, Akula:2012kk,Buchmueller:2012hv,Arbey:2012dq,Strege:2012bt}. Roughly speaking, the MSSM is in trouble, since a $\sim 125-126$ GeV Higgs generically requires rather large SUSY masses, which is in tension with the naturalness of the electroweak breaking at the correct scale. A big portion of the previously-acceptable CMSSM and NUHM parameter-space becomes excluded.
Some of these explorations present the results in the form of scatter plots, which illustrate the acceptable regions in slides of the parameter space, see e.g. \cite{Arbey:2012dq}.
While these analyses certainly reflect the situation, it is more difficult to tell from them the degree of tension of the CMSSM and the NUHM (or the general MSSM) with the experimental result $m_h\simeq 126$ GeV or what is the relative probability of the various regions of the associated parameter-space. 




There are two main strategies to attack these questions: the frequentist and
the Bayesian approaches. The frequentist one is based on the analysis of the
likelihood function in the parameter space: one performs a scan of the
parameter space, evaluating the likelihood. This leads to zones of estimated probability (inside contours of constant likelihood) around the best fit points. In doing so, the unplotted variables are optimized to obtain the best likelihood. A nice feature of the frequentist approach is that it does not require to establish a prior to analize the parameter space. Essentially, it corresponds to a search of the points with best likelihood, and therefore gives relevant information about the goodness of different regions of the parameter space. However, a frequentist approach does not necessarily give the fair probability-map (i.e. the probability of the parameters for given experimental data) of the parameter space of a model. This is obtained using a Bayesian approach. 
 
As an example of this, let's take the notion of fine-tuning. Fine-tuning has to do with statistical weight and a frequentist analysis is based entirely on likelihood, i.e. the ability to reproduce the experiment, and thus cannot ``see" the fine-tuning. E.g. a point (o a region) in the parameter space can present an optimal likelihood, but only after an extreme tuning of the unplotted parameters, involving cancellations. In the usual point of view, this feature makes this point very implausible or disfavored in the parameter space, since, a priori, cancellations are not likely unless there exists some known theoretical reason for them. However, since the point is capable to reproduce the experimental data, the fine-tuning considerations do not affect its privileged condition in a frequentist approach. On the other hand, a correct Bayesian analysis incorporates automatically (without any ad hoc consideration) a penalization of the fine-tuned regions, since they have little statistical weight.

Therefore the Bayesian approach can offer a representative diagnosis of the relative plausibility of the different regions of the parameter space associated to a given scenario and its future prospects of detection if nature has chosen it.
The main goal of this paper is to perform such Bayesian diagnosis in full detail, incorporating the impact of a possible Higgs at $\simeq 126$ GeV, as well as the last XENON100 dark-matter constraints and other recent experimental data. We will also incorporate significant improvements over previous Bayesian explorations. They will allow us to avoid the arbitrariness associated to the choice of the ranges where the parameters live. We will show that this ambiguity is the origin of important apparent discrepancies between previous literature. From the analysis we obtain the most probable regions in the CMSSM and NUHM parameter space and study the capacity of the LHC and XENON of probing those regions in the future.

Using Bayesian techniques, we will also explore in more general terms how the Higgs discovery affects the naturalness of SUSY, quantifying the tension between $m_h\simeq 126$ GeV and the supersymmetric predictions, thus giving a sound indication of the general health of the model after the Higgs discovery.



In sect. 2 we establish our set up, reviewing the Bayesian approach and how it can be rigorously implemented in the context of the MSSM. We discuss also the choice of priors issue. In sect. 3 we list the observables used in the analysis in order to calculate the likelihood of the different points of the parameter space. We also expose the different codes and computational techniques used in the numerical calculations. Sect. 4 is devoted to the results for the CMSSM. We present them with and without dark matter constraints. We also distinguish between the possibility that the dark matter is purely supersymmetric or it has other components. We also quantify the discrepancy between present experimental data on $g-2$ and the Higgs mass in the supersymmetric framework. This quantifiable tension is the reason why we have chosen not to include $g-2$ data in the analysis until the theoretical situation becomes fully clarified. In sect. 5 we perform a similar analysis for the NUHM and
  present the results. A comparison of the relative evidence between the different models, especially between the CMSSM and the NUHM, is given in sect. 6.  In sect. 7 we study the issue of the naturalness of SUSY after the Higgs discovery, using Bayesian techniques. Sect. 8 is devoted to the comparison with previous literature on this subject and to discuss the reasons of the apparent discrepancies. The conclusions are presented in sect. 9. In addition we describe some meaningful theoretical aspects of the Focus-Point region in the CMSSM and the NUHM.

\section{Set up}


We will consider two versions of the minimal supersymmetric standard model (MSSM), namely the so-called constrained MSSM (CMSSM) and the non-universal Higgs masses model (NUHM). In the CMSSM the soft parameters are assumed universal at a high scale ($M_X$), where the supersymmetry (SUSY) breaking is transmitted to the observable sector, as happens e.g. in the gravity-mediated SUSY breaking scenario. The CMSSM parameter-space is then defined by the following parameters:
\bea
\label{MSSMparameters}
\{\theta_i\}\ =\ \{m_0,m_{1/2},A,B,\mu,s\}.
\eea
Here $m_0$, $m_{1/2}$ and $A$ are the universal scalar mass, gaugino mass and trilinear scalar coupling; $B$ is the bilinear scalar coupling; $\mu$ is the usual Higgs mass term in the superpotential; and $s$ stands for the SM-like parameters of the model, i.e. essentially
gauge and Yukawa couplings (in particular the top one), $s\equiv{g_i, y_t, \cdots}$. All these initial parameters are understood to be defined at $M_X$.
The NUHM is defined by the same parameters as the CMSSM but allowing for a different scalar mass for the two Higgs doublets, $m_H$.

It is usual to trade the $B$ and $\mu$ parameters by
  $\tan\beta = \langle H_u\rangle/ \langle H_d\rangle$, sign$(\mu)$ and the
  requirement of correct radiative electroweak breaking (or, equivalently, the
  correct value of $M_Z$); which is very convenient for phenomenological
  studies. However, from a theoretical point of view $\tan\beta$ is a
  parameter derived from the previous ones (and $M_Z$ is an observable), so it
  is sensible to start with the set of parameters of
  eq.~(\ref{MSSMparameters}).  In subsect. 2.2 we will address the way to consistently pass from a set of parameters to the other within the Bayesian approach.

\subsection{The Bayesian approach}

The goal of the Bayesian approach is to generate a map of the relative probability of the different regions of the parameter space of the model under consideration (CMSSM and NUHM in our case), using all the available (theoretical and experimental) information. This is the so-called {\em posterior} or probability density function (pdf), $p(\theta_i|{\rm data})$, where "data" stands for all the experimental information and $\theta_i$ represent the various parameters of the model. The posterior is given by the Bayes' Theorem
\bea
\label{Bayes}
p(\theta_i|{\rm data})\ =\ p({\rm data}|\theta_i)\ p(\theta_i)\ \frac{1}{p({\rm data})},
\eea
where $p({\rm data}|\theta_i)$ is the likelihood (sometimes denoted by ${\cal L}$), i.e. the probability density of measuring the given data for the chosen point in the parameter space (this is the quantity used in frequentist approaches); $p(\theta_i)$ is the prior, i.e. the ``theoretical" probability density that we assign a priori to the point in the parameter space; and finally, $p({\rm data})$ is a normalization factor which plays no role unless one wishes to compare different classes of models, so for the moment it can be dropped from the previous formula. 
One can say that in eq.~(\ref{Bayes}) the first factor (the likelihood) is objective, while the second (the prior) contains our prejudices about 
how the probability is distributed a priori in the parameter space, given all our previous knowledge about the model. 

Ignoring the prior factor is not necessarily the most reasonable or ``free of prejudices" attitude. Such procedure amounts to an implicit choice for the prior, namely 
a completely flat prior in the parameters. However, one needs some theoretical basis to establish, at least, the parameters whose prior can be reasonably taken as flat.

Besides, note that a choice for the allowed ranges of the various parameters is necessary in order to make statistical statements. Often one is interested in showing the probability density of one (or several) of the initial parameters, say $\theta_i,\ i=1,...,N_1$, but not in the others, $\theta_i,\ i=N_1+1,...,N$.
Then, one has to {\em marginalize} the latter, i.e. integrate in the parameter space:
\bea
\label{marg}
p(\theta_i,\ i=1,...,N_1|{\rm data})\ = \int d\theta_{N_1+1},...,d\theta_N\ p(\theta_i,\ i=1,...,N|{\rm data}).
\eea
This procedure is very useful and common to make predictions about the values of particularly interesting parameters. Now, in order to perform the marginalization, we need an input for the prior functions {\em and} for the range of allowed values of the parameters, which determines the range of the definite integration (\ref{marg}). A choice for these ingredients is therefore inescapable in trying to make Bayesian LHC forecasts.

In consequence, an important indicator of the reliability of the results of a Bayesian analysis is the dependence of the results on the choice of the prior and on the ranges established for the unknown parameters, something that is addressed in the present paper (and very seldom done for the issue of the ranges).

\subsection{Model-independent incorporation of naturalness criteria in the MSSM}

The main motivation of low-energy SUSY is the nice implementation of the electroweak (EW) breaking, since the EW scale (or, equivalently, the $Z$ mass) does not suffer from enormous (quadratic) radiative corrections that occur in the SM. Actually, in the MSSM the EW breaking occurs naturally in a substantial part of the parameter space. This success is greatly due to the SUSY radiative contributions to the Higgs potential. Of course, in our analysis, the points of the parameter space that do not have a correct EW breaking are to be discarded, as usual.

It is common lore that the parameters of the MSSM, $\{m_0,m_{1/2},A,B,\mu\}$, should not be far from the experimental EW scale in order to avoid unnatural fine-tunings to obtain the correct size of the EW breaking. Previous Bayesian studies of the MSSM have attempted to incorporate this criterion by implementing some penalization of the fine-tuned regions or (more usually) by restricting the range of the soft parameters to $\lsim $ few TeV. This is more or less reasonable, but makes the results dependent on the actual ranges considered (although normally such test is not presented).

On the other hand, since the naturalness arguments are deep down statistical arguments, one might expect that an effective penalization of fine-tunings
should arise from the Bayesian analysis itself, with no need of introducing 
``naturalness priors"  or restricting the soft terms to the low-energy scale. It was shown in ref. \cite{Cabrera:2008tj} that this is indeed the case (see also \cite{Strumia:1999fr}
for a previous observation in this sense). 
The key point is the following. Normally, the value of $\mu$ (up to a sign) is deduced from the value of the other parameters by the condition that the Higgs vacuum expectation value is the correct one, or equivalently the $Z-$mass is the experimental one. However, this procedure does not incorporate the fact that such value of $\mu$ may involve an extreme fine-tuning, in which case should become statistically penalized. This shortcoming can be surmounted by simply 
considering $M_Z^{\rm exp}$ as experimental data on a similar foot to the others, entering the total likelihood, ${\cal L}$. For the sake of simplicity let us approximate the likelihood associated to the $Z$ mass as a Dirac delta, so
\bea
\label{likelihood}
p({\rm data}|s, m_0, m_{1/2}, A, B, \mu)\ \simeq\ \delta(M_Z-M_Z^{\rm exp})\ {\cal L}_{\rm rest},
\eea
where ${\cal L}_{\rm rest}$ is the likelihood associated to all the physical observables, except $M_Z$. Now, we can take advantage of this Dirac delta to marginalize the posterior
in one of the initial parameters, e.g. $\mu$, performing a change of variable $\mu\rightarrow M_Z$:
\bea
\label{marg_mu}
p(s, m_0, m_{1/2}, A, B| \ {\rm data} )& = &\int d\mu\ p(s, m_0, m_{1/2}, A, B, \mu | 
{\rm data} )
\nonumber\\ 
&\simeq&\ {\cal L}_{\rm rest}  \left|\frac{d\mu}{d M_Z}\right|_{\mu_Z}
p(s, m_0, m_{1/2}, A, B, \mu_Z),
\eea
where $\mu_Z$ is the value of $\mu$ that reproduces the experimental value
of $M_Z$ for the given values of $\{s, m_0, m_{1/2}, A, B\}$, and $p(s, m_0, m_{1/2}, A, B, \mu)$ is the prior in the initial parameters (still undefined). Note that the Jacobian factor, $\frac{d\mu}{d M_Z}$ can be written as $\frac{2\mu}{M_Z}\ \frac{1}{c_\mu}$, where 
$c_\mu = \left|\frac{\partial \ln M_Z^2}{\partial \ln \mu}\right|$ is the
conventional Barbieri-Giudice measure \cite{Ellis:1986yg, Barbieri:1987fn} of
the degree of fine-tuning. Thus, the above Jacobian factor has built-in the
desired fine-tuning penalization \footnote{If instead the
    Dirac-delta approximation for the $M_Z$ likelihood,
    eq.~(\ref{likelihood}), we had used the true Gaussian
    likelihood ${\cal L}_{M_Z} = \frac{1}{\sqrt{2\pi}\sigma_Z}\
    \exp\left\{(M_Z-M_Z^{\rm exp})^2/2\sigma_Z^2\right\}$, where $\sigma_Z$ is
    the uncertainty in the $Z-$mass, the results would be analogous. In
    particular the integration of eq.~(\ref{marg_mu}) can be
    done using the Laplace's method, getting $p(s, m_0, m_{1/2}, A, B| \ {\rm data} )= 
\ {\cal L}_{\rm rest}  \left|\frac{d\mu}{d M_Z}\right|_{\mu_Z} p(s, m_0,
m_{1/2}, A, B, \mu_Z)\left\{1+{\cal O}(\frac{\sigma_Z}{M_Z})^2\right\} $. Since $\sigma_Z\ll M_Z^{\rm exp}$ the correction is negligible in this case, though this technique could be useful in other physical cases. }, with no ad hoc assumptions.

Actually this is enough to make the high-scale region of the parameter space, say soft terms $\gsim\ {\cal O} (10)\ {\rm TeV}$, statistically insignificant; which allows in turn to consider a wide range for the soft parameters (up to the very $M_X$). In consequence, the results of our analysis are essentially independent on the upper limits of the MSSM parameters, in contrast with previous studies.

Beside the above $\mu\rightarrow M_Z$ change of variable, it is highly advantageous for phenomenological studies to trade the Yukawa couplings by the physical fermion masses (which are easily marginalized out) and the initial $B-$parameter by the derived $\tan\beta$ parameter. Consequently we should compute the whole Jacobian, $J$, of the transformation
\bea
\label{change_3}
\{\mu,y_t,B\}\ \rightarrow\  \{M_Z,m_t,\tan\beta\}.
\eea
The initial prior (still undefined) is therefore always affected by this model-independent Jacobian factor, thus resulting in an {\em effective prior} in the new variables
\bea
\label{eff_prior}
p_{\rm eff}(g_i, m_t, m_0, m_{1/2}, A, \tan\beta)\  \equiv\ 
J|_{\mu=\mu_Z}\  p(g_i, y_t, m_0, m_{1/2}, A, B, \mu=\mu_Z) ,
\eea
where we have already marginalized $M_Z$ using the associated likelihood $\sim \delta(M_Z-M_Z^{\rm exp})$ (recall that $\mu_Z$ is the value of $\mu$ that reproduces the experimental $M_Z$.) In eqs.(\ref{change_3}, \ref{eff_prior}) we have made explicit the dependence just on the top Yukawa coupling and mass, but for other fermions goes the same. In addition, the gauge couplings, $g_i$, are marginalized in the usual way.

So, to prepare the scan in the new variables we need an explicit evaluation of
the Jacobian factor. For the numerical analysis we have evaluated $J$ using
the \texttt{SoftSusy} code \cite{softsusy} which implements the full one-loop
contributions and leading two-loop terms to the tadpoles for the electroweak
breaking conditions with parameters running at two-loops. This essentially
corresponds to the next-to-leading log approximation; see
ref.~\cite{Kastening:1991gv,Ford:1992mv,Bando:1992np,Casas:1994us} for details. 

However, it is possible to give an analytical and quite accurate expression of $J$, and thus of the effective prior, by working with the tree-level potential with parameters running at one-loop (i.e. essentially the leading log approximation). At this level, the relation of $\{\mu,y_t,B\}$ with $\{M_Z,m_t,\tan\beta\}$ is given by the minimization conditions and the expression for the (running) top mass,
\bea
\label{mu}
\mu_{\rm low}^2= \frac{m_{H_d}^2 -\tan^2\beta\ m_{H_u}^2}{\tan^2\beta-1} - \frac{M_Z^2}{2},
\eea
\bea
\label{B}
B_{\rm low}= \frac{s_{2\beta}}{2\mu_{\rm low}}(m_{H_1}^2 + m_{H_2}^2+2\mu_{\rm low}^2),
\eea
\bea
\label{y}
y_{\rm low}=\frac{m_t}{v\ s_\beta},
\eea
with $s_X\equiv \sin(X)$. Here $y$ denotes the top Yukawa coupling and the ``low'' subscript indicates that the quantity is evaluated at low scale (more precisely, at a representative supersymmetric mass, such as the geometric average of the stop masses). The soft masses $m_{H_u}^2, m_{H_d}^2$ are also understood at low scale. For notational simplicity, we have dropped the subscript $t$ from the Yukawa coupling. (Note that all these low-energy quantities contain an implicit dependence on the top Yukawa coupling through the corresponding RG equations.) Then
\bea
\label{J}
J\propto\left[\frac{E}{R_\mu^2}\right]\ 
\frac{y}{y_{\rm low}} \frac{\tan^2\beta -1}{\tan\beta(1+\tan^2\beta)} \frac{B_{\rm low}}{\mu} ,
\eea
where $R_\mu$ and $E$ are defined through the renormalization group equations of $\mu$ and $y$,
\bea
\label{muBLH}
R_\mu\equiv \frac{\mu}{\mu_{\rm low}},\;\;\;\; y_{\rm low}^2\simeq \frac{y^2 E(Q_{\rm low})}{1+(3/4)y^2F(Q_{\rm low})}.
\eea
Here $Q$ is the renormalization scale, $F = \int_{Q_{\rm low}}^{Q_{\rm high}} E \ln Q$, and $E(Q)$ is a definite positive function that depends just on the gauge couplings and that was defined in ref. \cite{Ibanez:1983di}. Similarly, $R_\mu$  is a positive quantity which depends on the top-Yukawa coupling and the gauge couplings. The 1-loop expression for $R_\mu$ can be also found in ref. \cite{Ibanez:1983di}. 

The prefactor in eq.~(\ref{J})  
is just an (irrelevant) numerical factor times $({1}/{\sin \beta})$. The origin of the latter is the fact that in SUSY there is not a one-to-one correspondence between the top mass and the top Yukawa coupling, since the relation (\ref{y}) depends on $\tan\beta$. To see this, let us approximate (just for the sake of the argument) the top-mass likelihood as ${\cal L}_{\rm top}=\delta(m_t-m_t^{\exp})$. Then the marginalization in the top Yukawa-coupling gives $\int dy_t \ p(y_t, m_0, m_{1/2}, ...) {\cal L}_{\rm top}\ \sim (\sin\beta)^{-1} p(y_t^*, m_0, m_{1/2}, ...)$, where $y_t^*$ is the value of the Yukawa coupling reproducing the experimental top mass (which depends on $\tan\beta$). Now, assuming a logarithmically-flat prior for the Yukawa coupling, i.e. $p(y_t)\propto y_t^{-1}\propto \sin\beta$, this $({1}/{\sin \beta})$ factor in the Jacobian (\ref{J}) is cancelled by the Yukawa prior in eq.~(\ref{eff_prior}). This is in fact equivalent in practice to consider $m_{\rm top}$ as an initial parameter, in which case the shape of the associated prior is irrelevant (due to the precision in the experimental top mass). Implicitely this is the usual practice in MSSM Bayesian scans, but we see that if one takes the Yukawa couplings as the true initial parameters, there is a certain dependence on their prior. E.g. taking flat priors for the top and bottom Yukawas, we get a global factor $(\sin\beta \cos\beta)^{-1}$ in the pdf. On the other hand, it does not seem very reasonable to take flat priors for the Yukawas associated to the fermions masses. The fact that they take very assorted orders of magnitude suggests that the (unknown) underlying mechanism may produce Yukawa couplings of different orders with similar efficiency; and this is the meaning of a logarithmic prior for the Yukawas (which we will assume from now on) \footnote{A similar result is obtained postulating flat priors for the Yukawas but with a range of definition of the order of the actual value of each Yukawa coupling, i.e. ${\cal O} (m_f/v)$, where $f$ denotes any fermion. This is very reasonable since otherwise the actual values of e.g. the electron and the muon masses would be extremely unnatural. Then there is a $\sim 1/y_f^*$ normalization factor for the prior of the Yukawas, which also cancels the above-mentioned $(\sin\beta )^{-1}$, $(\cos\beta)^{-1}$ factors.}.

The important point is that the Jacobian (\ref{J}) is a model-independent factor, valid for any MSSM, which must be multiplied by whatever prior is chosen for the initial parameters. It {\em cannot} be ignored. Hence, the effective prior in 
the standard variables $\{m_t, m_0, m_{1/2}, A, \tan\beta\}$ reads
\bea
\displaystyle
\label{approx_eff_prior}
\hspace{-0.7cm} p_{\rm eff}(m_t, m_0, m_{1/2}, A, \tan\beta)\, &\propto& \,   \left[\frac{E}{R_\mu^2}\right]\, 
\frac{y}{y_{\rm low}} \frac{\tan^2\beta-1}{\tan\beta(1+\tan^2\beta)} 
\frac{B_{\rm low}}{\mu_Z}  \ 
\nonumber\\
&\times& \, p(m_0, m_{1/2}, A, B, \mu_Z), \phantom{\left[\frac{E}{R_\mu^2}\right]}
\eea
where all the subjectivity lies in the $p(m_0, m_{1/2}, A, B, \mu)$ piece, i.e. the prior in the initial parameters, for which we have still to make a choice (we will do it in the next subsection). This Jacobian (model-independent) prefactor
contains the above-discussed penalization of fine-tuned regions. Thus, using it automatically incorporates the fine-tuning penalizations with no ad hoc assumptions. Note that, besides the penalization for large $\mu$, the Jacobian factor 
contains a penalization of large $\tan\beta$, reflecting the smaller statistical weight of this possibility. Actually, the implicit fine-tuning associated to a large $\tan\beta$ was already noted in ref.~\cite{Hall:1993gn, Nelson:1993vc}, where it was estimated to be of order $1/\tan\beta$, in agreement with eq.(\ref{approx_eff_prior}).

\subsection{Priors in the initial parameters}

The choice of the prior in the initial parameters, $\{m_0, m_{1/2}, A, B, \mu\}$  must reflect our knowledge about them, before consideration of the experimental data (to be included in the likelihood piece). Four of them, $\{m, M, A, B\}$, are soft SUSY-breaking parameters. They typically go like $\sim F/\Lambda$, where $F$ is the SUSY breaking scale, which corresponds to the dominant VEV among the auxiliary fields in the SUSY breaking sector (it can be an $F-$term or a $D-$term) and $\Lambda$ is the messenger scale, associated to the interactions that transmit the breaking to the observable sector. Since the soft-breaking terms share a common origin it is logical to assume that their sizes are also similar. Of course, there are several contributions to a particular soft term, which depend on the details of the superpotential, the K\"ahler potential and the gauge kinetic function of the complete theory (see e.g. ref. \cite{Kaplunovsky:1993rd}). So, it is reasonable to assume
  that a particular soft term can get any value (with essentially flat probability) of the order of the typical size of the soft terms or below it. The $\mu-$parameter is not a soft term, but a parameter of the superpotential. However, it is desirable that its size is related (e.g. through the Giudice-Masiero mechanism \cite{Giudice:1988yz}) to the SUSY breaking scale. Otherwise, one has to face the so-called $\mu-$problem, i.e. why should be the size of $\mu$ similar to the soft terms', as is required for a correct electroweak breaking [see eq.(\ref{mu})]. Thus, concerning the prior, we can consider $\mu$ on a similar foot to the other soft terms.

Let us denote $M_S$ the typical size of the soft terms in the observable sector, $M_S\sim F/\Lambda$. Then, we define the ranges of variation of the initial parameters as
\bea
\label{rangos_mMABmu}
-qM_S \leq &B&\leq qM_S
\nonumber\\
-qM_S \leq &A&\leq qM_S
\nonumber\\
0 \leq  &m_0&\leq qM_S
\nonumber\\
0 \leq &m_{1/2}&\leq qM_S
\nonumber\\
0 \leq &\mu&\leq qM_S,
\eea
where $q$ is an ${\cal O}(1)$ factor. We have considered here the branch of positive $\mu$. For the negative one we simply replace $\mu\rightarrow -\mu$.
We can take $q=1$ with no loss of generality, provided $M_S$ is allowed to vary in the range $0 \leq M_S\leq \infty$.
In practice, to avoid divergences in the priors, we have to take a finite range for $M_S$, say
\bea
\label{rango_MS}
M_S^0 \leq M_S\leq M_X\,, \;\;\;\; M_S^0 \sim 10\; {\rm GeV}.
\eea
Nevertheless, the values of the upper and lower limits of the $M_S$ range are {\em irrelevant}. 
A small value for $M_S$ (as the above lower limit) is in practice excluded by the LHC exclusion limits. A high value for $M_S$ (as the above upper limit) is in practice statistically irrelevant thanks to the Jacobian prefactor in eq.(\ref{eff_prior}) or eq.(\ref{approx_eff_prior}), i.e. by the naturalness of the EW breaking. So the results of the present approach are certainly independent of the ranges established for the parameters, something very satisfactory.

Concerning the shape of the priors, as already stated, we find reasonable to assume (conveniently normalized) flat priors for the soft parameters inside the ranges (\ref{rangos_mMABmu}), i.e. 
\bea
\label{init_priors}
p(m_0)=p(m_{1/2})=p(\mu)=\frac{1}{M_S}\;,\;\;\;\;p(A)=p(B)=\frac{1}{2M_S}.
\eea
The next step is to choose a prior for $M_S$. Admittedly, this is the less objective part of the statistical analysis, but, as discussed above, one cannot simply ignore the prior; it is an unavoidable piece when one wants to determine the probability distribution in the parameter space. The most conservative attitude is to use two different, though still reasonable, priors, and then compare the results. This gives a fair measure of the prior-dependence.

For that matter we have considered two somehow standard types of prior: flat and logarithmic. In a flat (logarithmic) prior one assumes that, in principle, the typical size (order of magnitude) of the soft terms can be anything, say from $M_S^0$ up to $M_X$, with equal probability. In our opinion, a logarithmic prior is probably the most reasonable option, since it amounts to consider all the possible magnitudes of the SUSY breaking in the observable sector on the same foot (this occurs e.g. in conventional SUSY breaking by gaugino condensation in a hidden sector). However, we will consider log and flat priors at the same level throughout the paper, in order to compare the results and thus evaluate the prior-dependence.

Finally, we can marginalize $M_S$, which thus disappears completely from the subsequent analysis, leaving a prior which depends just on the $\{m_0, m_{1/2}, A, B, \mu\}$ parameters\footnote{This procedure is a ``hierarchical Bayesian technique", first used in ref. \cite{Allanach:2007qk}, but using complicated functions that were not possible to integrate analytically.}. Using eqs.(\ref{init_priors}), one gets, for the logarithmic case, 
\bea
\label{MS_marg}
p(m_0, m_{1/2}, A, B,\mu) &\propto& \int^{M_X}_{\max\{m_0, m_{1/2}, |A|, |B|,\mu, M_S^0\}} 
\frac{1}{M_S^6}\ dM_S
\nonumber\\
&\simeq& \frac{1}{[\max\{m_0, m_{1/2}, |A|, |B|,\mu, M_S^0\}]^5}
\eea
which is the prior to be plugged in eq.(\ref{eff_prior}) (or in the approximated expression (\ref{approx_eff_prior})) to get the effective prior in the scan parameters. For the flat case, the analogous computation gives
\bea
\label{MS_marg_2}
p(m_0, m_{1/2}, A, B,\mu) \propto \ \frac{1}{[\max\{m_0, m_{1/2}, |A|, |B|,\mu, M_S^0\}]^4}.
\eea
The fact that eqs.(\ref{MS_marg}, \ref{MS_marg_2}) correspond, respectively, to logarithmic and
flat priors in the size of the soft terms can be verified by marginalizing all
the soft parameters but one, say $m_{1/2}$, before incorporating any
experimental information. Then one does obtain logarithmic and flat behaviors,
respectively, as is clear from the exponents in the above denominators (for a
more throughout discussion see ref.~\cite{Cabrera:2009dm}).

\section{Physical observables and computational methods used in the analysis}


A first group of physical observables used in the analysis are the SM-like parameters, e.g. masses and couplings of SM particles. Some of them, in particular the top and bottom masses and the electromagnetic and strong coupling constants, have a strong influence in the analysis. We have treated them as nuisance variables (marginalizing them)
using the central values and uncertainties given in Table \ref{tab:nuis_params}.

\begin{table*}
\begin{center}
\begin{tabular}{l l l l }
\hline
\hline
\multicolumn{4}{c}{SM nuisance parameters} \\
\hline
 & Gaussian prior  & Range scanned & ref. \\
\hline
$M_t$ [GeV] & $173.2 \pm 0.9$  & (167.0, 178.2) &  \cite{topmass:1} \\
$m_b(m_b)^{\bar{MS}}$ [GeV] & $4.20\pm 0.07$ & (3.92, 4.48) &  \cite{pdg07}\\
$[\alpha_{em}(M_Z)^{\bar{MS}}]^{-1}$ & $127.955 \pm 0.030$ & (127.835, 128.075) &  \cite{pdg07}\\
$\alpha_s(M_Z)^{\bar{MS}}$ & $0.1176 \pm  0.0020$ &  (0.1096, 0.1256) &  \cite{Hagiwara:2006jt}\\
\hline
\end{tabular}
\end{center}
\caption{\fontsize{9}{9}\selectfont Nuisance parameters adopted in the scan of the CMSSM parameter space, indicating the mean and standard deviation adopted for the Gaussian prior on each of them, as well as the range covered in the scan.}\label{tab:nuis_params} 
\end{table*}

Furthermore, the full set of experimental data used to compute the likelihood function 
is listed in Table \ref{tab:exp_constraints}. The full likelihood function is the product of the individual likelihoods associated to each piece of experimental data. As discussed in the next section, not all those observables are included in the computation of the full likelihood all the time. In particular, we plug and unplug the dark matter constraints to check its impact.
For the quantities for which positive measurements have been made (as
listed in the upper part of Table~\ref{tab:exp_constraints}), we assume a
Gaussian likelihood function with a variance given by the sum of the
theoretical and experimental variances, as motivated by eq.~(3.3) in
ref.~\cite{deAustri:2006pe}. For the observables for which only lower or 
upper limits are available (as listed in the bottom part of
Table~\ref{tab:exp_constraints}) we use a smoothed-out version of the
likelihood function that accounts for the theoretical error in the
computation of the observable, see eq.~(3.5) and fig.~1 in
ref.~\cite{deAustri:2006pe}. 

\begin{table*}
\begin{center}
\begin{tabular}{|l | l l l | l|}
\hline
\hline
Observable & Mean value & \multicolumn{2}{c|}{Uncertainties} & ref. \\
 &   $\mu$      & ${\sigma}$ (exper.)  & $\tau$ (theor.) & \\\hline
$M_W$ [GeV] & 80.399 & 0.023 & 0.015 & \cite{lepwwg} \\
$\sin^2\theta_{eff}$ & 0.23153 & 0.00016 & 0.00015 & \cite{lepwwg} \\
$\brbsgamma \times 10^4$ & 3.55 & 0.26 & 0.30 & \cite{hfag}\\
$R_{\Delta M_{B_s}}$ & 1.04 & 0.11 & - & \cite{Aaij:2011qx} \\
$\RBtaunu$   &  1.63  & 0.54  & - & \cite{hfag}  \\
$\DeltaO  \times 10^{2}$   &  3.1 & 2.3  & - & \cite{Aubert:2008af}  \\
$\RBDtaunuBDenu \times 10^{2}$ & 41.6 & 12.8 & 3.5  & \cite{Aubert:2007dsa}  \\
$\Rl$ & 0.999 & 0.007 & -  &  \cite{Antonelli:2008jg}  \\
$\Dstaunu \times 10^{2}$ & 5.38 & 0.32 & 0.2  & \cite{hfag}  \\
$\Dsmunu  \times 10^{3}$ & 5.81 & 0.43 & 0.2  & \cite{hfag}  \\
$\Dmunu \times 10^{4}$  & 3.82  & 0.33 & 0.2  & \cite{hfag} \\
$\brbsmumu$ & $3.2 \times 10^{-9}$  & $1.5\times 10^{-9}$ & 10\% & \cite{:2012ct}\\
$\Omega_\chi h^2$ & 0.1109 & 0.0056 & 0.012 & \cite{wmap} \\
$\mhl$ [GeV] & 125.85 & 0.4 & 2 & \cite{CMS} \\
\hline\hline
   &  Limit (95\%~$\cl$)  & \multicolumn{2}{r|}{$\tau$ (theor.)} & ref. \\ \hline
Sparticle masses  &  \multicolumn{3}{c|}{As in table~4 of
  ref.~\cite{deAustri:2006pe}.}  & \\
$m_0, m_{1/2}$ & \multicolumn{3}{c|}{LHC exclusion limits, see text} & \cite{LHCSUSY} \\
$m_A, tan \beta$ & \multicolumn{3}{c|}{LHC exclusion limits, see text} & \cite{LHCSUSYNUHM} \\
$m_\chi - \sigmaSI$ & \multicolumn{3}{c|}{XENON100 2012 limits (224.6 $\times$ 34 Kg days)} &\cite{Aprile:2012nq} \\
\hline
\end{tabular}
\end{center}
\caption{\fontsize{9}{9} \selectfont Summary of the observables used
for the computation of the likelihood function
For each quantity we use a
likelihood function with mean $\mu$ and standard deviation $s =
\sqrt{\sigma^2+ \tau^2}$, where $\sigma$ is the experimental
uncertainty and $\tau$ represents our estimate of the theoretical
uncertainty. Lower part: Observables for which only limits currently
exist. The explicit form of the likelihood function is given in
ref.~\cite{deAustri:2006pe}, including in particular a smearing out of
experimental errors and limits to include an appropriate theoretical
uncertainty in the observables. \label{tab:exp_constraints}}
\end{table*}


Let us comment  briefly on some of the observables and the theoretical method used to compute the theoretical prediction for them and the corresponding likelihood. 

To calculate the MSSM spectrum we have used \texttt{SoftSusy} \cite{softsusy}, where SUSY masses are computed at full one-loop level and the Higgs 
sector includes two-loop leading corrections \cite{Dedes:2003km}. We discard 
points suffering from unphysicalities: no self-consistent solutions to the 
RGEs, no EW breaking, tachyonic states or charged lightest-supersymmetric-particle. In our treatment of the radiative corrections to
the electroweak observables $M_W$ and $\sineff$ we include full two-loop and 
known higher order SM corrections as computed in 
ref.~\cite{Awramik:2003rn, Awramik:2004ge}, as well as
gluonic two-loop MSSM corrections obtained in~\cite{dghhjw97}.

The most important piece of experimental information is by far the Higgs mass, $m_h$, for which we use the value obtained from the combination of the ATLAS and CMS experimental results, expressed in eq.(\ref{mhATLASCMScomb}). At each point of the supersymmetric parameter-space we evaluate the theoretical prediction for $m_h$ through \texttt{SoftSusy}, which has the two-loop computation incorporated. The estimated theoretical uncertainty is $\pm 2$ GeV. 

Concerning B-physics, the branching ratio for the 
$B \rightarrow X_s \gamma$ decay (the most important one) has been computed with the numerical code 
\texttt{SusyBSG} \cite{Degrassi:2007kj} using the full NLO QCD contributions, 
including the two-loop calculation of the gluino contributions presented 
in \cite{Degrassi:2006eh} and the results of \cite{D'Ambrosio:2002ex} for the 
remaining non-QCD 
$\tan\beta$-enhanced contributions. The supersymmetric contributions to $b\rightarrow s \gamma$ grow with decreasing masses of the supersymmetric particles {\em and} with increasing $\tan\beta$. 
The other B(D)-physics observables summarized in Table \ref{tab:exp_constraints}
have been computed with the code \texttt{SuperIso} (for details
on the computation of the observables see \cite{Mahmoudi:2008tp} and references 
therein). Among them is the $\brbsmumu$ for which we include the new 
LHCb measurement   $(3.2^{+1.5}_{-1.2}) \times 10^{-9}$, derived from
a combined analysis of 1 fb$^{-1}$ data at $\sqrt{s} = 7$ TeV
collision energy and 1.1 fb$^{-1}$ data at $\sqrt{s} = 8$ TeV
collision energy \cite{:2012ct}. We implement this 
constraint as a Gaussian distribution with a conservative experimental
error of $\sigma = 1.5 \times 10^{-9}$, and a $10\%$ theoretical error. 
Both codes have been integrated into the forthcoming \texttt{SuperBayes-v2.0} 
code \cite{superbayes}. 

We also include recent data from the XENON100 direct detection experiment, 
obtained with an effective volume of $34$ kg and an exposure of $224.6$ days 
between February 2011 and March 2012 ~\cite{Aprile:2012nq}. Two candidate scattering 
events were detected within the signal region, with an 
expected number background of $b = 1.0 \pm 0.2$ events, resulting in  
exclusion limits in the $(m_{\neut},\sigmaSI)$ plane. The corresponding likelihood 
function is described in detail in ref.~\cite{Strege:2012bt}, which we 
refer the reader to for further details.

In addition, we incorporate the most recent constraints from SUSY searches at the 
LHC. We follow the analysis presented by the ATLAS collaboration in 
ref.~\cite{Aad:2011hh}, since it is the one associated to the more stringent 
constraints: a data sample corresponding to a luminosity of 5.8 fb$^{-1}$ for a 
center-of-mass energy of $\sqrt{s}=8$ TeV in the 0-lepton channel.  
As no significant deviations have been so far found with respect to SM predictions, this
has been used to derive the constraints on the ($m_0,m_{1/2}$) plane
\cite{LHCSUSY}. We note that while 
this 95\% exclusion line has been obtained assuming values of 
$\tan\beta = 10$ and $A_0 = 0$ GeV, the final result is not expected 
to change much for different values of these two parameters since the
production and decay of squarks and gluinos are quite insensitive to the values of 
$\tan\beta$ and $A_0$~\cite{Allanach:2011ut} . Besides, we have
applied this limit to the NUHM since the SUSY signal for gluino and
squarks production does not depend significantly on other parameters
than $m_0$ and $m_{1/2}$.  Additionally we have included  exclusion
limit in the ($m_A$,$\tan \beta$) plane from a CMS search for the
decay of neutral Higgs bosons into final states containing two muons
and missing $E_T$, based on 4.5 fb$^{-1}$ integrated luminosity of
data collected at $\sqrt{s} = 7$ TeV collision energy \cite{LHCSUSYNUHM}. 
We include those exclusion limits by
assigning a likelihood equal to zero to samples falling below the exclusion 
line.

As a scanning algorithm we adopt the MultiNest \cite{Feroz:2007kg} algorithm
as implemented in \texttt{SuperBayeS-v2.0}. It is based
on the framework of Nested Sampling, invented by Skilling
\cite{SkillingNS,Skilling:2006}. MultiNest has been developed in such a way as
to be an extremely efficient sampler even for likelihood functions defined
over a parameter space of large dimensionality with a very complex structure
as it is the case of the CMSSM. The main purpose of the Multinest is the
computation of the Bayesian evidence and its uncertainty but it produces
posterior inferences as a by--product. For the marginalization procedure we
have used the above-discussed ranges for our priors.  Besides, we have
considered $2 < \tan\beta < 62$, although the precise limits of this range are
irrelevant. A too low value of $\tan\beta$ is excluded by the Higgs mass
value, unless the supersymmetric masses are extremely high, which is
drastically disfavored by the naturalness of the EW breaking through the
above-discussed Jacobian factor. (More precisely, $\tan\beta=2$ requires stop
masses above $10^3$ TeV \cite{Cabrera:2011bi,Giudice:2011cg}). On the other hand, the
region of very large $\tan\beta$ is also strongly suppressed by the EW
mechanism, as is clear by the prefactor in eq.(\ref{approx_eff_prior}).

\section{Results for the CMSSM}

We have organized the presentation of the results in the following way. 

In subsect. 4.1 we show the plots of the probability distribution function (in one and two dimensions) using only the most robust group of experimental data, namely LHC exclusion limits, Higgs mass, and EW- and B(D)-physics observables. At this stage we leave aside dark matter (DM) constraints. Certainly supersymmetry offers a good candidate for DM, namely the lightest supersymmetric particle (LSP), usually a neutralino. One may assume that  the LSP makes up the whole content of DM or that, due to an efficient annihilation at early times, the LSP component of the DM is subdominant. Both scenarios are interesting and are considered later, in subsect. 4.2. However, there is still the possibility that the early cosmology is not ``standard", in the sense that there are mechanisms that dilute the potential supersymmetric DM. For example, it might happen that 
the early thermal production of LSPs is diluted by 
electroweak inflation. Admittedly, the latter is not a most natural or 
popular scenario of inflation, but mechanisms for it have been explored 
\cite{Knox:1992iy}. Alternatively, the LSP could be
unstable assuming tiny violations of R-parity, see e.g. \cite{Ibarra:2008jk}.
In these cases the observed dark matter should be
  provided by other candidates,  e.g. axions; and the experimental constraints on DM would be irrelevant for supersymmetry. This is the reason why we, conservatively, start in subsect. 4.1 not considering DM constraints in the analysis.
In subsect. 4.2 we include the dark matter constraints in two different fashions: either requiring that the production of supersymmetric dark matter is
consistent with the actual dark matter observed abundance or less than it. Note that the latter, more conservative, case could be realistic if there is an additional source of dark matter (such as axions). In subsect. 4.3 we discuss the g-2 issue. Along this section we present the results just for the CMSSM scenario. The results for the NUHM one are presented in section 6 following a similar scheme.

\subsection{Results with no dark matter constraints}

At this stage we have used all the observables, except those related to dark matter. 

\begin{figure}[t]
\begin{center}
\includegraphics[angle=0,width=0.35\linewidth]{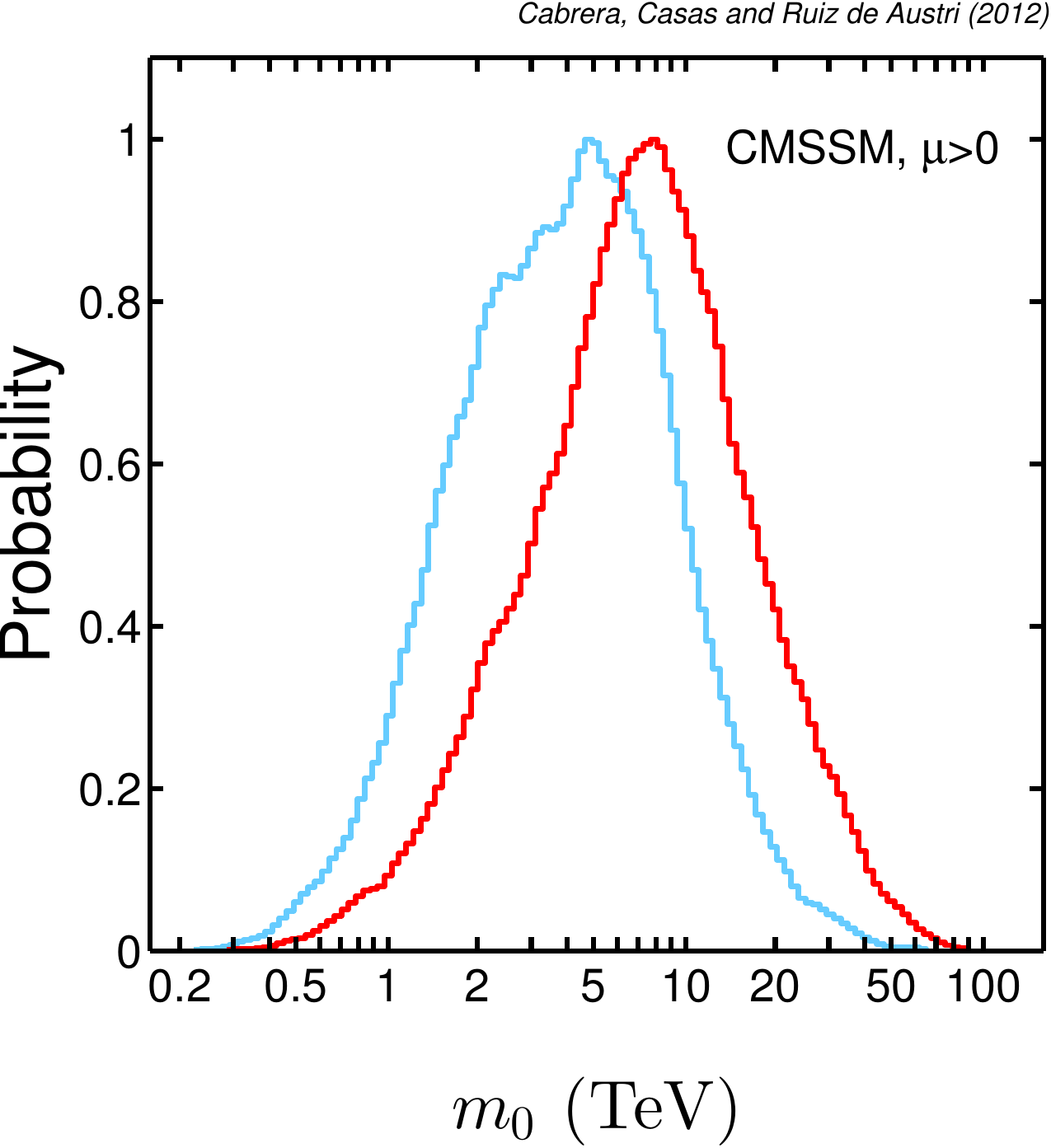} \hspace{1.2cm}%
\includegraphics[angle=0,width=0.35\linewidth]{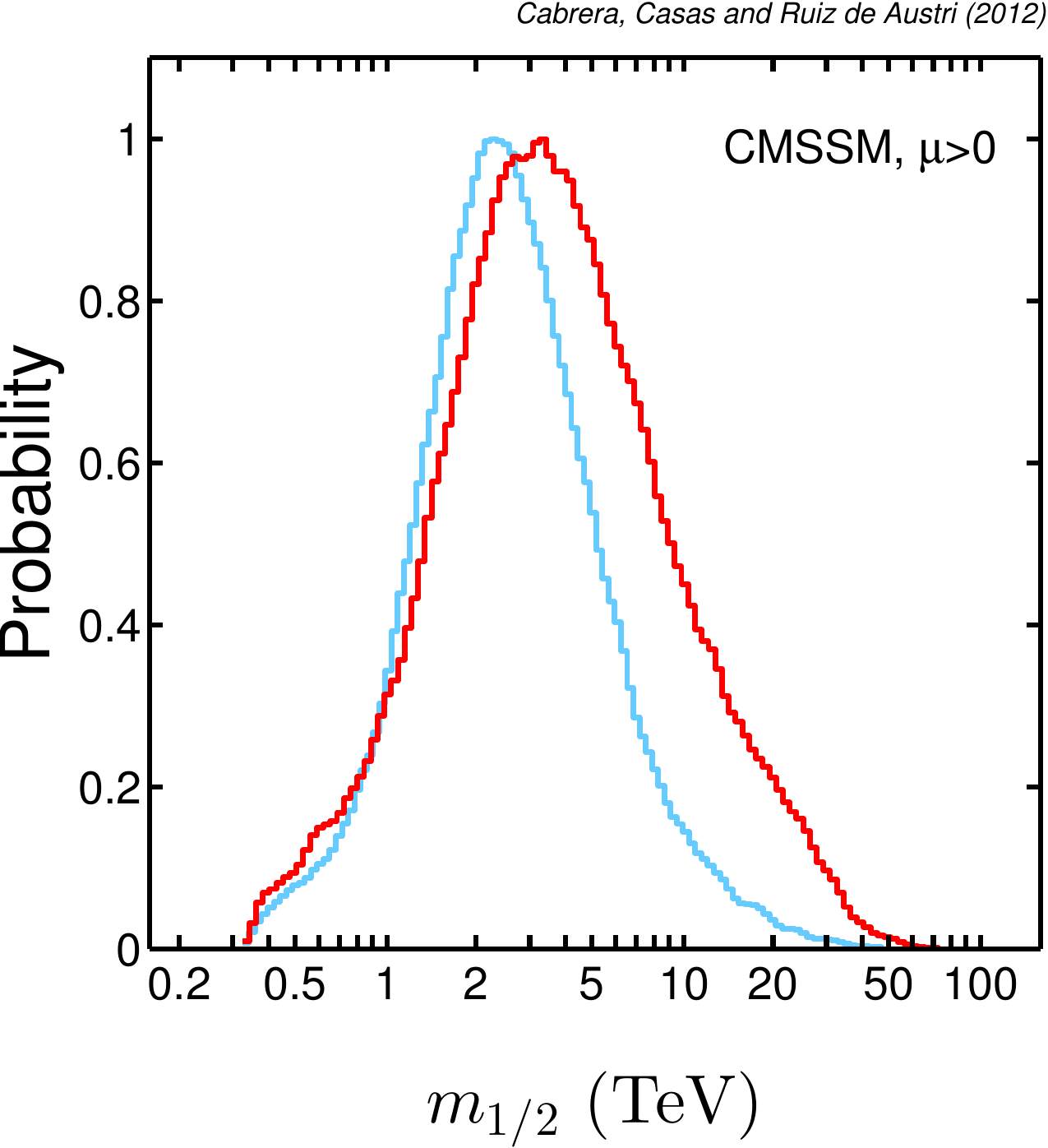} \\ \vspace{0.4cm}
\includegraphics[angle=0,width=0.35\linewidth]{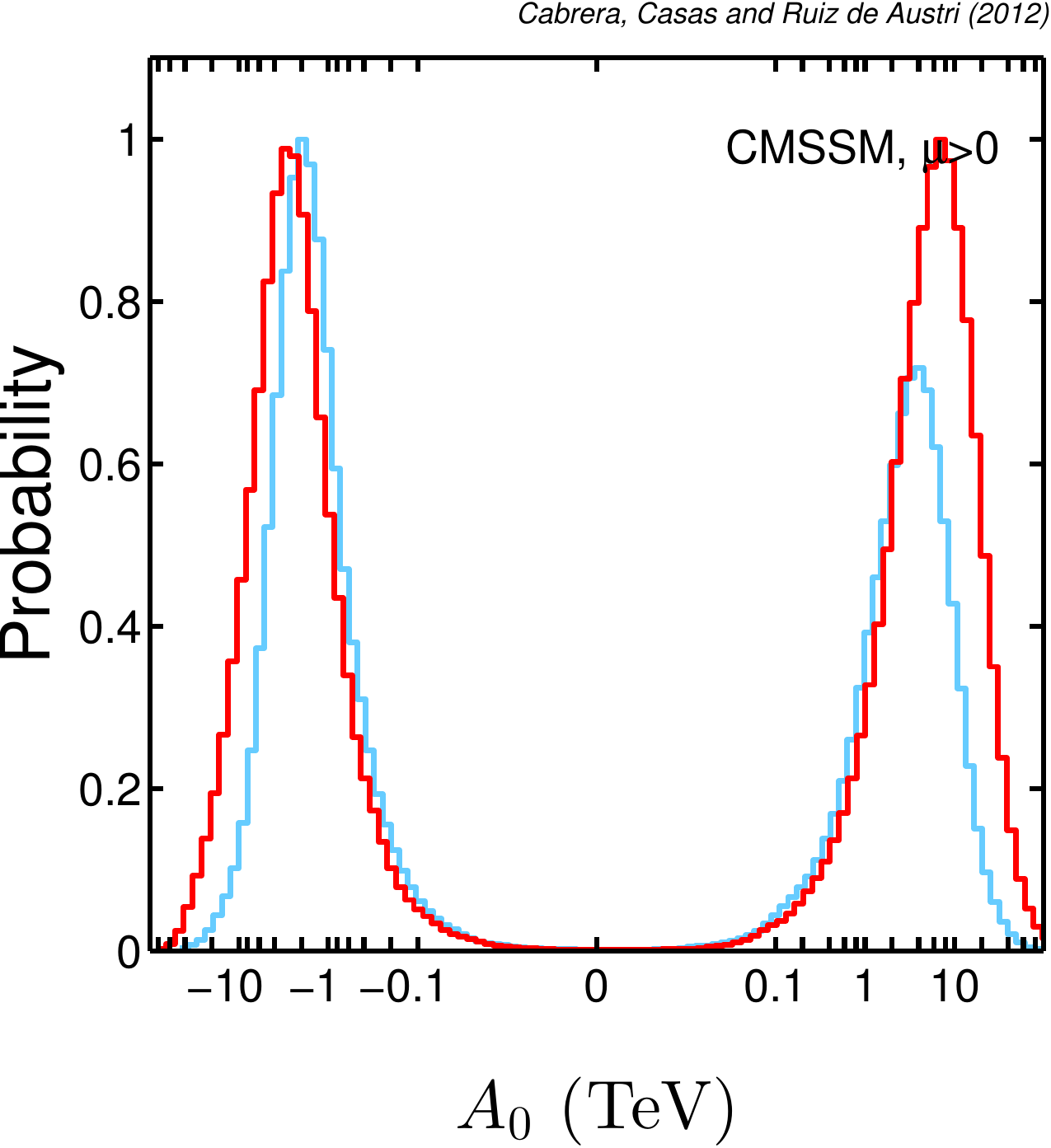} \hspace{1.2cm}
\includegraphics[angle=0,width=0.35\linewidth]{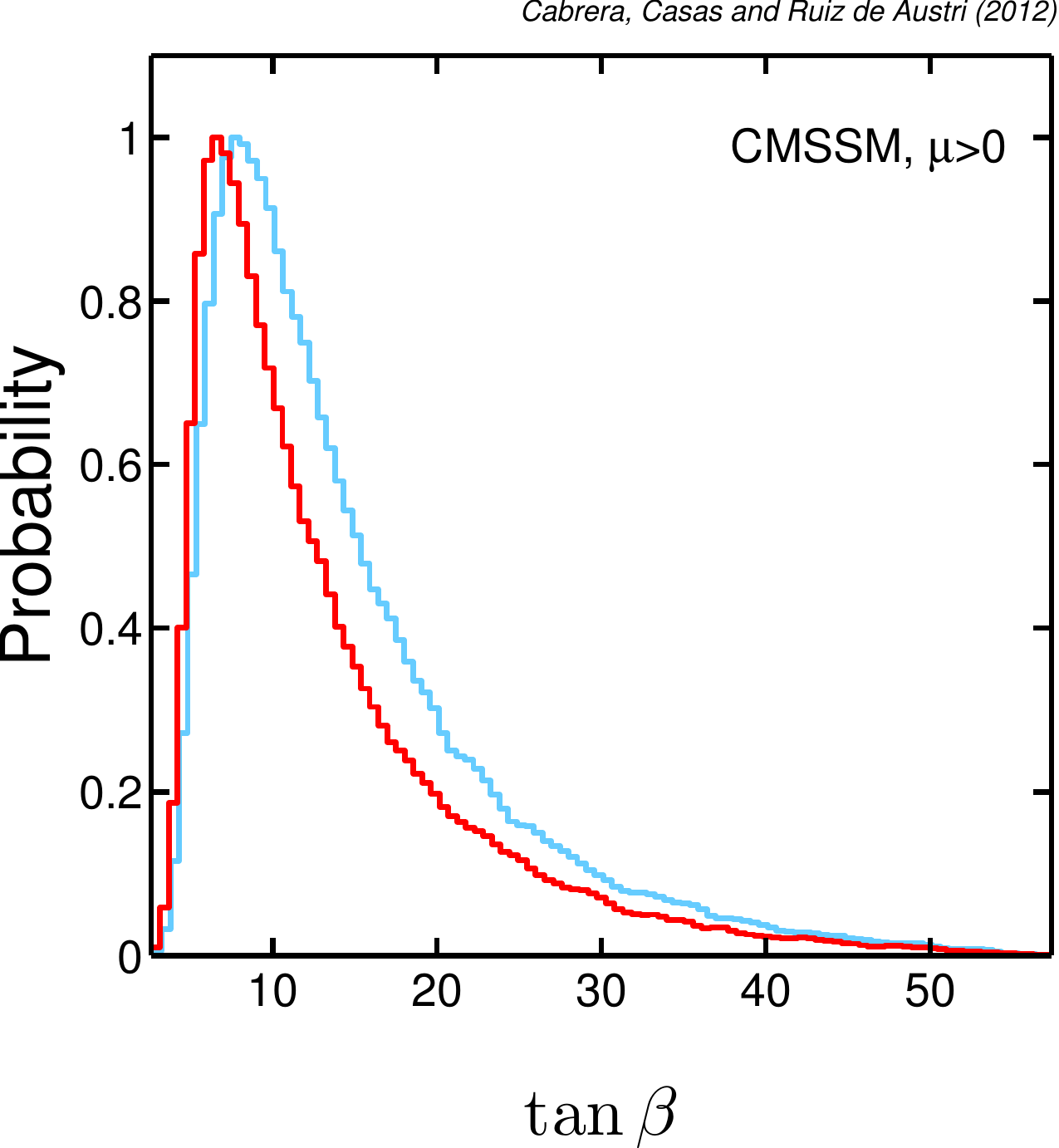} \\ \vspace{0.4cm}
\includegraphics[angle=0,width=0.35\linewidth]{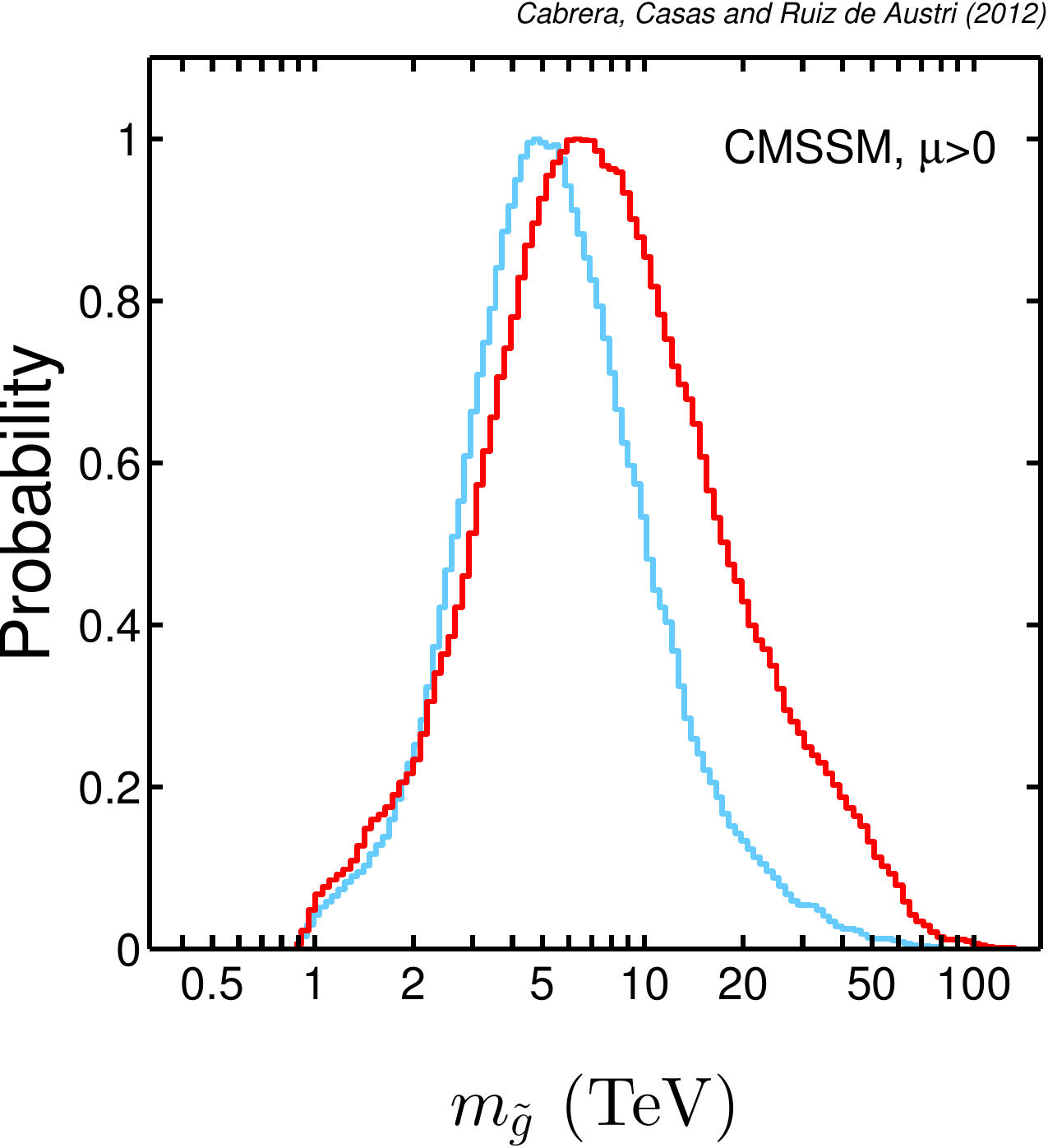} \hspace{1.2cm}%
\includegraphics[angle=0,width=0.35\linewidth]{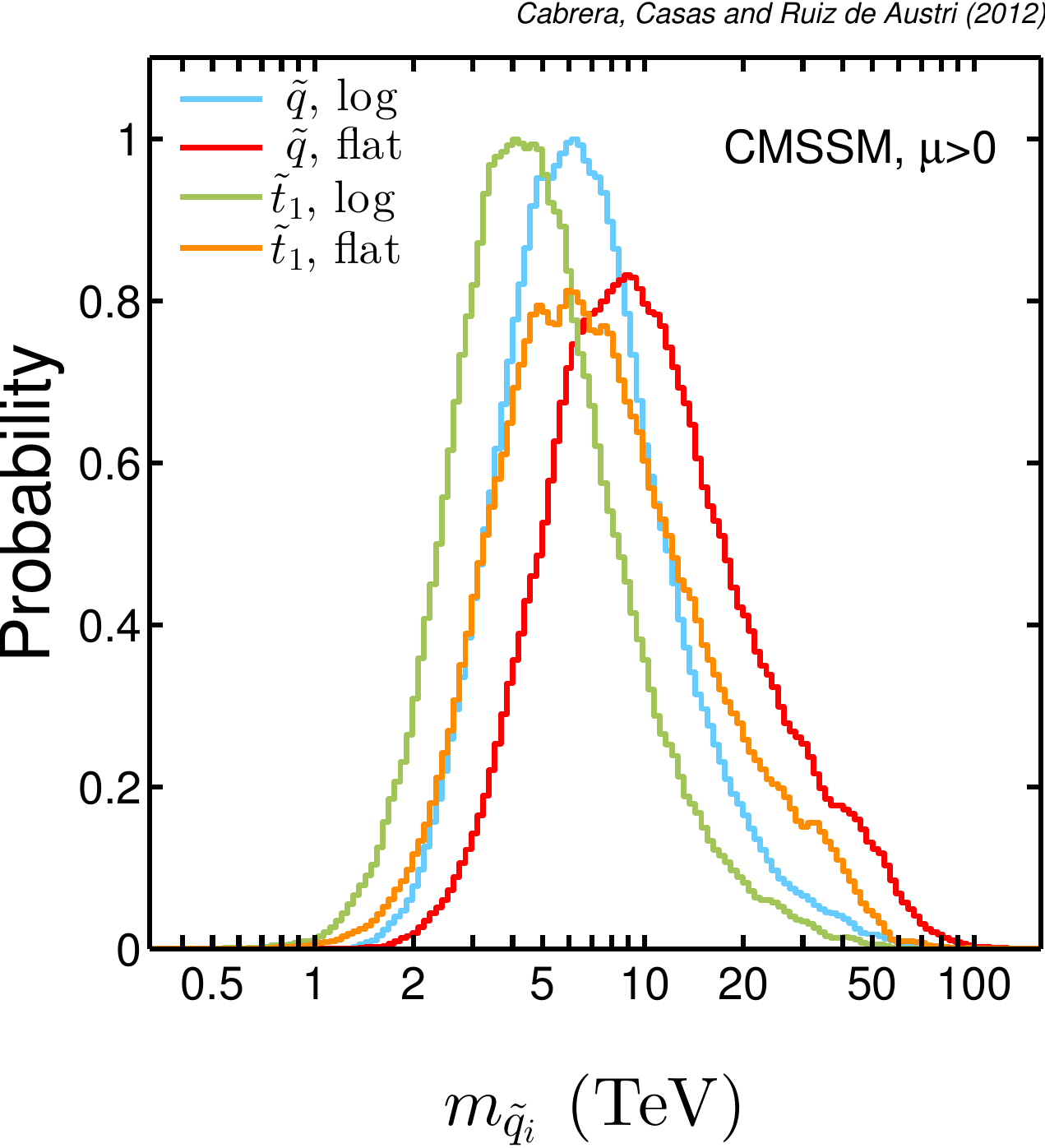}
\caption[test]{1D marginalized posterior probability distribution of the CMSSM parameters: $m_0,m_{1/2},A, \tan \beta$, and the physical gluino and  squark masses (the first two generations and the stops). The red(cyan) [green(orange) for stops] corresponds to flat(logarithmic) priors. \label{fig:cmssm_all_nodm_1d}}
\end{center}
\end{figure}

In fig.~\ref{fig:cmssm_all_nodm_1d} we present the probability distribution function (posterior) for the four CMSSM parameters, $m_0, m_{1/2}, A, \tan\beta$, and the physical gluino and (the average of first and second generation) squark masses. For each case, the non-plotted parameters have been marginalized. The red(cyan) lines show the results for flat(logarithmic) priors. Stop masses are also represented in a different color.
All the plots in fig.~\ref{fig:cmssm_all_nodm_1d} correspond to positive $\mu$. The results for negative $\mu$ are almost identical, so we do not show them to avoid proliferation of figures. 

\begin{figure}[t]
\begin{center}
\includegraphics[angle=0,width=0.35\linewidth]{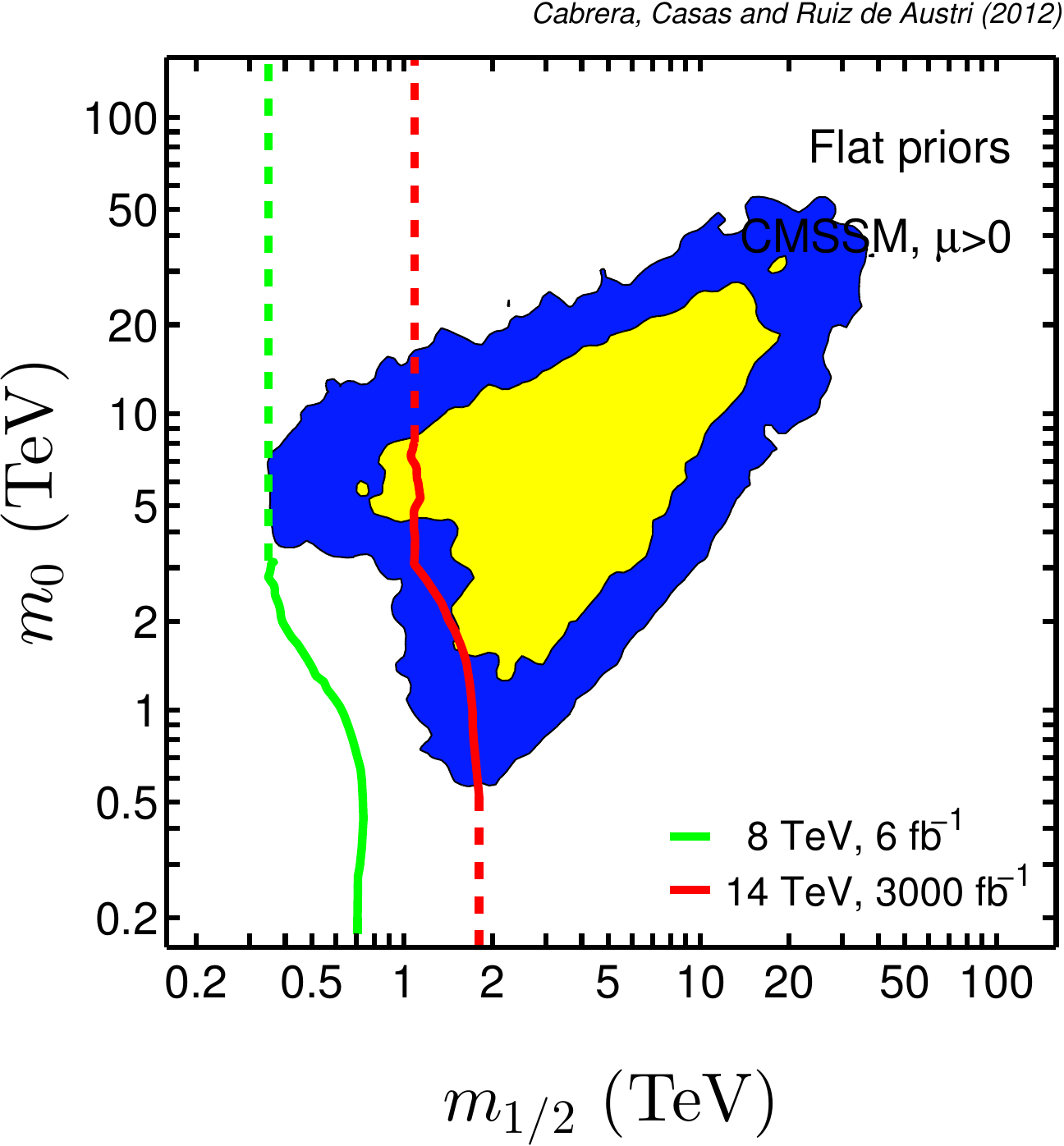} \hspace{1.2cm}
\includegraphics[angle=0,width=0.35\linewidth]{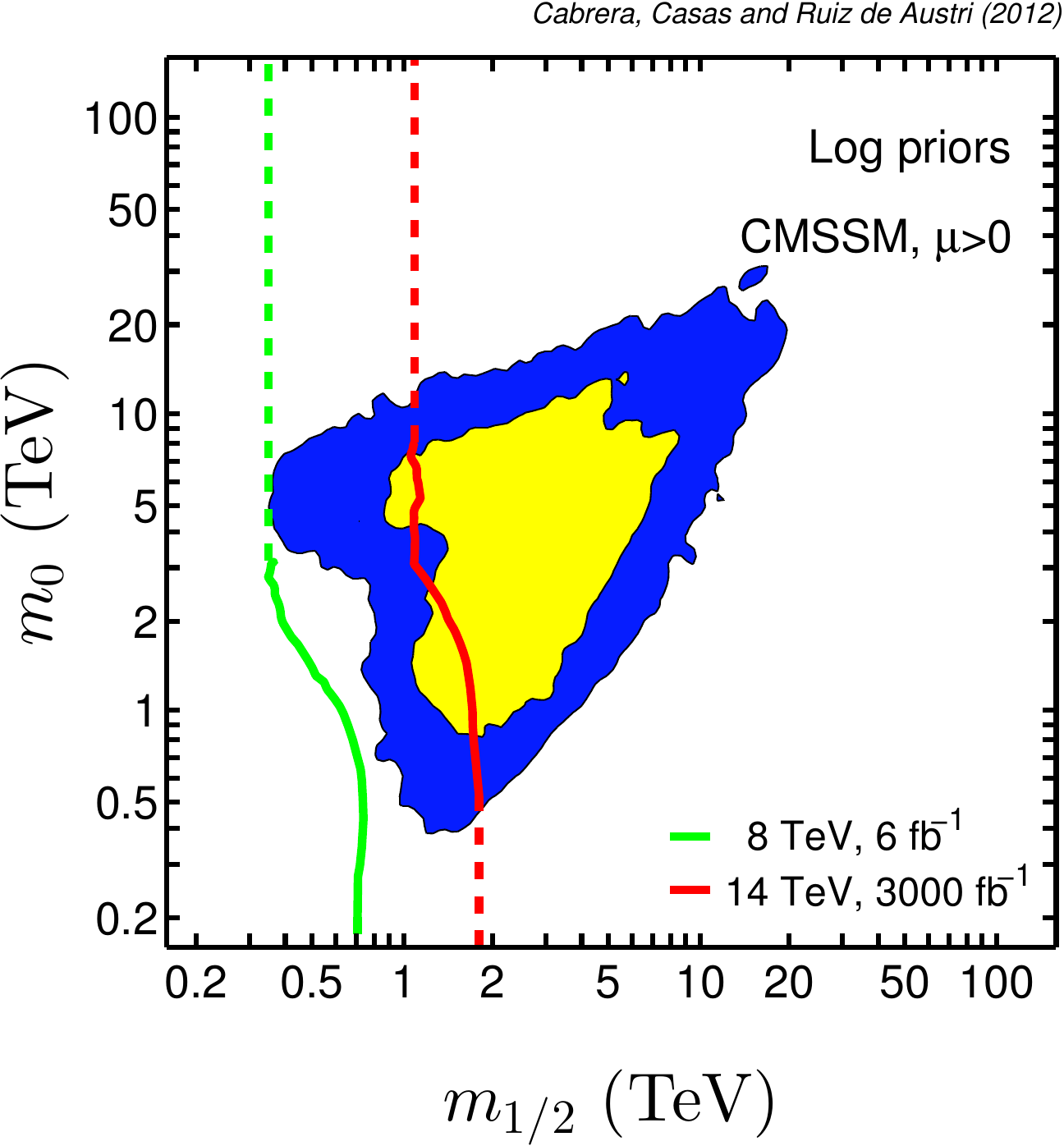}\\ \vspace{0.5cm} 
\includegraphics[angle=0,width=0.35\linewidth]{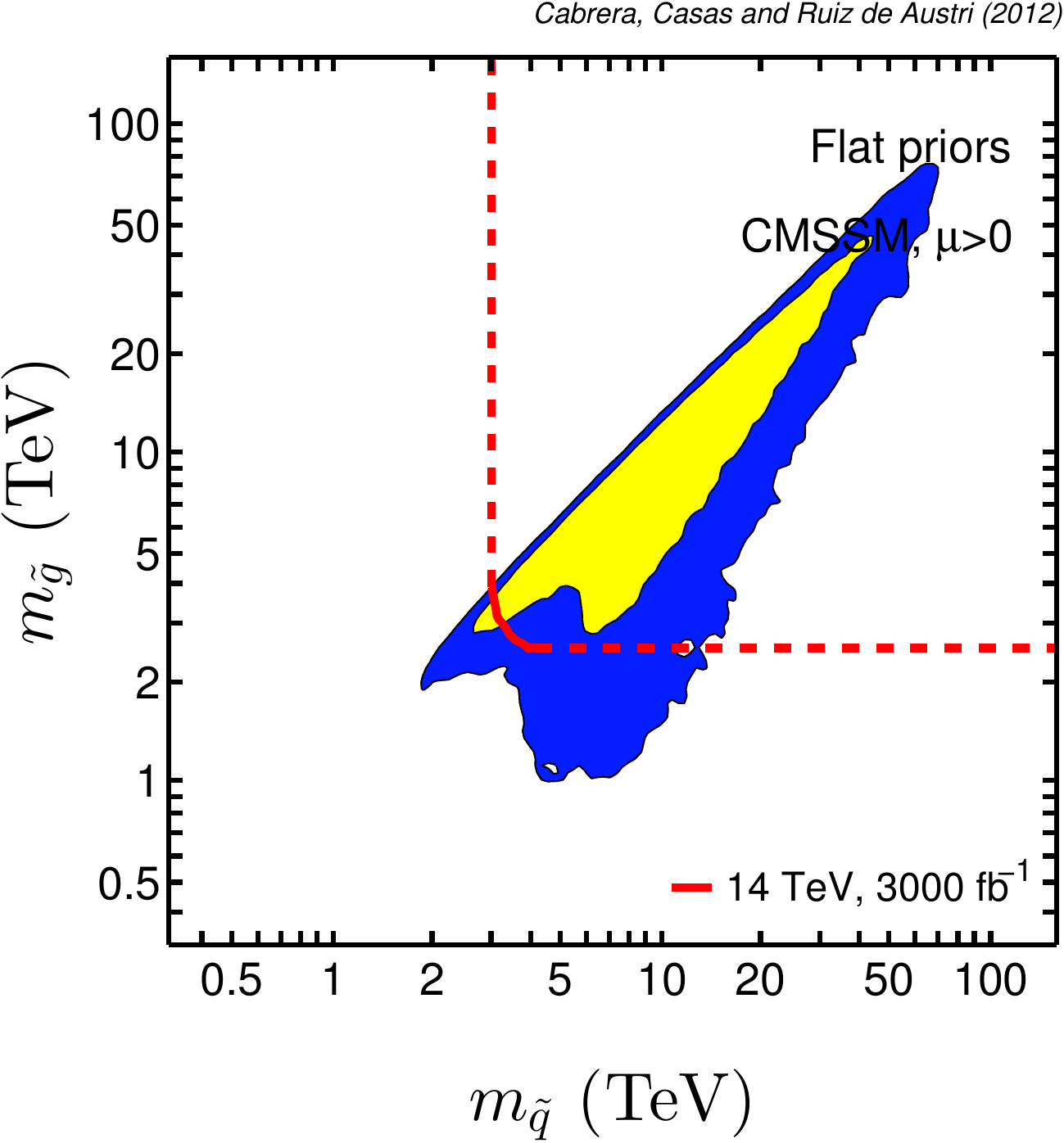} \hspace{1.2cm} 
\includegraphics[angle=0,width=0.35\linewidth]{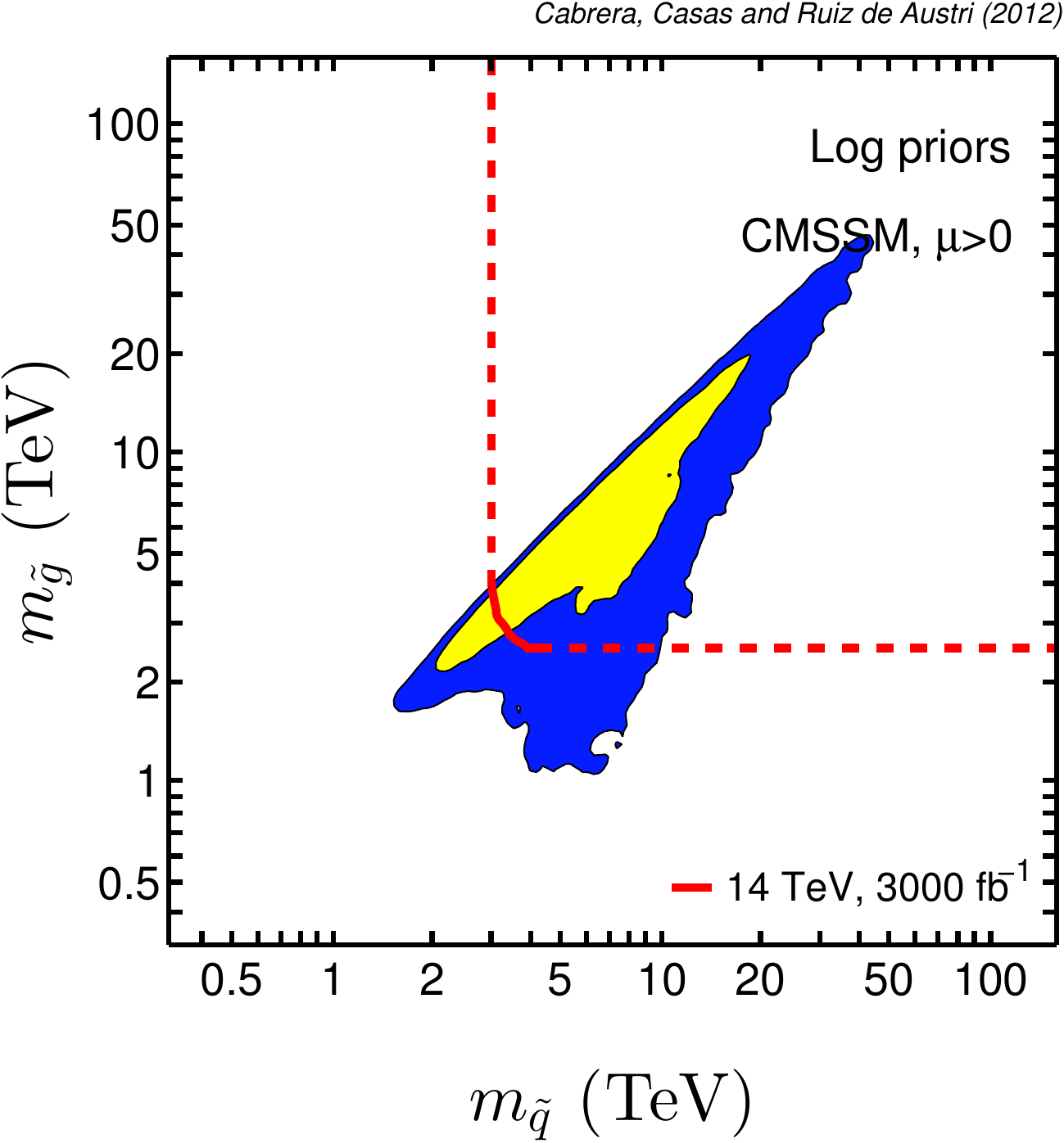}
\caption[test]{2D marginalized posterior probability distribution of the CMSSM in the $m_{1/2}-m_0$ and the $m_{\tilde g}-m_{\tilde q}$ planes.
Left panels show the flat priors case and right panels the logarithmic one. 
Both for $\mu>0$.
The inner and outer contours enclose respective 68\% and 95\% joint regions.
\label{fig:cmssm_all_nodm_2d}}
\end{center}
\end{figure}

Fig.~\ref{fig:cmssm_all_nodm_2d} shows the probability distribution functions for flat and log priors when all parameters except two are marginalized. This gives rise to the two-dimensional plots in the $m_0-m_{1/2}$ and $m_{\tilde q}-m_{\tilde g}$ planes. The present LHC limits (at 5.8 fb$^{-1}$)\cite{LHCSUSY} and the expected ones in the future (3000 fb$^{-1}$)\cite{HL_LHC} are shown for reference.

Clearly, the bulk of the probability lies in the high-energy region which is not directly accessible to the LHC at present. 
The rather high Higgs mass reported by ATLAS and CMS is the main responsible
of this effect. As it is well-known, in the MSSM the tree-level Higgs-mass in
the MSSM is bounded by the mass of the $Z$-boson, so large radiative
corrections are needed in order to reconcile theory and experiment. An
approximate analytic formula for $m_h$ \cite{Ellis:1990nz, Ellis:1991zd,
    Okada:1990vk, Okada:1990gg, Haber:1990aw, Barbieri:1990ja, Casas:1994us,
    Carena:1995bx}, useful for discussions, reads
\bea
\label{mhaprox}
m_h^2\simeq M_Z^2\cos^2 2\beta + \frac{3}{4\pi^2}\frac{m_t^4}{v^2}\left[\log\frac{M^2_{\rm SUSY} }{m^2_t} +\frac{X_t^2}{M^2_{\rm SUSY}} \left (1-
 \frac{X_t^2}{12 M^2_{\rm SUSY}}
 \right)
 \right]+\cdots
\label{mass}
\eea
where $m_t$ is the top running mass, $M_{\rm SUSY}$ represents a certain average of the stop masses and $X_t=A_t-\mu\cot\beta$ \cite{Haber:1990aw,Casas:1994us,Carena:1995bx}. The first term of (\ref{mhaprox}) is the tree-level Higgs-mass and the second are the dominant radiative and threshold contributions. Note that the radiative corrections grow logarithmically with the stop masses while the threshold correction has a maximum for $X_t = \pm \sqrt{6}M_{\rm SUSY}$.

The important point is that, in order to achieve $m_h=125-126$ GeV, one typically needs (besides non-small $\tan\beta$) stop masses $\gsim 3$ TeV, unless $X_t$ happens to be close to the above-mentioned maximizing value (we will comment on the latter possibility below). This is what the probability distributions of Figs.~\ref{fig:cmssm_all_nodm_1d}, \ref{fig:cmssm_all_nodm_2d} show. It should be kept in mind that in the CMSSM all squark masses come from a universal value at $M_X$, so it is not possible to have large stop masses (required to reproduce the value of $m_h$) and lighter masses for the other squarks. In a general MSSM Figs.~\ref{fig:cmssm_all_nodm_1d}, \ref{fig:cmssm_all_nodm_2d} would approximately represent the constraints on the third-generation squark masses and trilinear coupling, and on the gluino mass (at $M_X$ and at low-energy). 

Of course, these kinds of results are known and expected; but the present Bayesian analysis allows to quantify the effect, showing the new probability distribution functions of the various parameters.

The results of Figs.~\ref{fig:cmssm_all_nodm_1d}, \ref{fig:cmssm_all_nodm_2d} show clearly that, with the present information, the probability of detecting the CMSSM at present at the LHC (even if it is the true model) are $\lsim 5\%$, at least with the current techniques to search for physics beyond the SM in LHC data. In the long-term future the probability is substantially higher, but still the chances of {\em not} discovering it are larger. 
To put this in optimistic/wishful-thinking terms, since $m_h\simeq 126$ GeV, if one believes in the CMSSM one should not be worried about the no-detection of SUSY at LHC up to now! But its detection in the future will be challenging too. This is the main conclusion of the Bayesian analysis at this stage. 

On the other hand, it should be kept in mind that in a Bayesian analysis one assumes that the scenario studied (in this case the CMSSM) is the true one, and then one determines the probability distribution in the parameter space. It does not give a direct measure of how likely is the scenario itself, unless one compares it with an alternative scenario using evidence-comparison techniques. We will come back to this point in the discussion of sect. 6 below. But it is already clear that the experimental results on the Higgs mass have pushed the probability to a region which was statistically disfavored (fine-tuned) before the discovery, which casts a shadow on the naturalness of the CMSSM. In sect. 6 we will make this point more quantitative.


Let us comment now on other interesting aspects of the results shown in Figs.~\ref{fig:cmssm_all_nodm_1d}, \ref{fig:cmssm_all_nodm_2d}. As already mentioned, the results for negative $\mu$ are almost the same. The experimental Higgs mass requires heavy supersymmetric particles, which makes SUSY decoupled for many purposes and makes all the sign $\mu$ effects in observables (e.g. electroweak and $B-$physics observables) rather irrelevant. An exception to this would be the $g-2$ data, which have not been included in the analysis for the reasons explained in subsect. 4.3. On the other hand it is quite remarkable the stability of the results against changes in the priors. Although the log priors prefer logically slightly lower values of the supersymmetric masses, the difference is not significant. This is quite satisfactory, showing that the experimental data are already able to determine pretty unambiguously the preferred supersymmetric regions in the parameter space. Still, however, the acceptable ranges for the supersymmetric parameters are quite wide, but systematically beyond 1 TeV. 

The probability distribution for $\tan\beta$, with a maximum around $\tan\beta \sim 6$,  is originated by two opposed effects. On the one hand, the large experimental Higgs mass favors large $\tan \beta$, since the tree-level contribution to $m_h$ is proportional to $\cos^2 2\beta$ and therefore grows with $\tan\beta$; although the effect is negligible beyond $\tan\beta\simeq 8$. On the other hand the Jacobian factor (\ref{J}) penalizes large $\tan\beta$, due to the implicit fine-tuning among the initial parameters required to obtain it. In addition, B-physics observables generally prefer low values of $\tan\beta$, to keep them within the experimental errors. However, since the favored regions in the parameter space have quite high supersymmetric masses, the supersymmetric effects on these observables are typically negligible, thus having little impact on the $\tan\beta$ pdf.

The $A_0$ parameter shows two wide peaks for positive and negative values, see fig.~\ref{fig:cmssm_all_nodm_1d}. This effect is in fact due to the logarithmic scale used for the representation. In a linear scale, the probability presents just one peak in the central region. On the other hand, it can be noticed that the preferred positive values of $A_0$ tend to be larger than the negative ones. This comes from the fact that the RG equations drive $A$ towards negative values, which increases (decreases) its absolute value if its initial sign is negative (positive). 
Thus a too large initial negative value of $A_0$ gives too large values of $A$ at low energy, leading to the appearance of tachyons. 

\begin{figure}[t]
\begin{center}
\includegraphics[angle=0,width=0.35\linewidth]{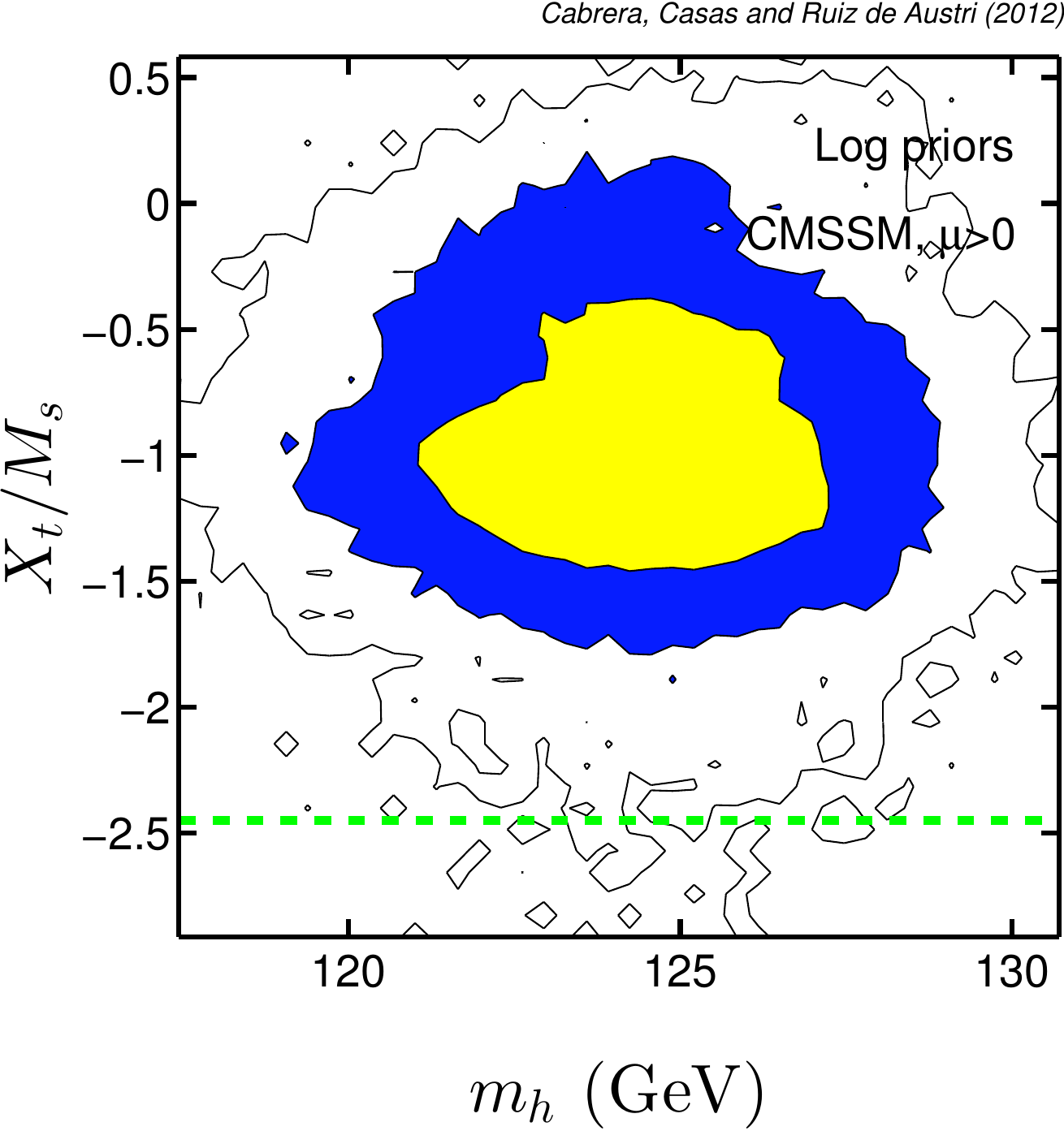}
\caption[test]{2D marginalized posterior probability
    distribution of the CMSSM in the $m_h$--$X_t/M_s$ plane, using log priors for
  $\mu>0$. $M_s\equiv M_{\rm SUSY}$ denotes a certain
      average of the stop masses. The inner and outer contours enclose 
  respective 68\% and 95\% joint regions. The white one
      corresponds to 99.9\% c.l.. The horizontal dashed line represents the
      maximizing value, $X_t = -\sqrt{6}M_{s}$. 
  \label{fig:X_T_m_h}}
\end{center}
\end{figure}

The discussion of the $A-$parameter raises another important point. As mentioned above, the value of $m_h$ is maximized for $X_t = \pm \sqrt{6}M_{\rm SUSY}$. In that case, lower masses for the stops are acceptable, which in principle would make SUSY less fine-tuned. This possibility is certainly interesting. However, it represents itself another kind of fine-tuning unless one has some theoretical reason to expect $X_t$ in the favorable range. Incidentally, note that for moderately large $\tan\beta$ the value of $X_t$ is essentially given by $A_t$, so one would essentially need a value of $A_t$ in the favorable range. In any case, this would amount to go beyond the MSSM, e.g. invoking a particular SUSY breaking and mediation mechanism that could lead to 
this result for $A$. \footnote{See ref.\cite{Aparicio:2012iw} for a paper arguing that this can actually occur in a class of superstring scenarios.} If we do not rely on a mechanism of that kind, the question is whether it pays or not 
 (in statistical weight) to have $X_t$ around the ``maximal" value. This question is answered by the Bayesian analysis. If the answer were positive, the value of $X_t$ should be concentrated around that privileged value. Fig.~\ref{fig:X_T_m_h} shows the probability distribution function in the plane $m_h-(X_t/M_S)$ (all the remaining parameters are marginalized). Clearly, the pdf does not show any preference for that value. 
 
Of course there are solutions in the CMSSM where $X_t$ is close to maximal, but in order to see them one has to plot the $99.9\%$ c.l. contour, as it is shown in Fig.~\ref{fig:X_T_m_h}. The reason why the maximal-$X_t$ region is disfavored is not that the likelihood is worse. Actually, the likelihood in that region is similar to the likelihood in the 68\% and 95\% c.l. regions. This is logical since the value of the Higgs mass is similar and SUSY is essentially decoupled for most of the low-energy observables. The important point here is that this region is statistically less important, an effect that is especially noticeable for positive $X_t$. To understand this, notice the following. The running of $A_t$ from high- to low-energy normally drives it more negative, an effect that is more important if $A_t$ is initially positive. This implies that large positive values of $A_t$ at low energy require extremely large $A_t$ at $M_X$. But this is penalized in several ways. The absolute value of $m_{H_u}^2$ at low energy becomes large, which implies a large value of $\mu^2$, and thus a more severe fine-tuning. Also, a very large $|A_t|$ is grievously penalized in the priors, see subsect.~2.3. So at low-energy one can expect larger absolute values for negative $A_t$ than for positive one, exactly as 
Fig.~\ref{fig:X_T_m_h} shows (amusingly, at the $M_X$ scale it happens the opposite, as already discussed). Consequently the negative maximal value, $X_t = -\sqrt{6}M_{\rm SUSY}$ is more favored than the positive one, as is clear in the figure. Nevertheless large negative values of $A_t$ are not very natural either. For $A_t$ negative and large (close to the maximal value), its RGE is dominated by $A_t$ itself, which makes it to run towards larger absolute values at high-energy, though the effect is not as strong as for positive $A_t$. Therefore, $X_t \simeq -\sqrt{6}M_{\rm SUSY}$ is also penalized for the same reasons as $X_t \simeq +\sqrt{6}M_{\rm SUSY}$, though in a milder way.

\subsection{Incorporation of dark matter}

There are different astrophysical and cosmological observations that offer 
impressive evidence of the existence of dark matter (DM) in the Universe 
(see Table~\ref{tab:exp_constraints} for a recent determination of $\Omega_{\rm DM}$). On the other 
hand, the consistency with the observed large structure of the Universe 
favors cold dark matter (CDM), i.e. non-relativistic matter at the beginning 
of galaxy formation. This leads to the hypothesis of a weakly interacting 
massive particle (WIMP) as the component of CDM. 

Supersymmetry offers a good candidate for such a WIMP, namely the LSP, which 
is stable in the standard (R-parity conserving) SUSY formulations 
(for a review see \cite{Jungman:1995df}). Although, depending on the models, 
there are several possibilities for the SUSY WIMP, the most popular and 
natural candidate is the lightest neutralino (from here on we will refer it 
as the neutralino), $\chi^0_1$, which is the LSP in 
most of the CMSSM parameter space. One can assume that the LSP makes up the whole content of DM required by the standard cosmology, which is of course an attractive scenario. However, it is also possible that, due to an efficient annihilation at early times, the LSP component of the DM is subdominant.
Motivated for this, we discuss next, separately, the cases where the DM 
is made only of LSPs and where it is partially made of LSPs.

\vspace{0.3cm}
\noindent
{\bf {\em Single-Component DM}} \\

\noindent
In this scenario we assume a Gaussian likelihood on $\Omega_\chi h^2$ 
(see Table~\ref{tab:exp_constraints}). Besides, we apply the constraints 
from DM-induced recoil events in the XENON100 experiment, as described above.
For the present analysis we neglect the effect of the 
astrophysical and hadronic uncertainties entering into the determination 
of the spectrum of recoil events. Namely, we fix the 
astrophysical parameters that describe the density and velocity distribution 
of DM particles at the commonly adopted benchmark values: local CDM 
density $\rho_{\odot,{\rm CDM}}=0.4\,$GeV\,cm$^{-3}$, 
circular velocity $v_0=235$ km s$^{-1}$ and escape velocity $v_{esc}=550$ 
km s$^{-1}$ (see, e.g., \cite{Pato:2010zk} and references therein for a 
recent discussion of the astrophysical uncertainties on these 
quantities). For the contribution of the light quarks to the nucleon form
factors we have  adopted the values $f_{Tu}=0.02698$, $f_{Td}=0.03906$ and 
$f_{Ts}=0.36$ \cite{Ellis:2008hf}. Incorporating the associated uncertainties  
of both the astrophysical and nuclear form factors parameters  
to the analysis as nuisance parameters (i.e. marginalizing) would only mildly change 
the numerical results presented here; the main conclusions would remain 
unchanged, in agreement with the findings of \cite{Bertone:2011nj}. 

\begin{figure}[t]
\begin{center}
\includegraphics[angle=0,width=0.35\linewidth]{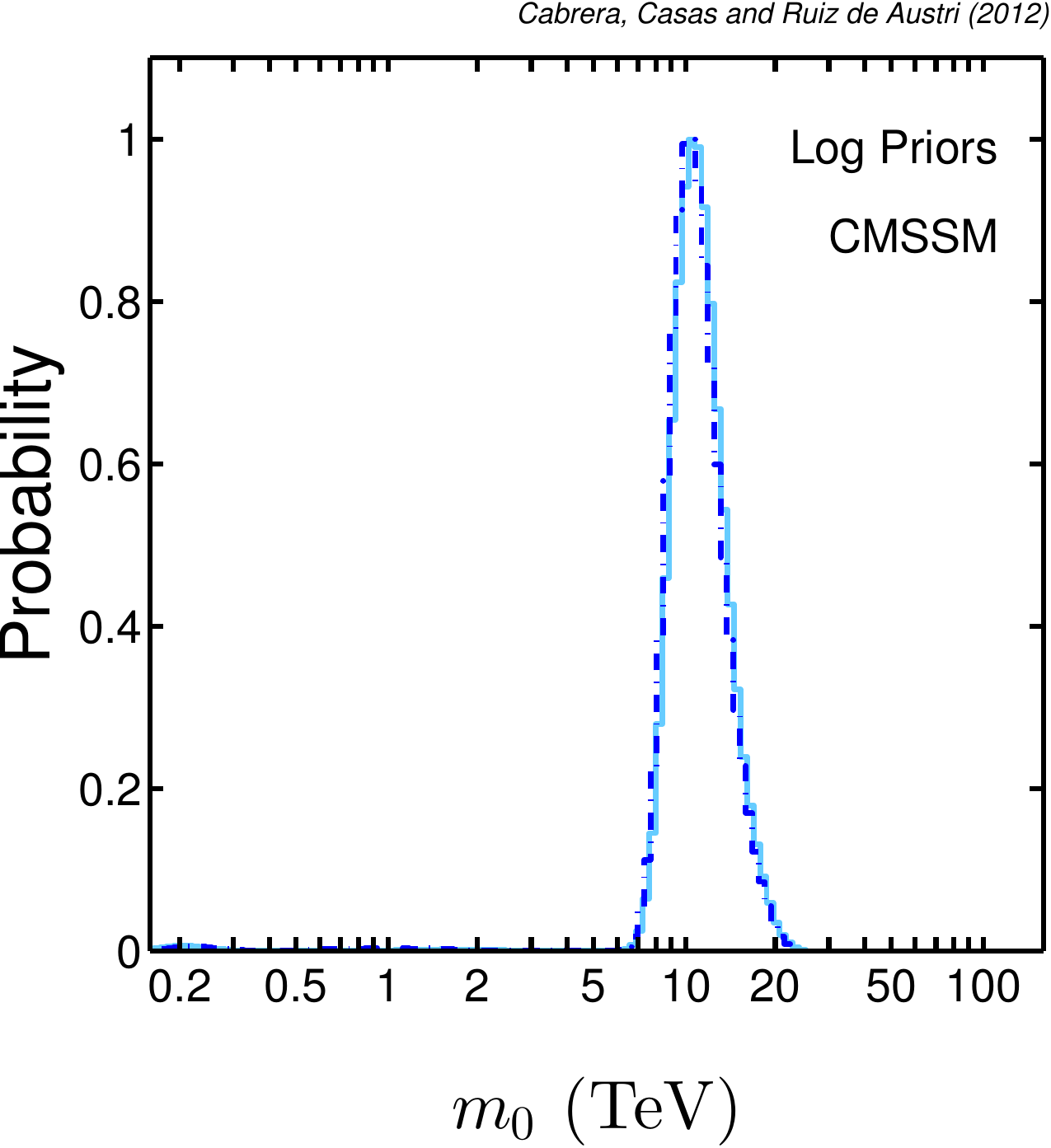} \hspace{1.2cm}
\includegraphics[angle=0,width=0.35\linewidth]{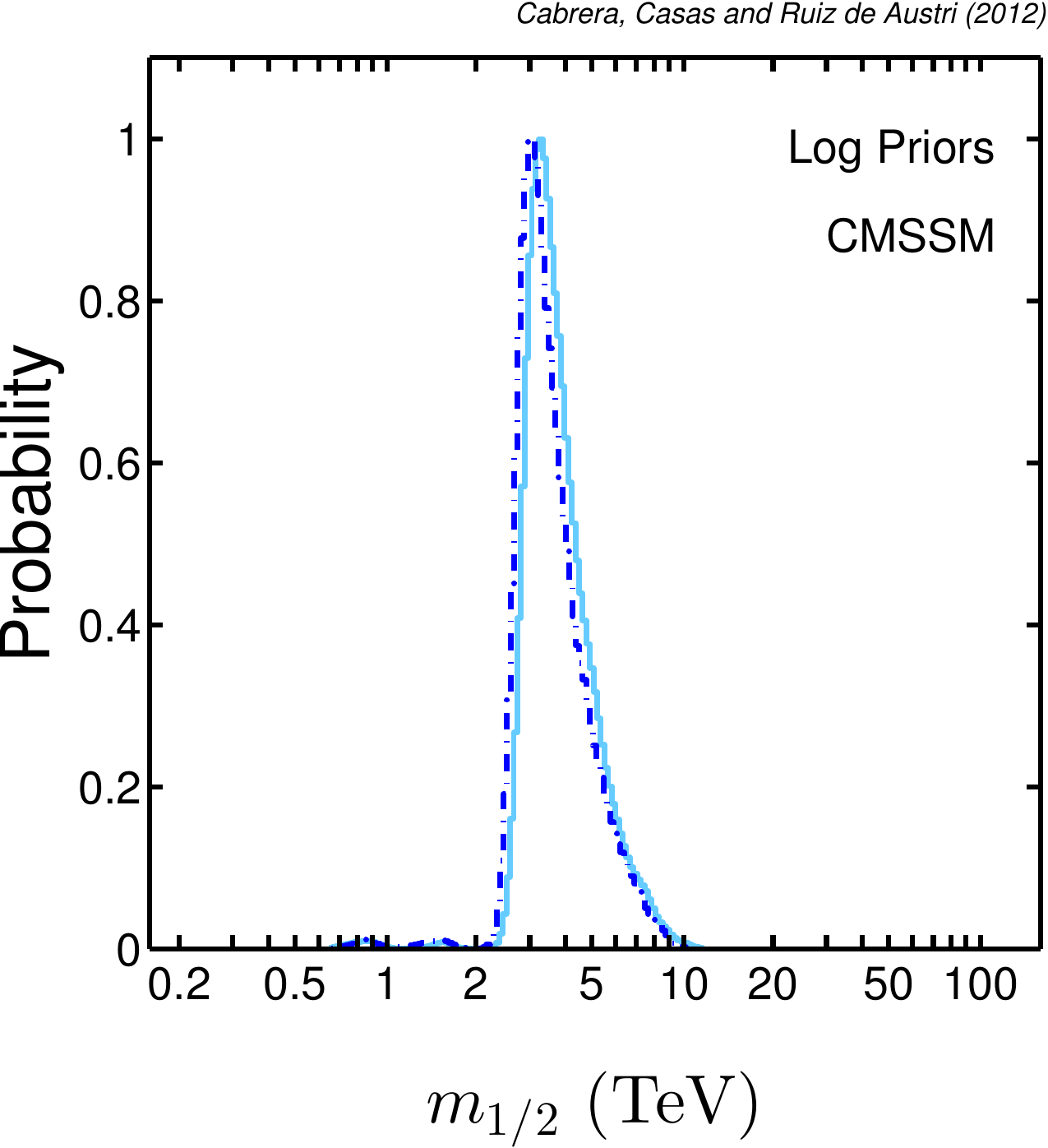} \\ \vspace{0.2cm}
\includegraphics[angle=0,width=0.35\linewidth]{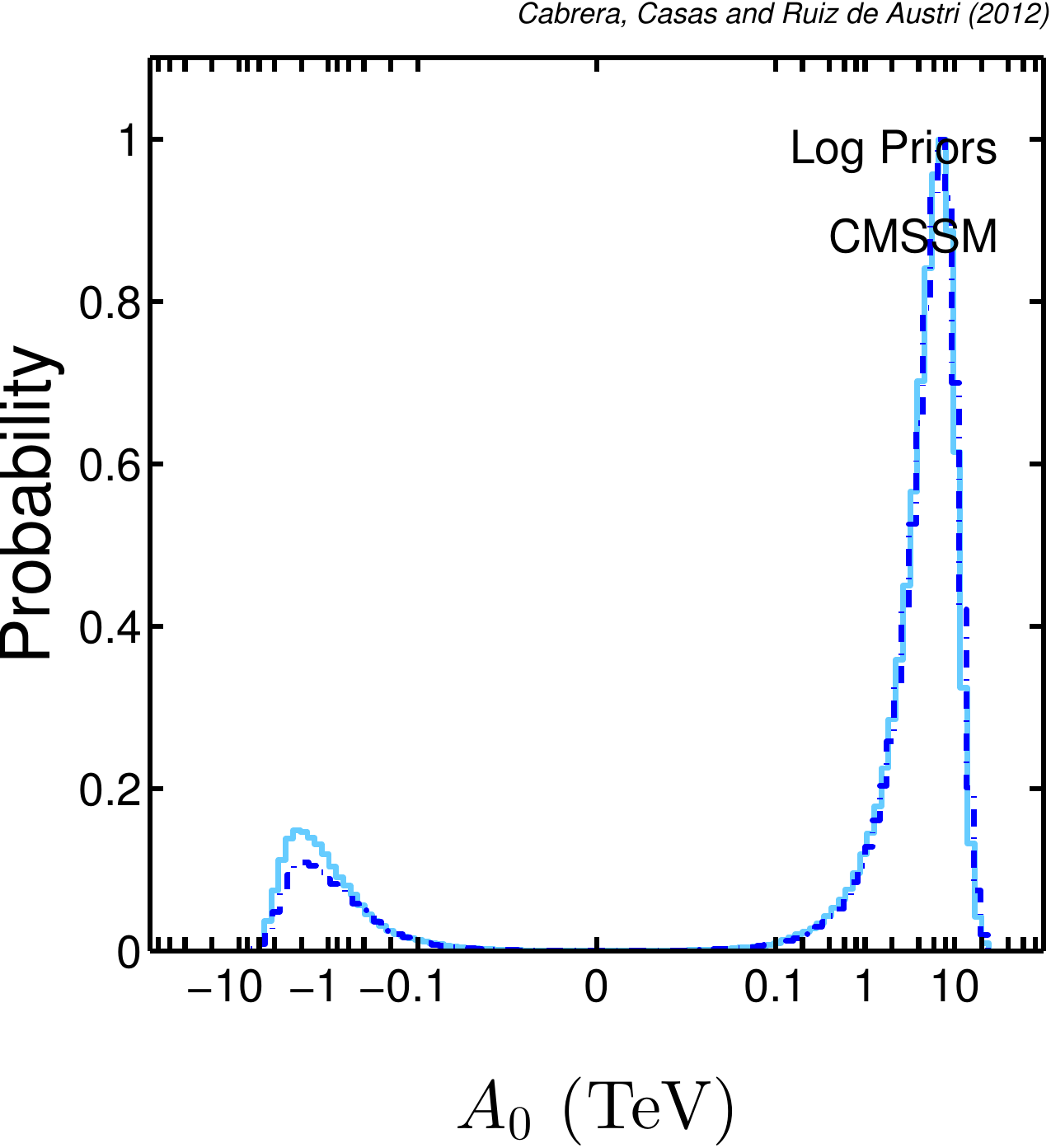} \hspace{1.2cm}
\includegraphics[angle=0,width=0.35\linewidth]{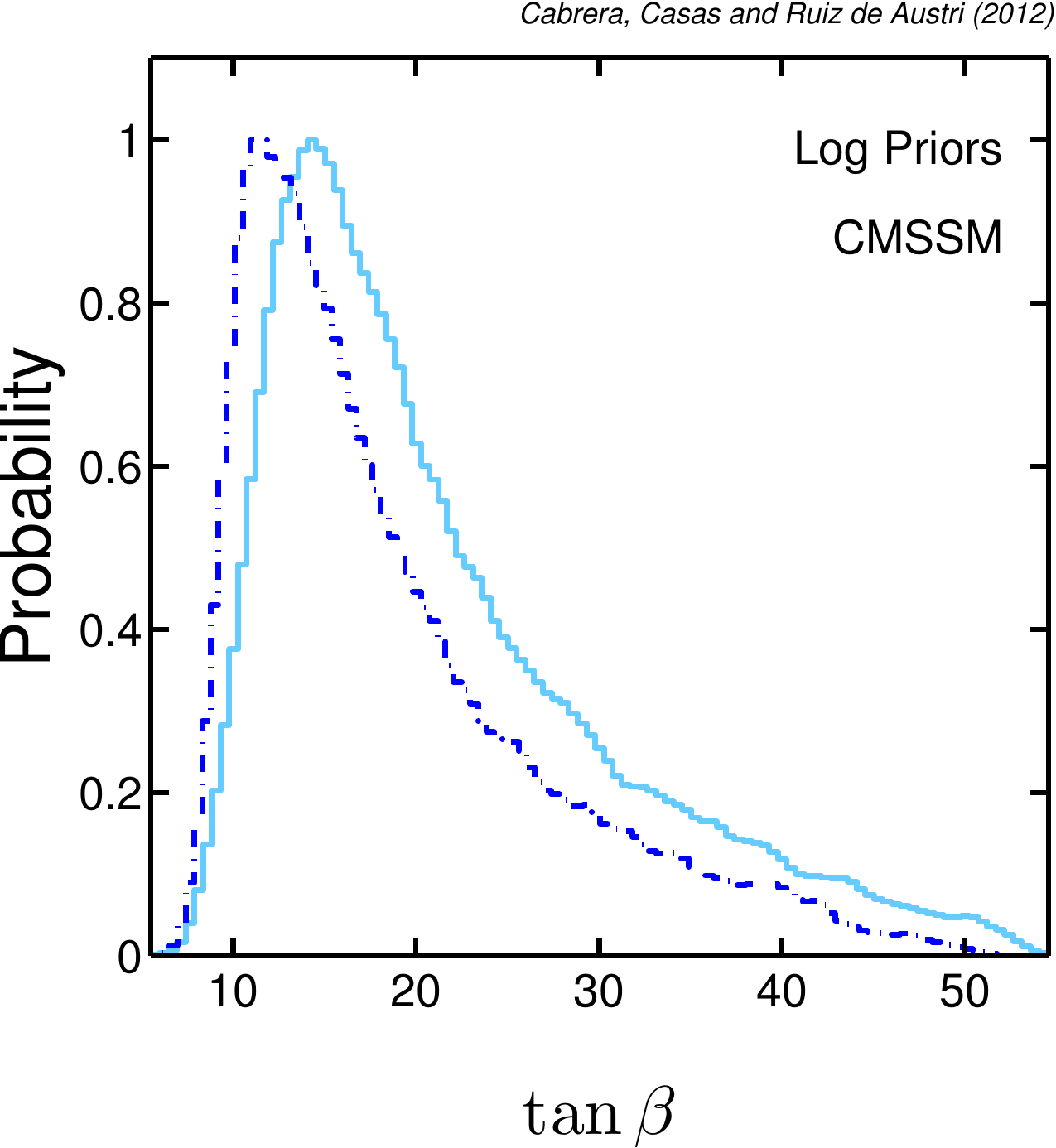} \\ \vspace{0.2cm}
\includegraphics[angle=0,width=0.35\linewidth]{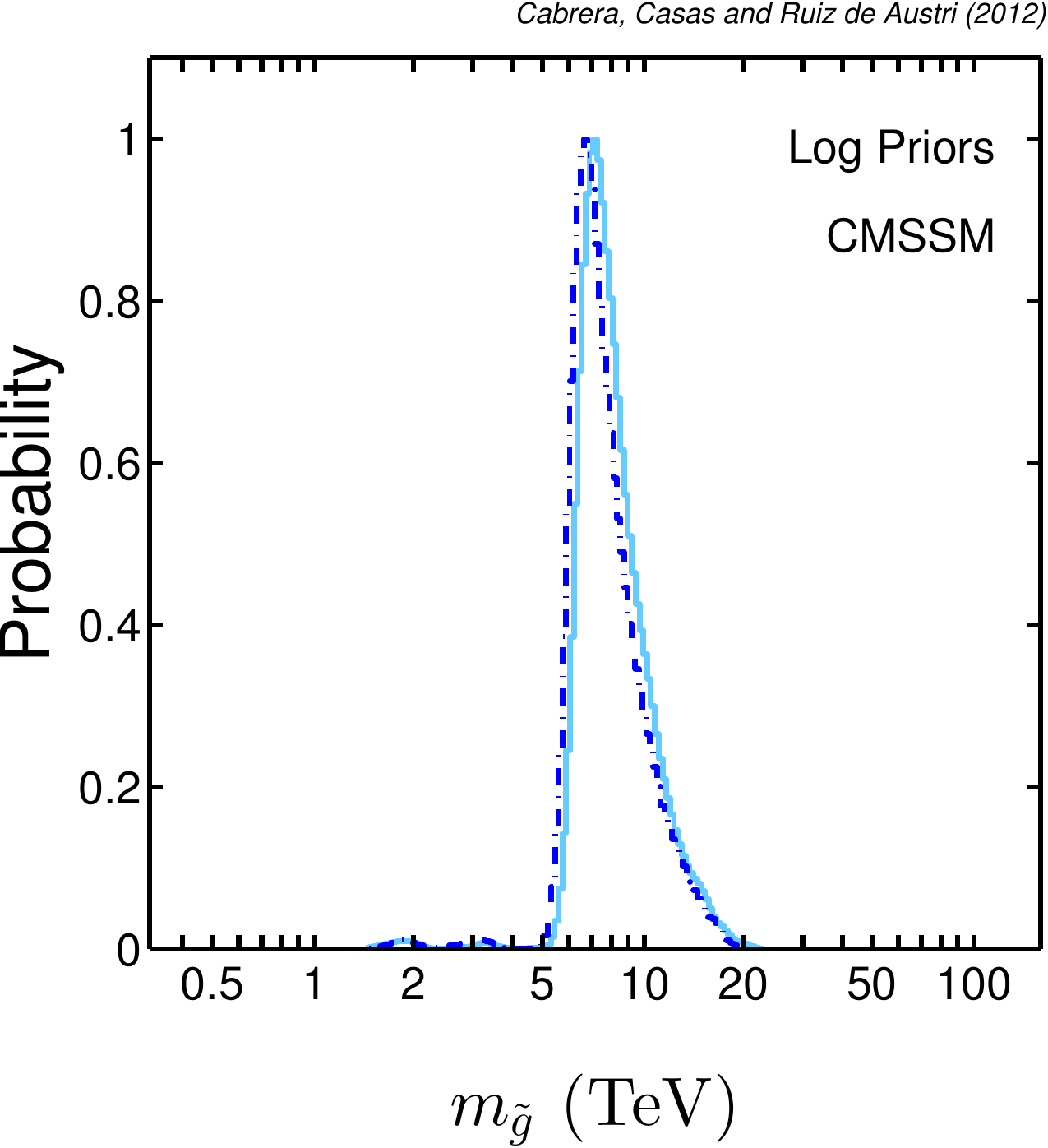} \hspace{1.2cm}
\includegraphics[angle=0,width=0.35\linewidth]{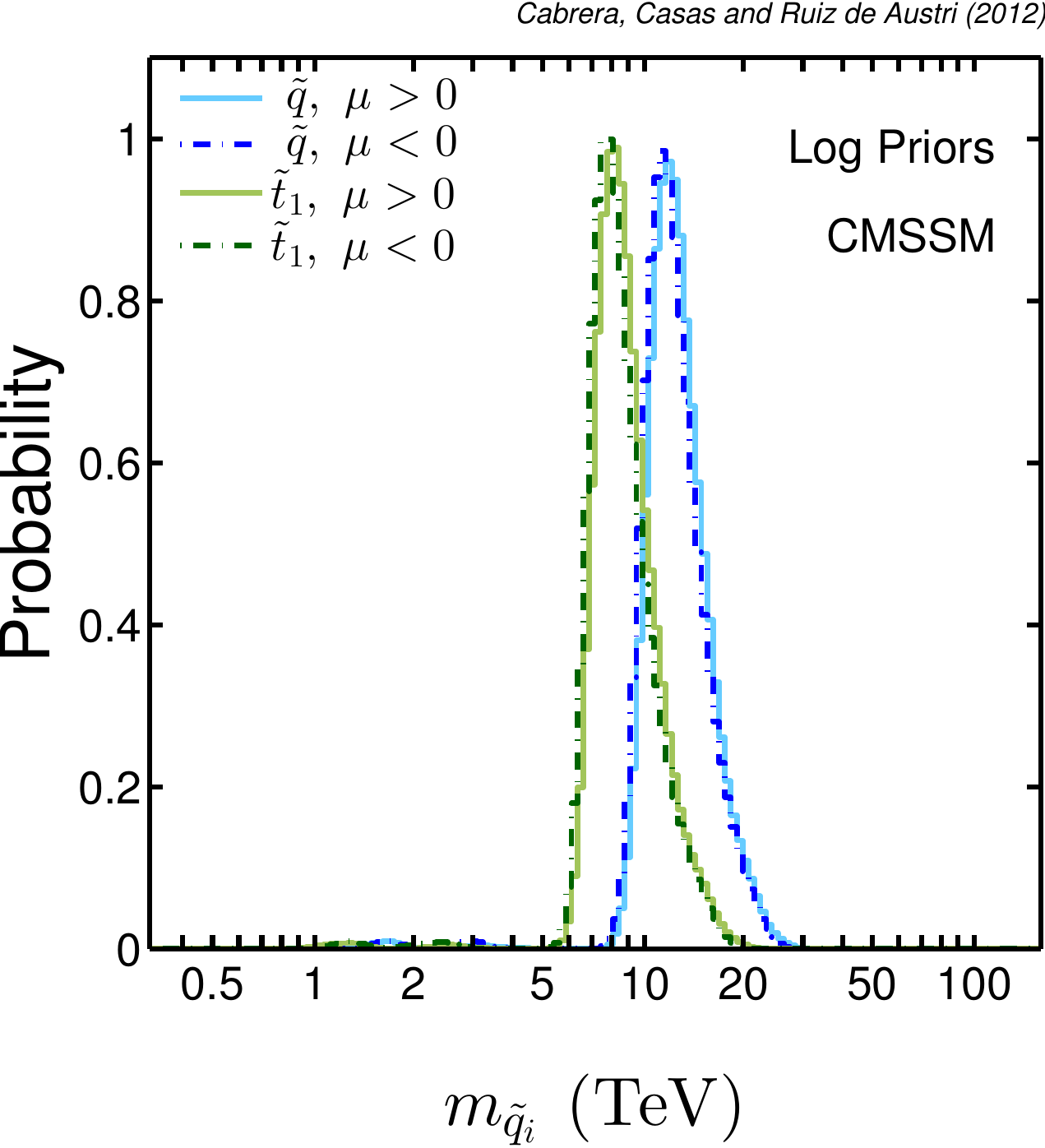}
\caption[test]{As in fig.~\ref{fig:cmssm_all_nodm_1d} but with an additional 
constraint from DM 
(WMAP 7-years and XENON100) considering that the lightest supersymmetric particle (LSP)
 is the solely DM component. The continuous (dashed) lines correspond to $\mu>0$  ($\mu<0$).\label{fig:cmssm_all_scdm_1d}}
\end{center}
\end{figure}

\begin{figure}[t]
\begin{center}
\includegraphics[angle=0,width=0.35\linewidth]{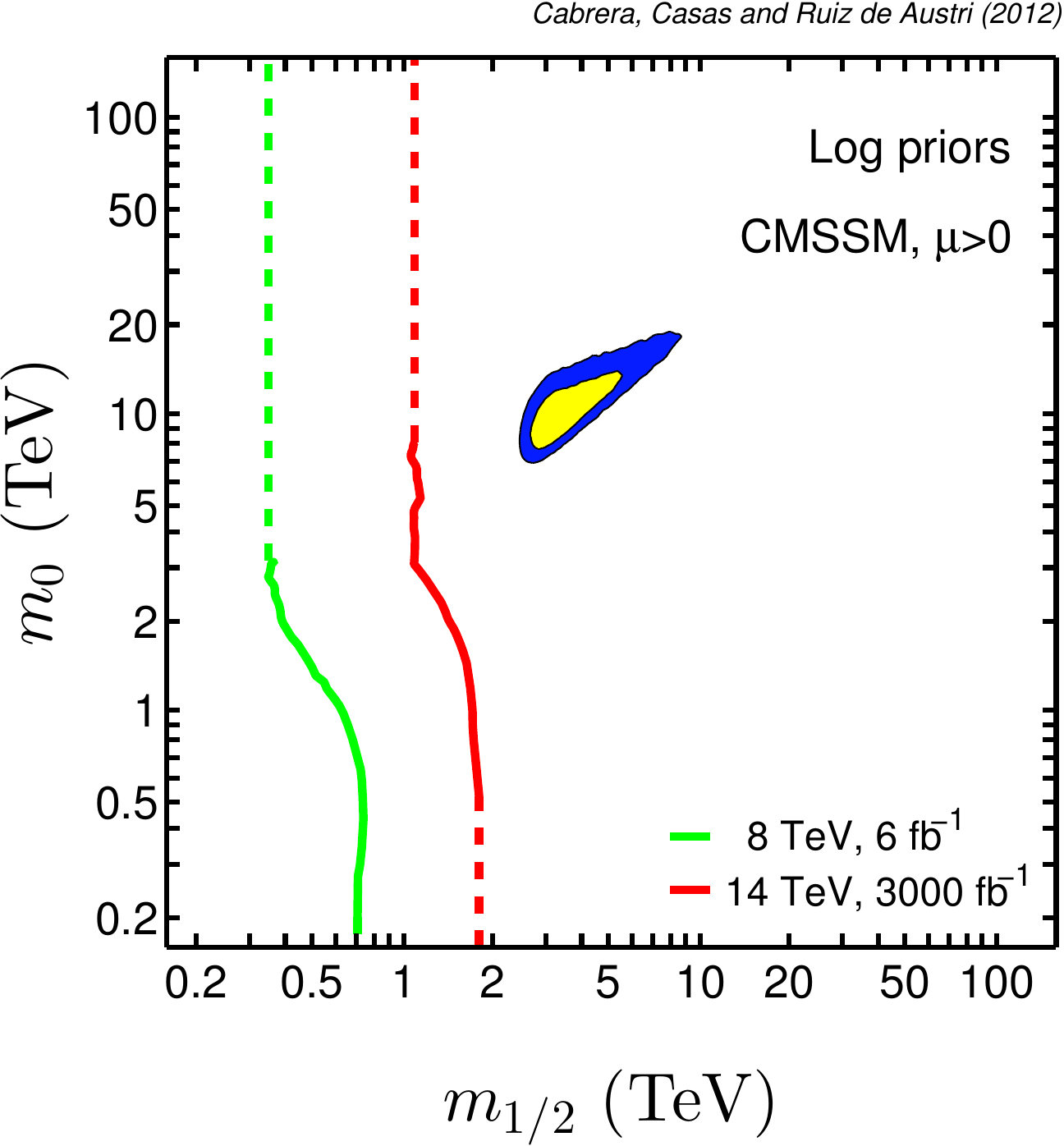} \hspace{1.2cm}
\includegraphics[angle=0,width=0.35\linewidth]{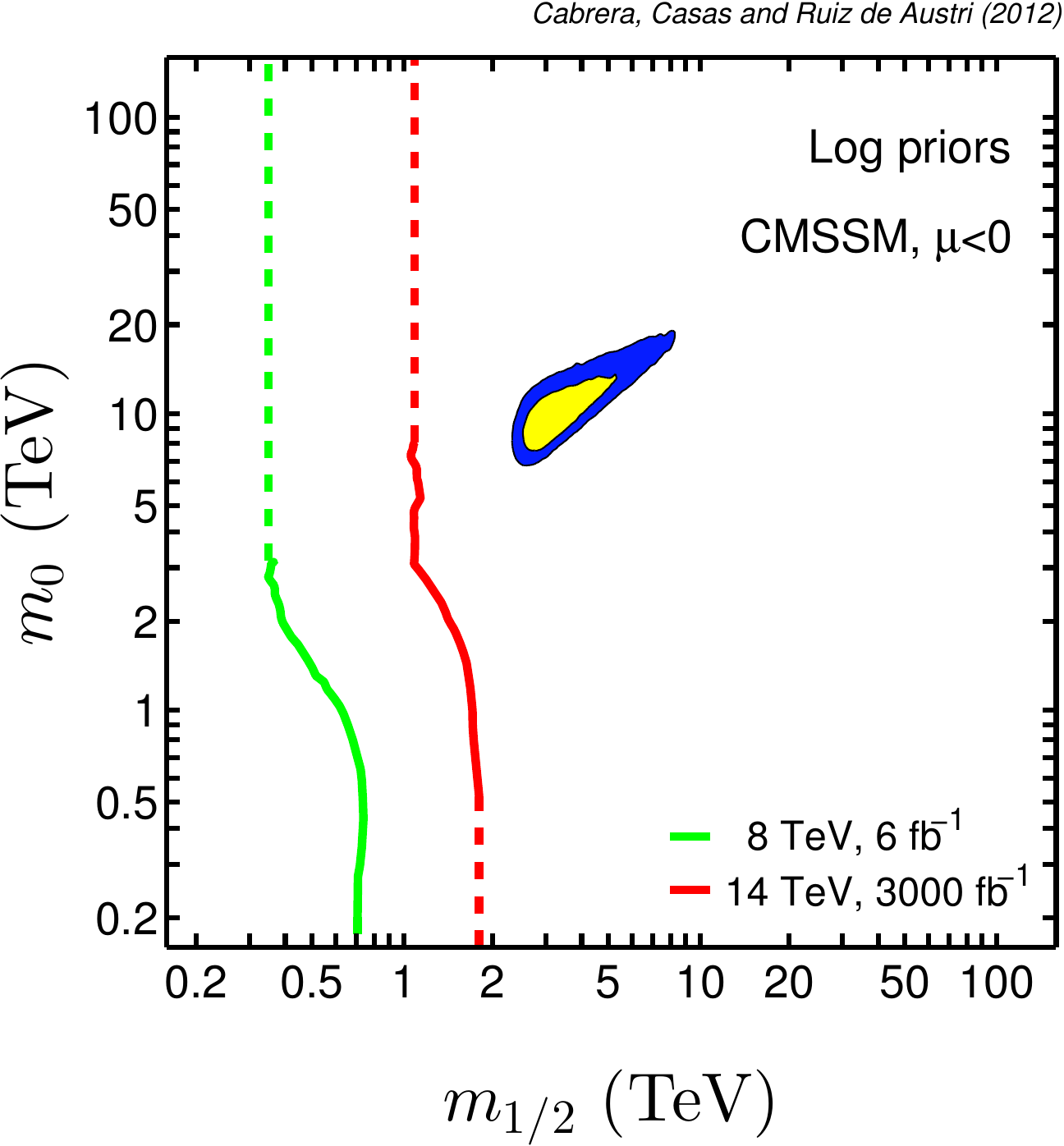} \\ \vspace{0.5cm}
\includegraphics[angle=0,width=0.35\linewidth]{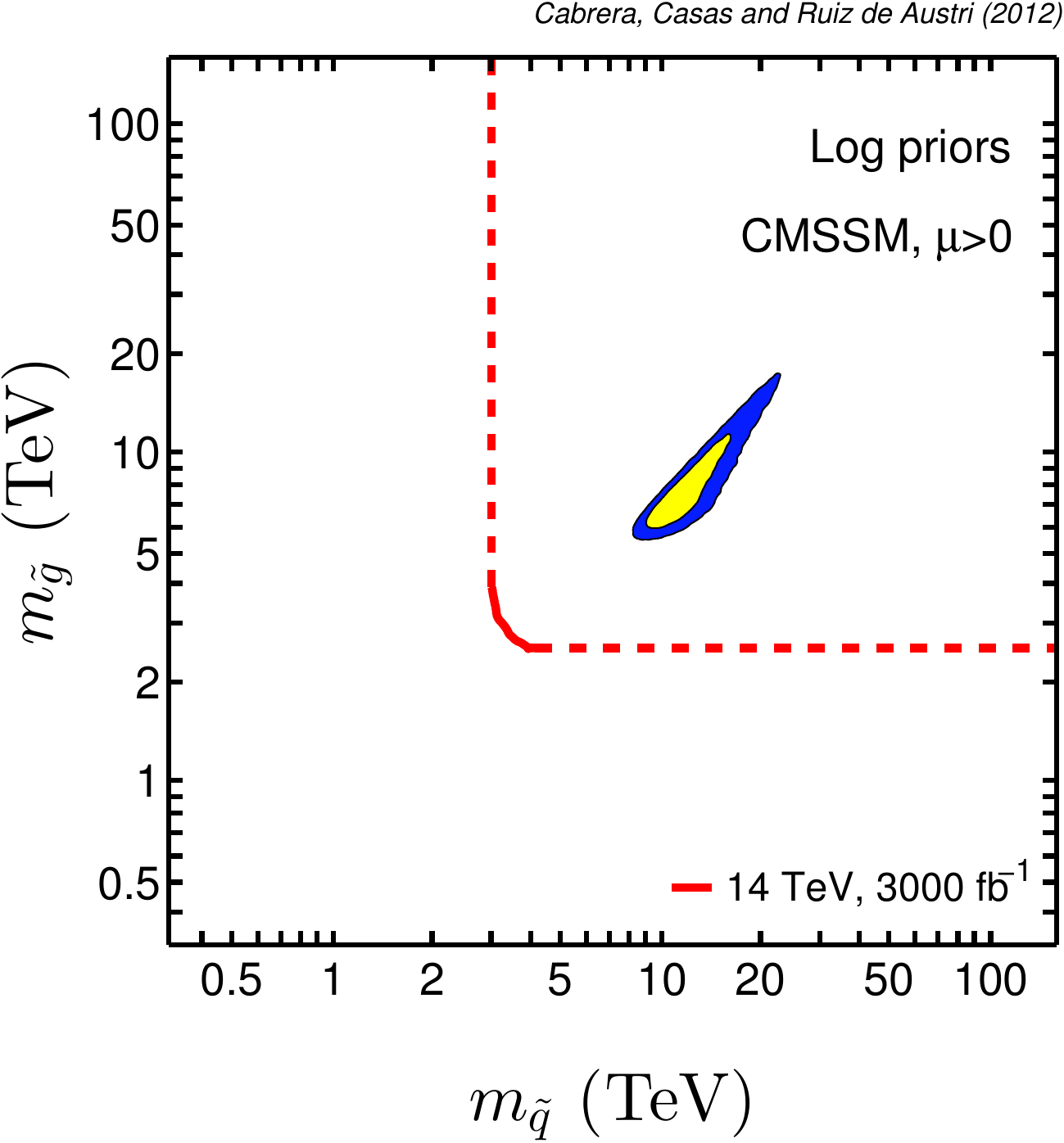} \hspace{1.2cm}
\includegraphics[angle=0,width=0.35\linewidth]{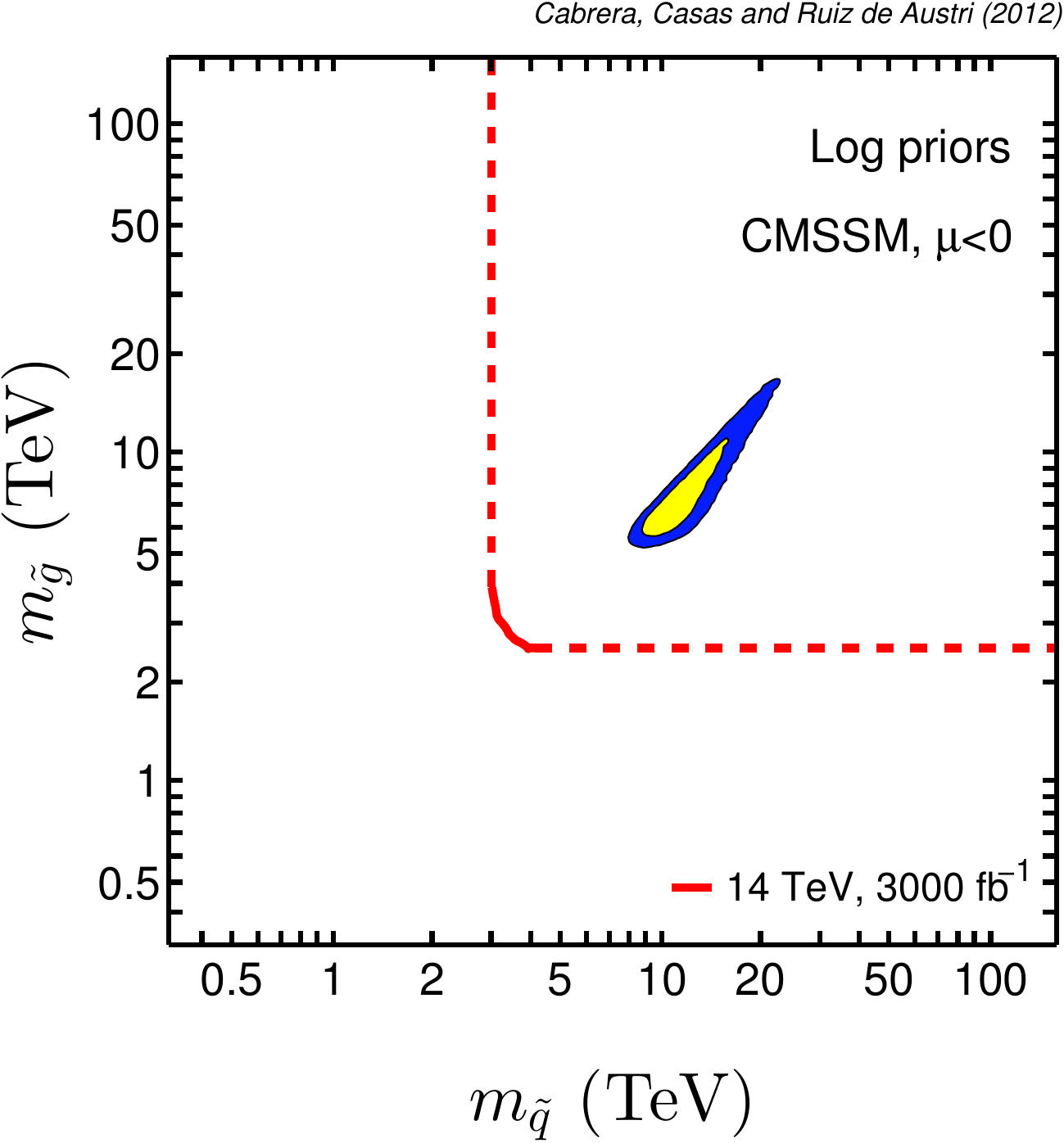}
\caption[test]{As in fig.~\ref{fig:cmssm_all_nodm_2d} but with an additional constraint from DM (WMAP 7-years and XENON100) considering that the LSP is the solely DM component.\label{fig:cmssm_all_scdm_2d_1}}
\end{center}
\end{figure}

In fig.~\ref{fig:cmssm_all_scdm_1d} we plot the 1D pdfs of various supersymmetric parameters and the physical gluino
and squark masses. The cyan-continuous (blue-dashed) line corresponds to $\mu>0$  ($\mu<0$).
This is complemented with the 2D pdfs of fig.~\ref{fig:cmssm_all_scdm_2d_1}. In all cases the non-displayed parameters have been marginalized. The results correspond to log priors (for flat priors they are very similar).
The first thing to notice is that, due to the DM constraints, the pdfs of $m_{1/2}$ and $m_0$ (an thus of $m_{\tilde g}, m_{\tilde q}$) have moved further
to large values. Indeed, more than the $95\%$ of the posterior probability is out of the LHC reach, even with an energy of $14$ TeV and a integrated luminosity of 3000 fb$^{-1}$ as projected for the HL-LHC (High Luminosity) phase. \cite{HL_LHC}. This conclusion holds for the flat priors case.
The reason is the following. There are in fact two additional regions where the required DM abundance can be reproduced: the stau-coannihilation and the $A-$funnel regions. However, the first one is disfavored, as it corresponds to rather small soft terms, difficult to reconcile with the Higgs mass. The second one requires intermediate/large $\tan\beta$ values, which are penalized by the Jacobian prefactor in the r.h.s. of eq.(\ref{approx_eff_prior}). In consequence, one is essentially left with the Focus-Point region (actually the other two regions survive but with very tiny statistical weight, as discussed in sect. 7).

In Appendix A we describe some characteristics of the Focus-Point in the CMSSM and the NUHM.
Let us now mention some of them, which will be useful to understand the shape of Figs.~\ref{fig:cmssm_all_scdm_1d}, \ref{fig:cmssm_all_scdm_2d_1}. The Focus-Point region corresponds to heavy scalars ($m_0$), less heavy gauginos ($m_{1/2}$) and $\tan\beta\gsim 8$. In that region the value of $m_{H_u}^2$ at low energy is quite insensitive to its value at high-energy, this is the reason why it can reach large values with no additional fine-tuning. For too-large $m_0$, however, the electroweak breaking solution gets lost. This occurs when the required value of $\mu^2$ gets negative. For slightly smaller values of $m_0$ the value of $\mu$ is small, and this is welcome for fine-tuning reasons, since the Jacobian prefactor in eq.(\ref{approx_eff_prior}) gets smaller. Hence, this is a theoretically interesting region by itself. But, besides, it is also appropriate to correctly reproduce the DM abundance. The reason is that if $\mu$ is smaller than (or similar to) $m_{1/2}$, the lightest neutralino is essentially a Higgsino (or a mixture of Higgsino and bino). That sizable fraction of Higgsino makes the neutralinos to self-annihilate efficiently (and also to coannihilate with the 
lightest chargino and the second lightest neutralino), which may lead to the
correct DM abundance. Actually, if the lightest neutralino is essentially a
pure Higgsino and its mass is too small, the annihilation is too
efficient. Then one needs a sizable bino component to reduce the annihilation
so that the correct DM abundance is reproduced. Therefore the largest
acceptable neutralino mass corresponds to the pure Higgsino case, and is
$M_\chi\simeq 1$ TeV \cite{ArkaniHamed:2006mb} (in that limiting case the neutralino, chargino, and second neutralino masses are all $\sim\mu$). Below that limit, the neutralino mass can vary in a wide range, provided it contains the suitable bino fraction. Indeed, this was the situation prior to the new XENON100 data. 

The point is that XENON is sensitive to the spin-independent cross-section of $O(100)$ GeV neutralinos with the nucleons, if the former contain both a Higgsino and bino components.
This occurs through diagrams exchanging a Higgs state. In this way XENON100
has been able to exclude all the above-mentioned Focus-Point region, whenever
there is a non-negligible Higgsino-bino mixture. In essence, only the pure
Higgsino possibility remains alive \cite{Farina:2011bh}. In consequence
the value of $m_\chi$ and $\mu$ are constrained around 1 TeV. In addition, the
bino mass cannot be smaller than that, otherwise the lightest neutralino would
get a sizable bino component, which is excluded by XENON. This is the reason
of the lower limit, $m_{1/2}>2$ TeV, clearly visible in
Figs.~\ref{fig:cmssm_all_scdm_1d}, \ref{fig:cmssm_all_scdm_2d_1} (the bino mass is approximately half the initial gaugino mass at $M_X$). 

\begin{figure}[t]
\begin{center}
\includegraphics[angle=0,width=0.35\linewidth]{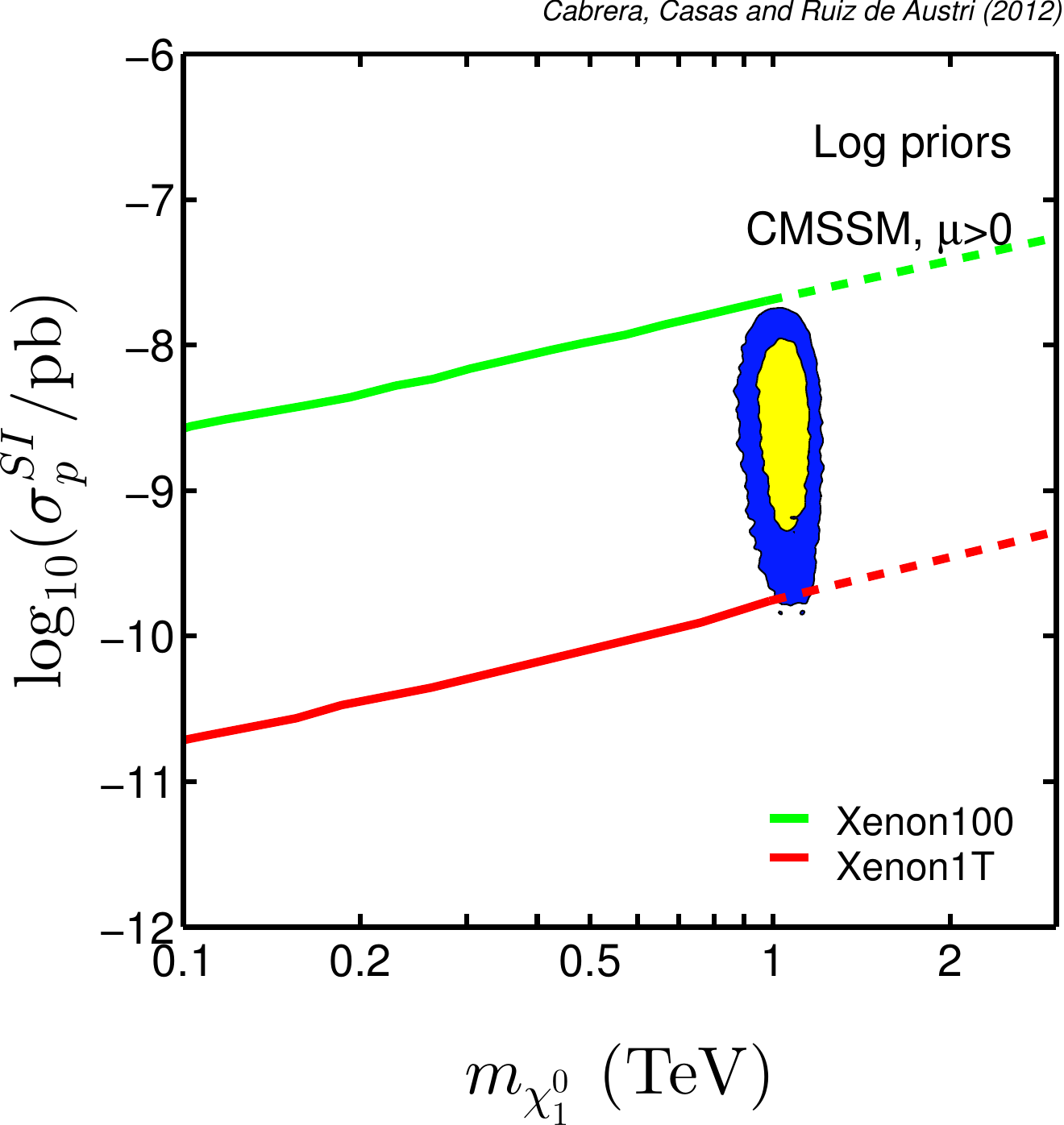} \hspace{1.2cm}
\includegraphics[angle=0,width=0.35\linewidth]{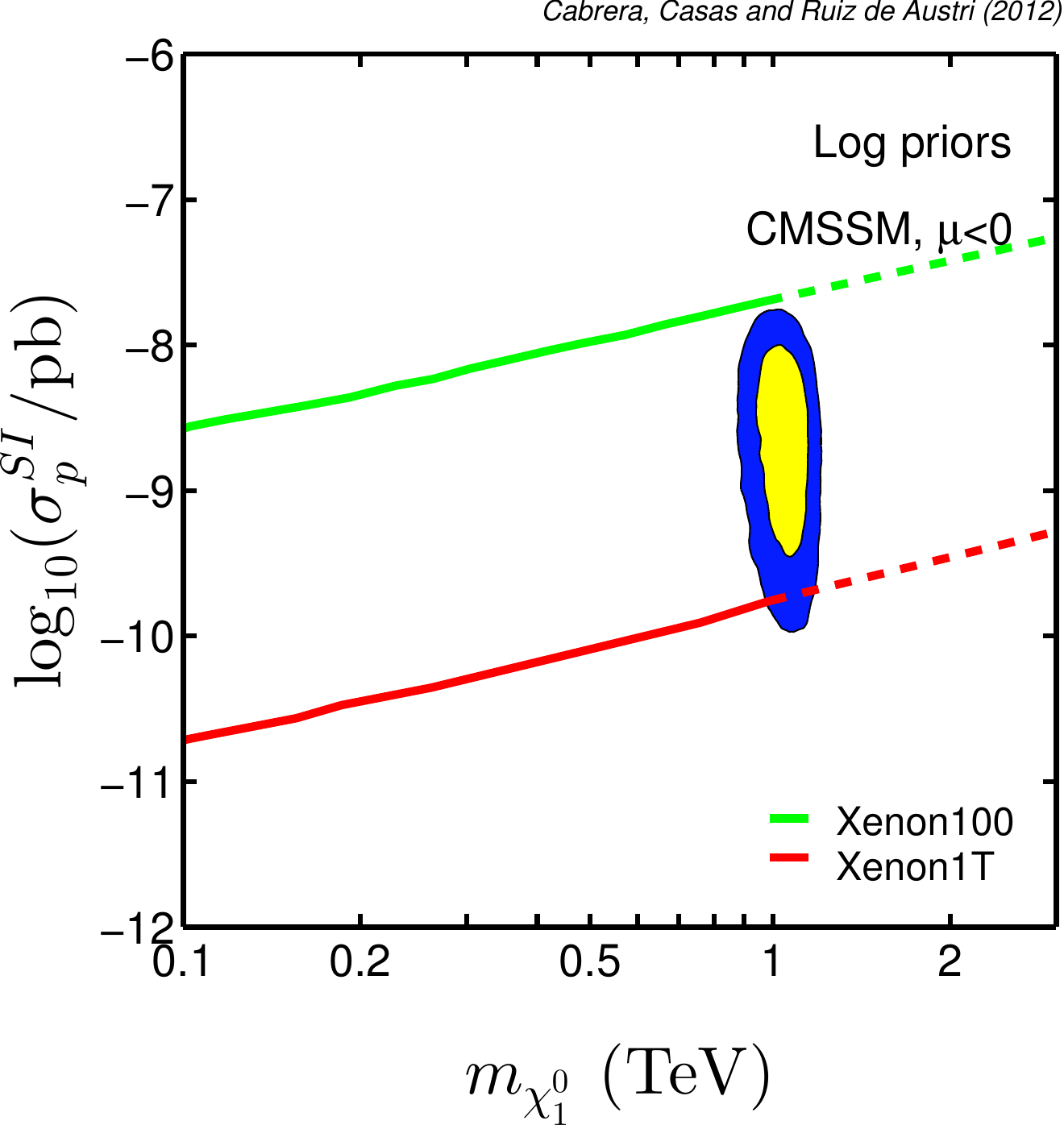}
\caption[test]{2D marginalized posterior probability distribution of the spin--independent scattering cross section of a neutralino off a proton versus neutralino mass being this 
the solely DM component for log priors, $\mu > 0$ (left panel) and $\mu<0$ (right panel). \label{fig:cmssm_all_scdm_2d_2}}.
\end{center}
\end{figure}

Although the prospects of LHC to discover or discard this scenario are clearly pessimistic, future XENON measurements could do the job.  In fig.~\ref{fig:cmssm_all_scdm_2d_2} we display the 2D joint posterior pdf of the spin--independent scattering cross section of a 
neutralino off a proton versus the neutralino mass for log priors and both signs of $\mu$. As expected from the previous discussion, the posterior pdf of the 
neutralino masses is confined around 1 TeV. In contrast, the spin-independent
scattering cross section covers several orders of magnitude. The present XENON100 and future XENON1T limits are also indicated in the plots. Clearly, most of the significant region favored at 
the 95 \% level will be within reach of the next generation of direct 
detection experiments, which are certainly good news for the CMSSM.

Let us finally notice the fact, that, contrary to what happened before including DM constraints, the pdf in $A_0$ shows a clear trend towards positive values. This has to do with the Focus-point regime. If $A_0$ is negative, its absolute value at low-energy increases. 
This increase is translates to the running of $m_{H_u}^2$, which becomes more negative at low-energy. As a consequence, the required value of $\mu^2$ to achieve successful electroweak breaking increases. This implies a larger fine-tuning and makes more difficult the implementation of the Focus-Point solution to the DM abundance. For positive $A_0$, the behavior is the opposite. Without the DM constraints the Focus-Point area was already inside the preferred region, but it was not the only one, as can be seen from fig.~\ref{fig:cmssm_all_nodm_2d}, thus the two signs of $A_0$ had approximately the same statistical weight.

\vspace{0.2cm}
\noindent
{\bf {\em Multi-Component DM}} \\

Relaxing the requirement that all of the DM is made of LSPs implies that 
the WMAP measurement is just an upper bound on its abundance. At it is shown in the Appendix 
of ref. \cite{Bertone:2010ww}, in this case, the correct effective likelihood is given by the 
expression 
\be \label{eq:upperbound}
\like_\text{WMAP}(\OhLSP) = \like_0 \int_{\OhLSP/\siW}^\infty  
e^{-\frac{1}{2}(x-r_\star)} x^{-1}{\rm d}x ,
\ee
where $ \like_0$ is an irrelevant normalization constant, 
$r_\star \equiv \muW/\siW$ and $\OhLSP$ is the predicted relic density of the 
LSP as a function of the CMSSM and SM parameters.

When the LSP is not the only constituent of DM, the rate of events in a direct-detection experiment is smaller, since it is proportional
to the local density of the LSP,  $\rho_\chi$, which is now smaller than the total local DM density, $\rho_\text{DM}$. The suppression is given by the factor $\xi \equiv \rho_\chi / \rho_{\rm DM}$. Following ref. \cite{Bertone:2010rv}, we assume that ratio of local LSP and total DM densities is equal to that for the cosmic abundances, thus $\xi \equiv \rho_\chi / \rho_{\rm DM}= \Omega_\chi / \Omega_{\rm DM}$. For $\Omega_{\rm DM}$ we adopt the central value of the WMAP determination
see Table 2, while for $\rho_{\rm DM}$ we adopt, following ref. \cite{Pato:2010zk}, the value
$0.4\,$GeV\,cm$^{-3}$. This allows to evaluate $\xi$ for each point in the CMSSM parameter space.

\begin{figure}[t]
\begin{center}
\includegraphics[angle=0,width=0.35\linewidth]{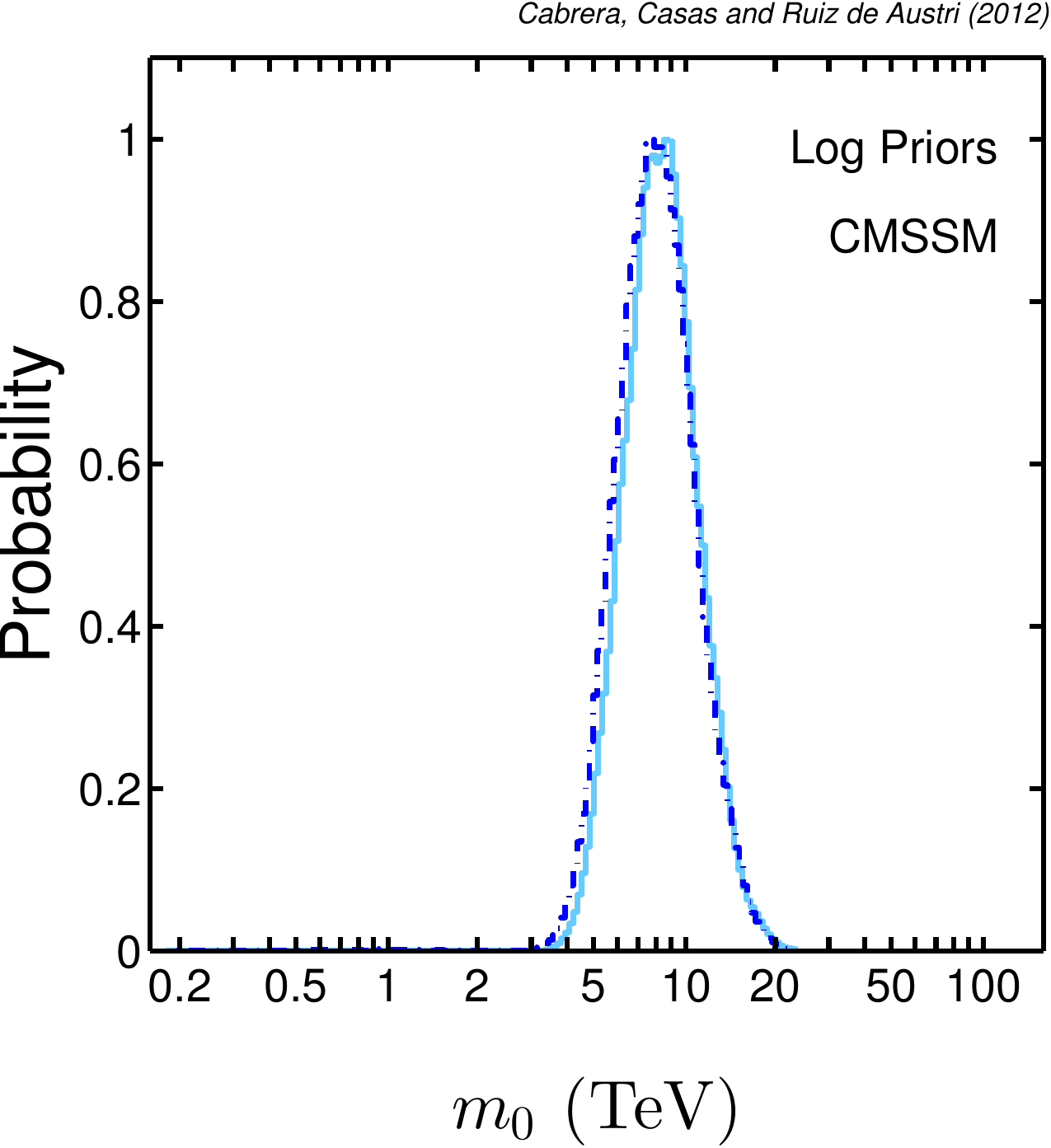} \hspace{1.2cm}
\includegraphics[angle=0,width=0.35\linewidth]{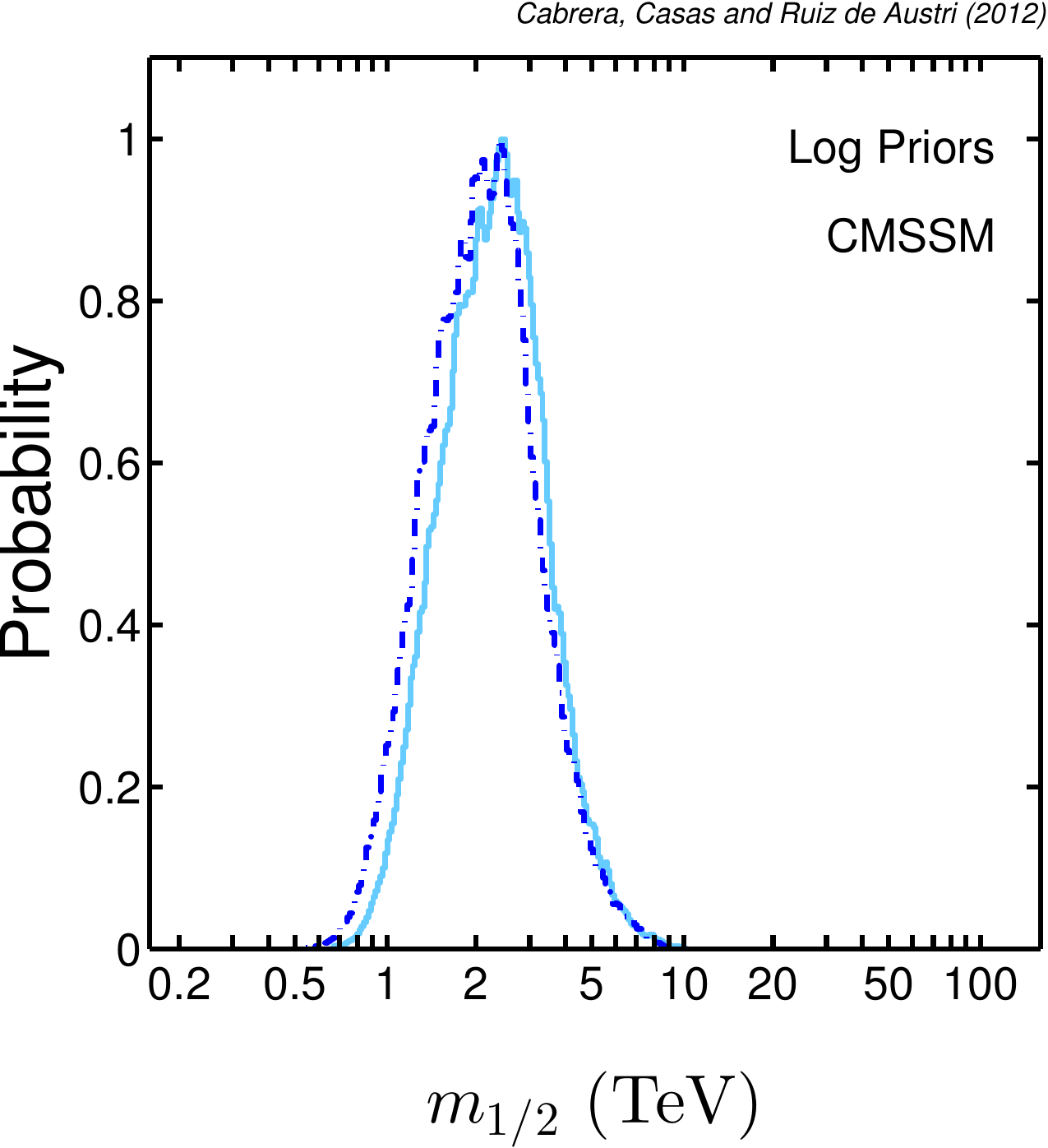} \\ \vspace{0.4cm}
\includegraphics[angle=0,width=0.35\linewidth]{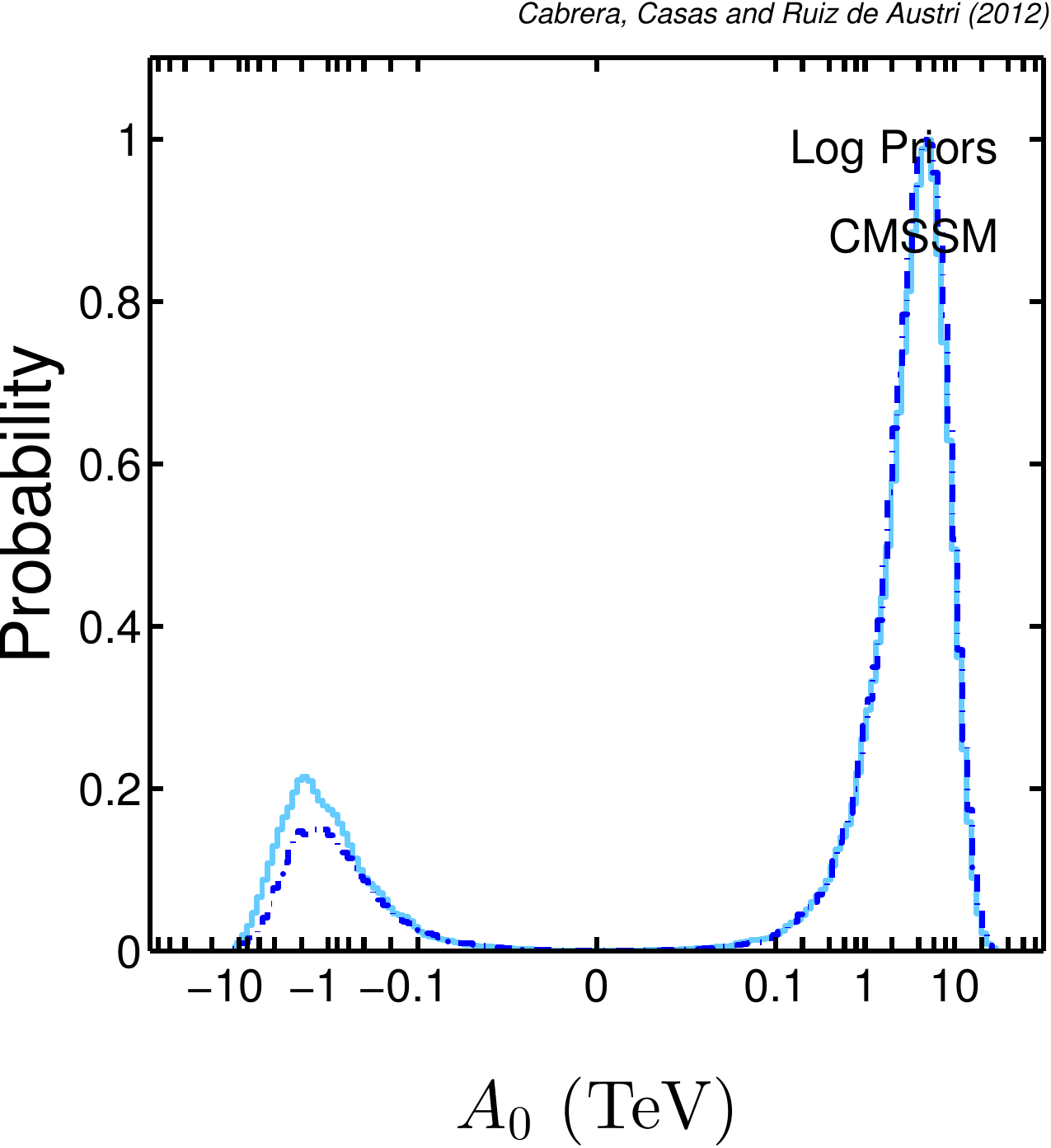} \hspace{1.2cm}
\includegraphics[angle=0,width=0.35\linewidth]{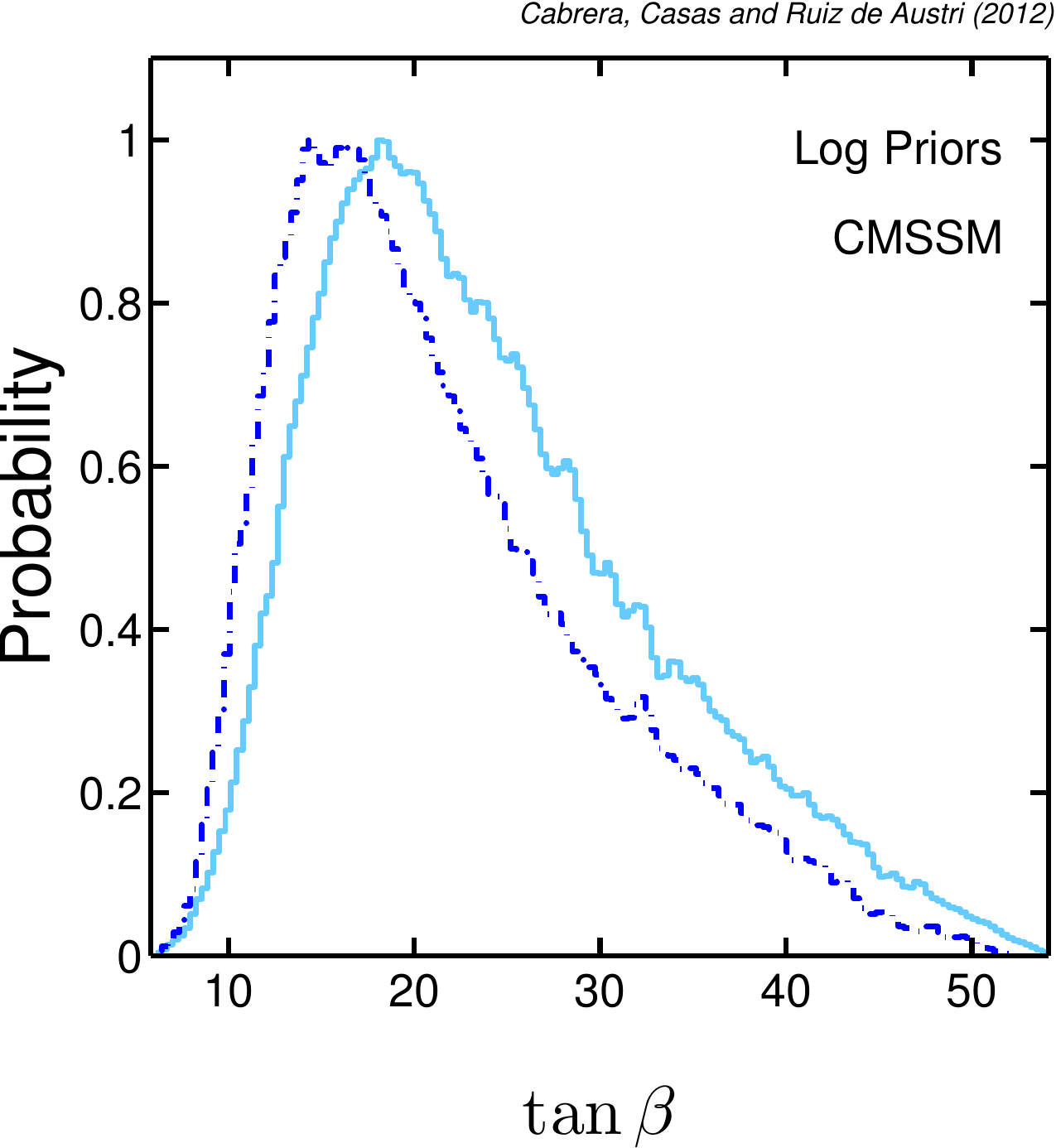} \\ \vspace{0.4cm}
\includegraphics[angle=0,width=0.35\linewidth]{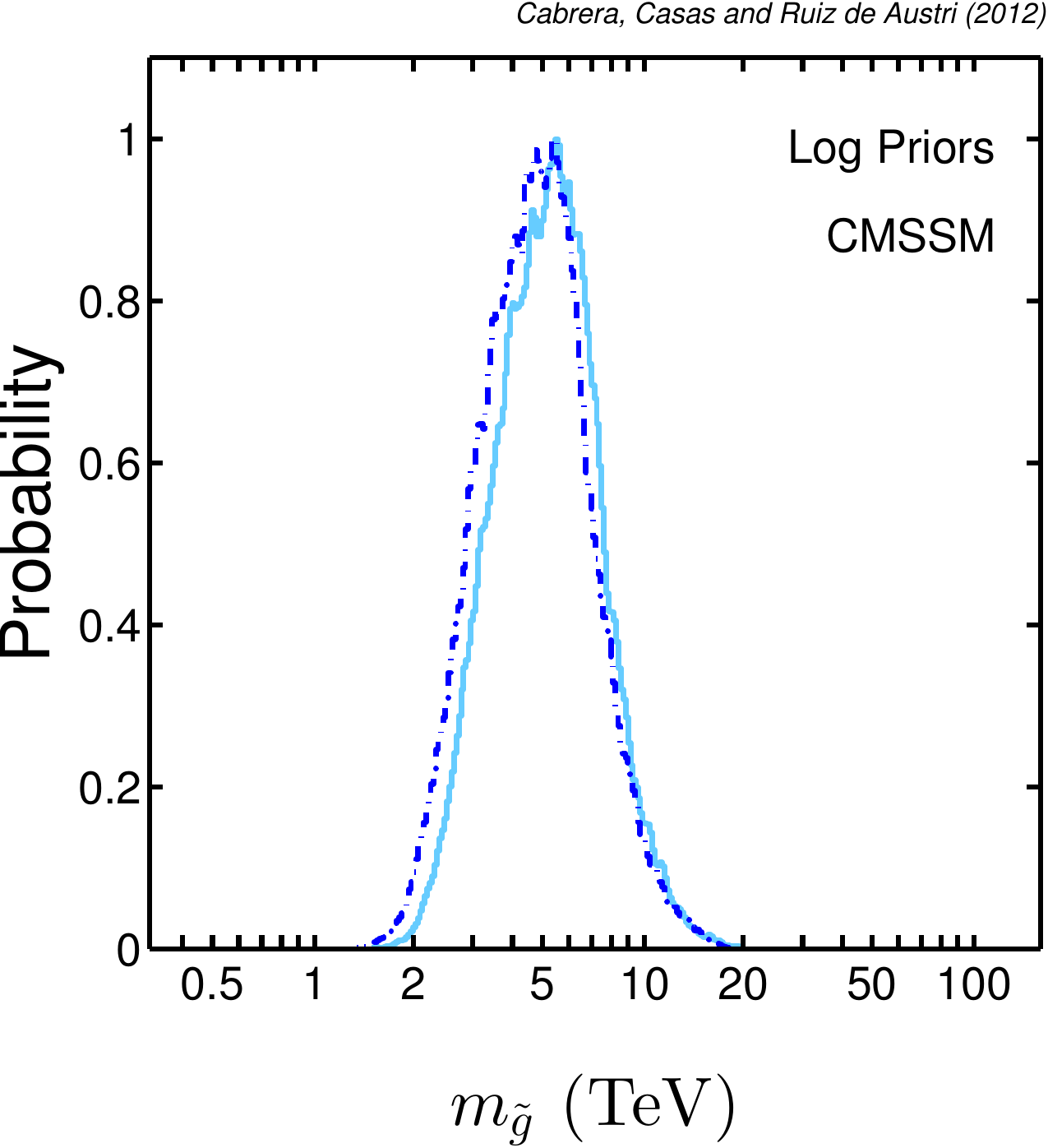} \hspace{1.2cm}
\includegraphics[angle=0,width=0.35\linewidth]{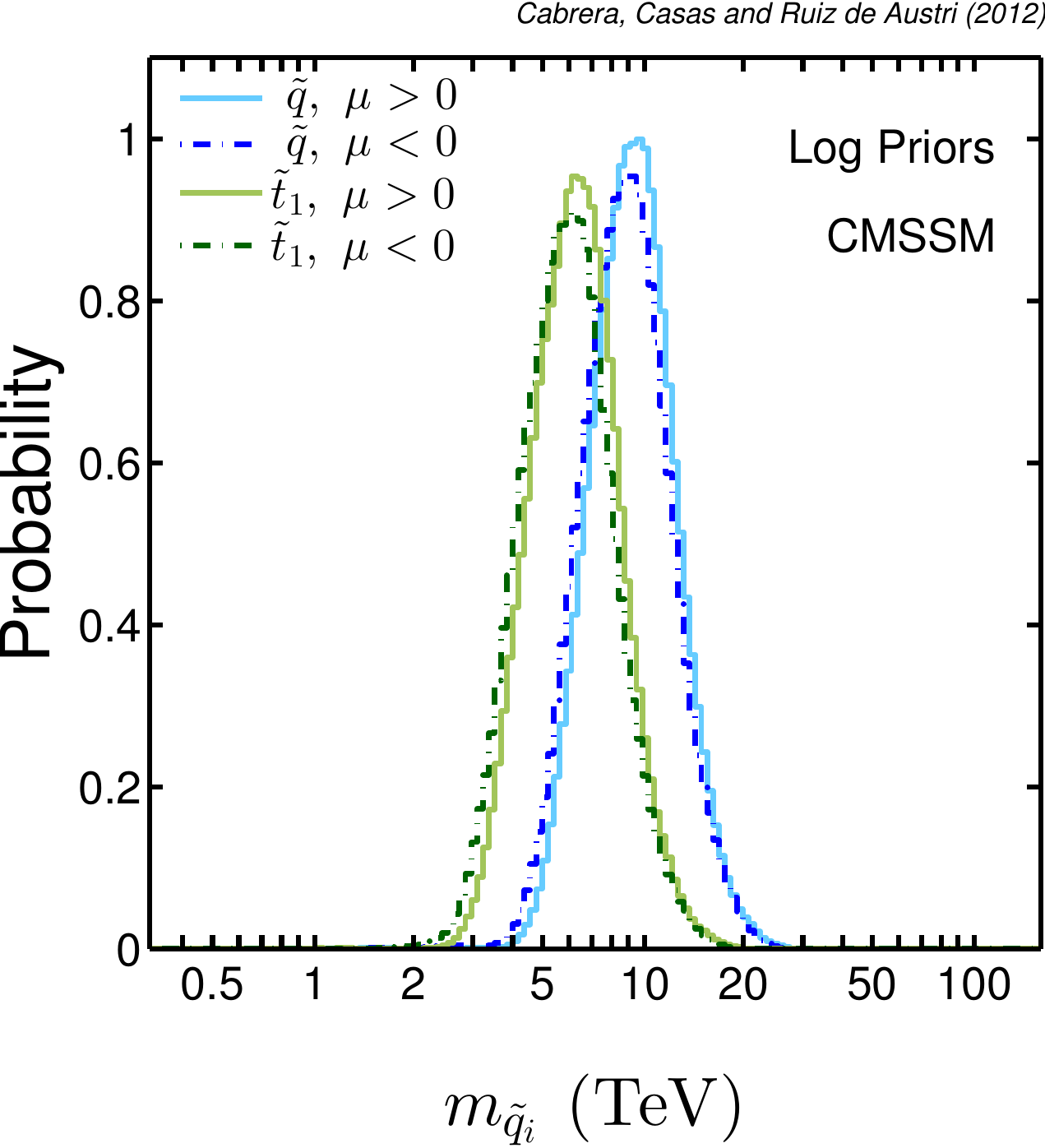}
\caption[test]{As in fig.~\ref{fig:cmssm_all_scdm_1d} but considering that the LSP is not the solely DM component.\label{fig:cmssm_all_mcdm_1d}}
\end{center}
\end{figure}

\begin{figure}[t]
\begin{center}
\includegraphics[angle=0,width=0.35\linewidth]{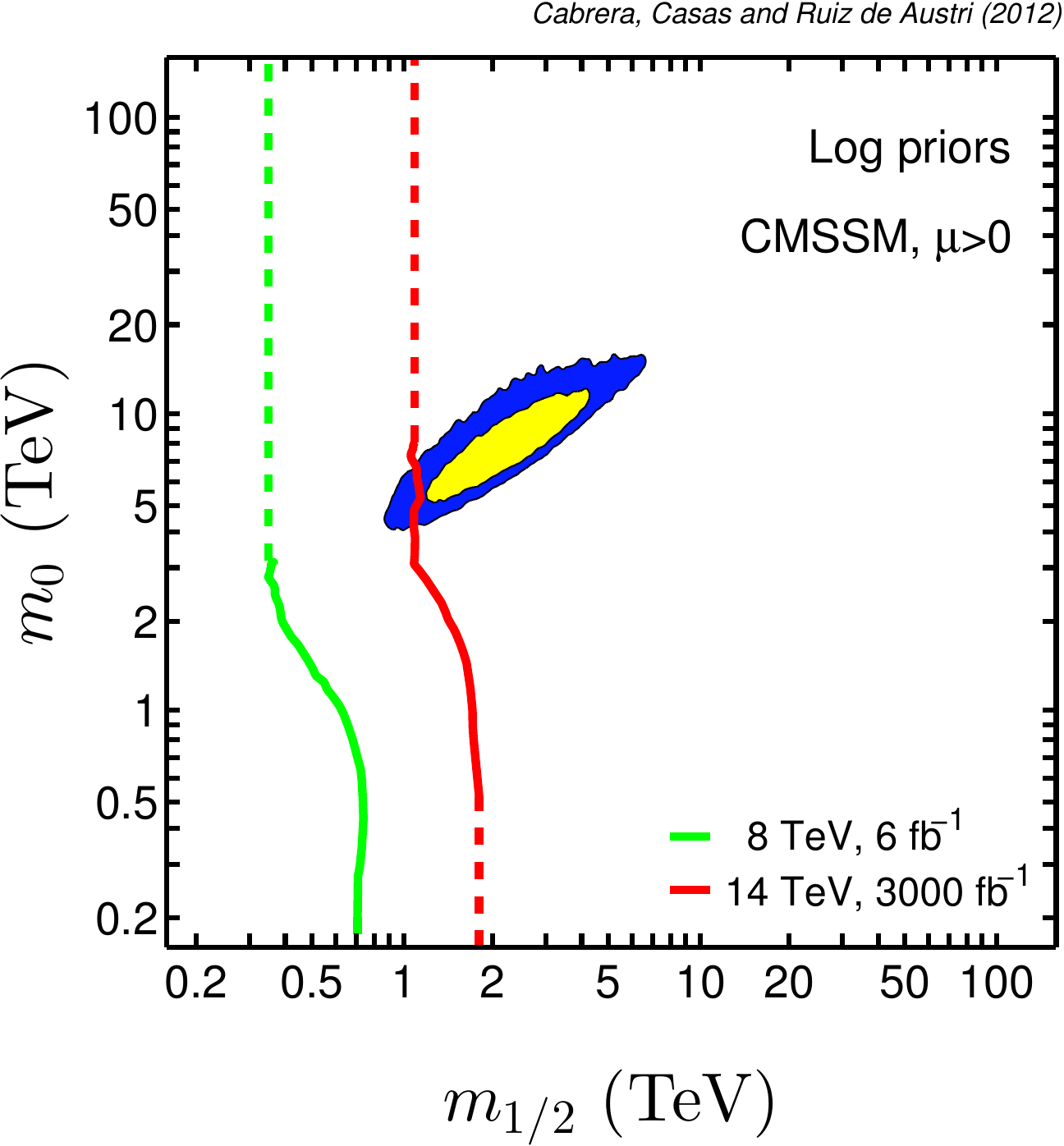} \hspace{1.2cm}
\includegraphics[angle=0,width=0.35\linewidth]{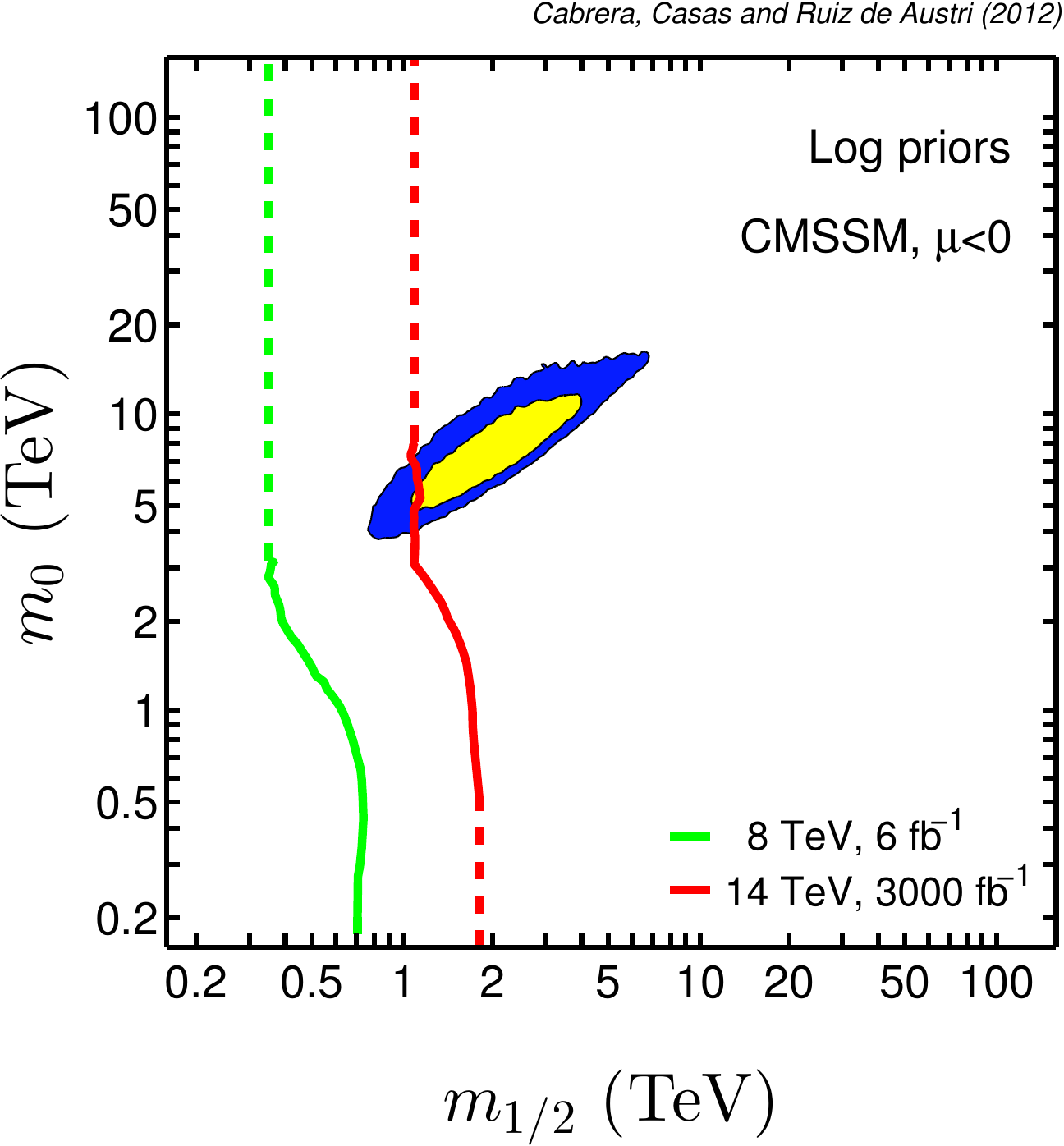} \\ \vspace{0.5cm}
\includegraphics[angle=0,width=0.35\linewidth]{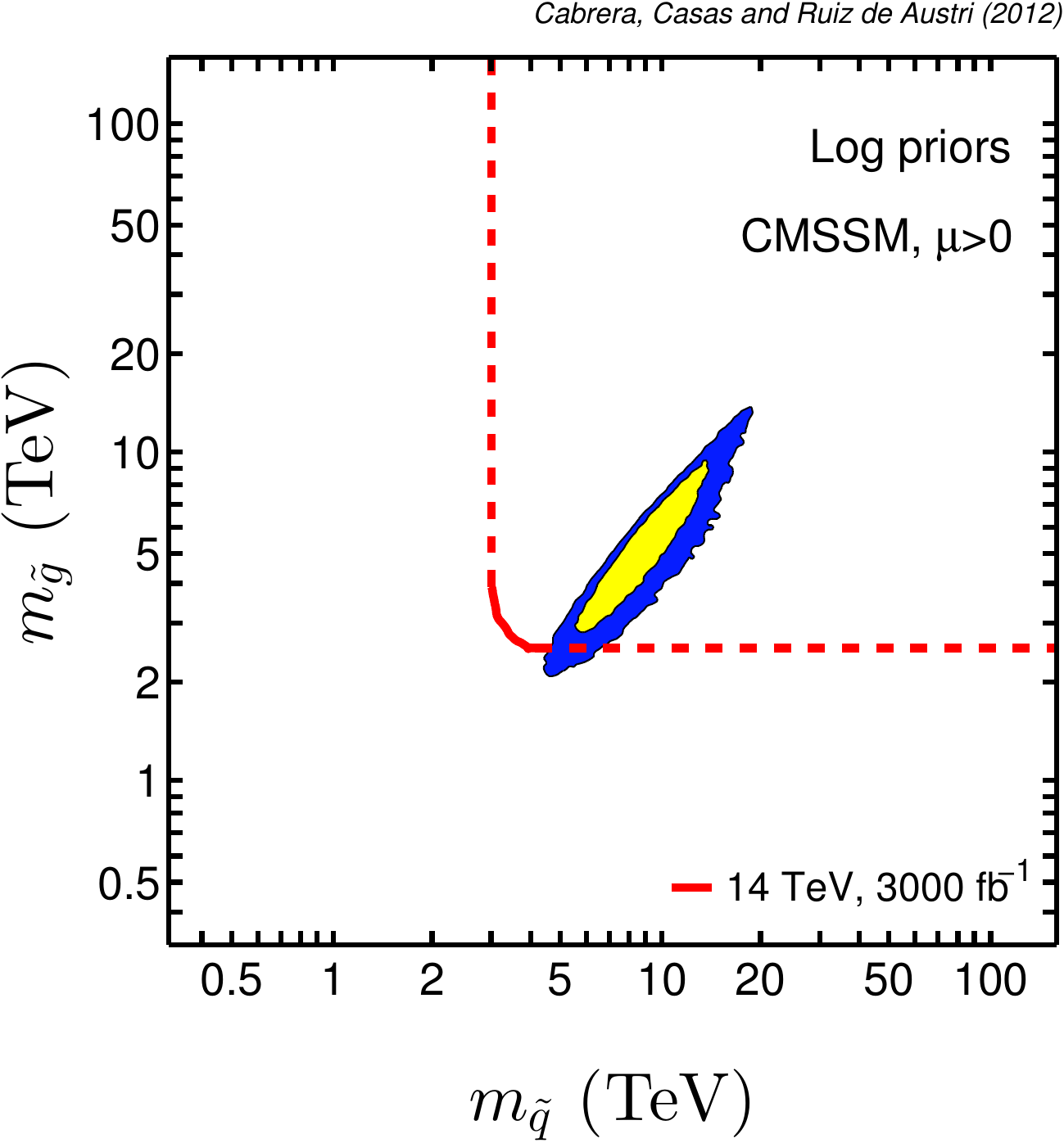} \hspace{1.2cm}
\includegraphics[angle=0,width=0.35\linewidth]{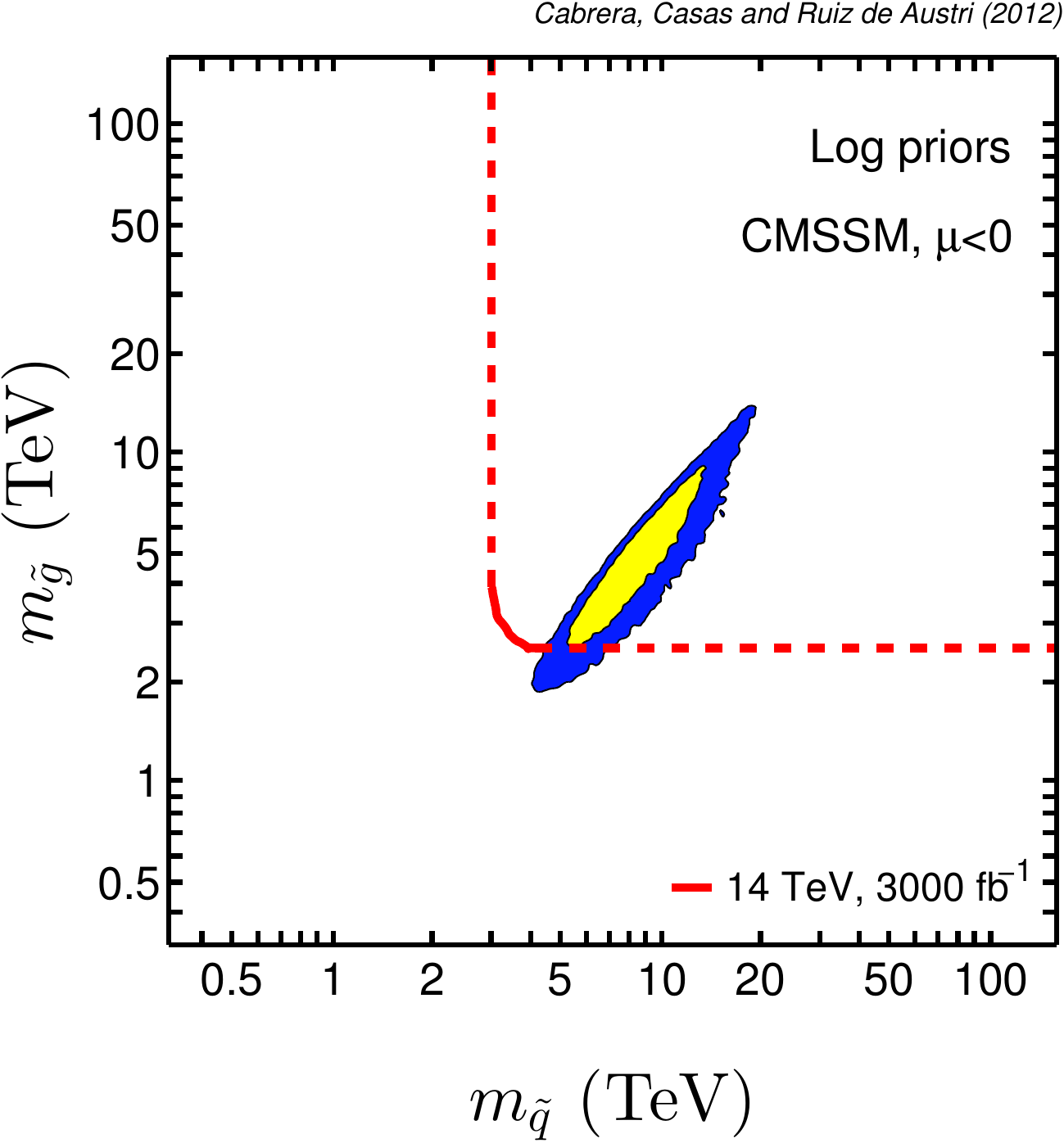}
\caption[test]{As in fig.~\ref{fig:cmssm_all_scdm_2d_1} but considering that the LSP is not the solely DM component.\label{fig:cmssm_all_mcdm_2d_1}}
\end{center}
\end{figure}

The effect of allowing a multi-component DM in the 1D posterior pdfs is displayed in 
Figs.~\ref{fig:cmssm_all_mcdm_1d}, \ref{fig:cmssm_all_mcdm_2d_1}, which should be compared with the previous Figs.~\ref{fig:cmssm_all_scdm_1d}, \ref{fig:cmssm_all_scdm_2d_1}, when the LSP was assumed to be the only component of DM. The most notable effect is that now large areas with lighter $m_{1/2}$ are recovered. This is easy to understand. As mentioned above, if the neutralino is pure Higgsino, for $m_\chi<1$ TeV, the LSP abundance is less than the observed DM abundance. This does not represent now any problem, as one assumes that the DM has other components. Hence, smaller values of $m_\chi$, or equivalently of $\mu$, are now rescued. The condition for the Higgsino to be the LSP is that the bino mass is not smaller than $\mu$. Consequently, $m_{1/2}$ (and thus $m_{\tilde g}$) can now be smaller, as reflected in Figs.~\ref{fig:cmssm_all_mcdm_1d}, \ref{fig:cmssm_all_mcdm_2d_1}. On the other hand, it is now possible for  $m_{1/2}$ to go even below that limit. The reason is 
 that, although in that case the neutralino becomes a mixture of Higgsino and bino, which is potentially dangerous from XENON limits, the rescaling factor $\xi$ reduces the predicted signal of recoil events. The actual value of  $\xi$ depends on the neutralino mass and the composition of the mixture.

The effect of all this is to bring $m_{1/2}$ to lower values, leading to gluino and squark masses of $\mathcal{O}(1)$ TeV 
within the LHC reach. This can be seen in the lower panel of fig.~\ref{fig:cmssm_all_mcdm_1d}. Therefore we conclude that, regarding  LHC searches, there are much better detection prospects in the case of multi-component DM than in the 
single-component DM scenario.

\begin{figure}[t]
\begin{center}
\includegraphics[angle=0,width=0.35\linewidth]{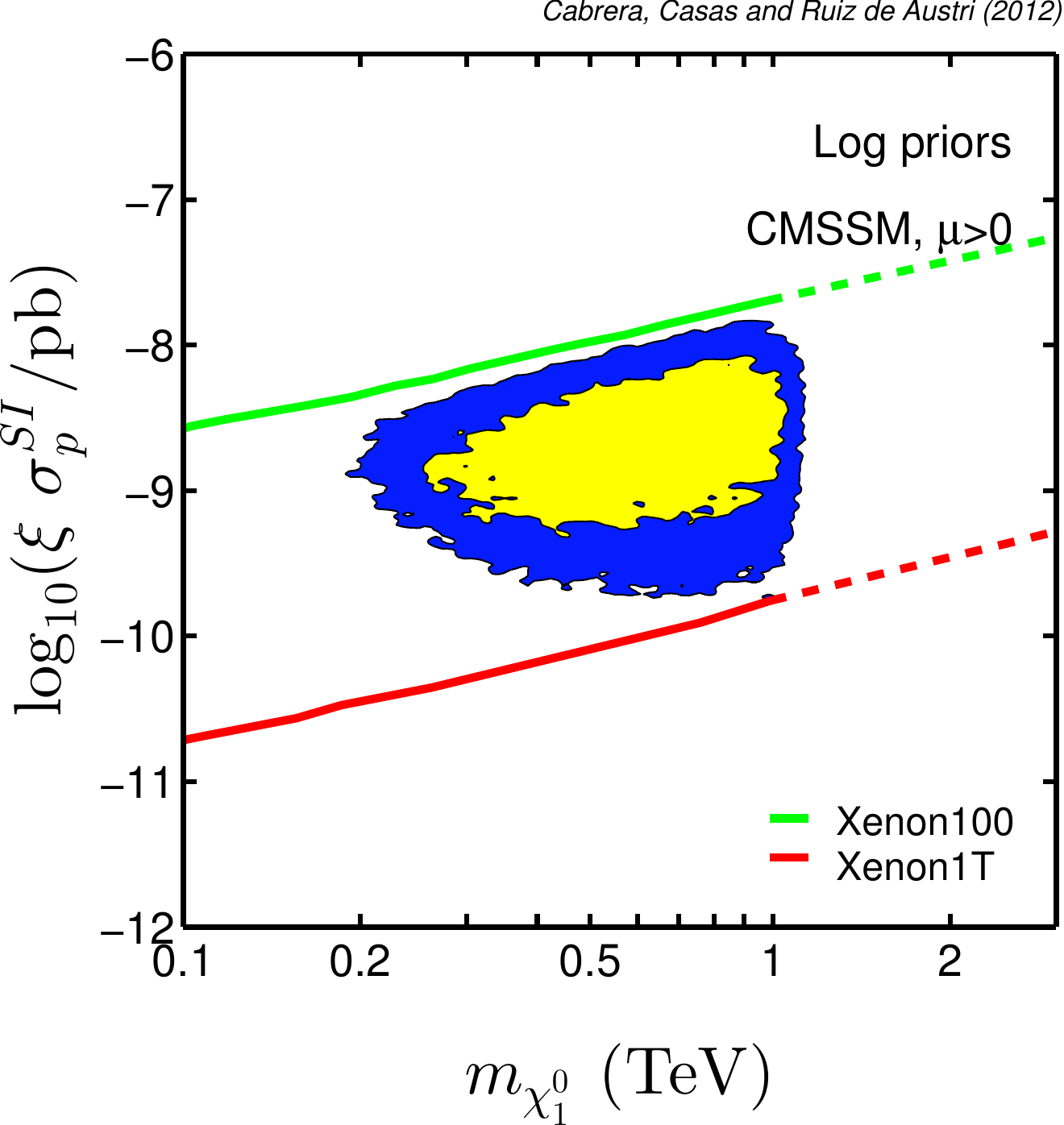} \hspace{1.2cm}
\includegraphics[angle=0,width=0.35\linewidth]{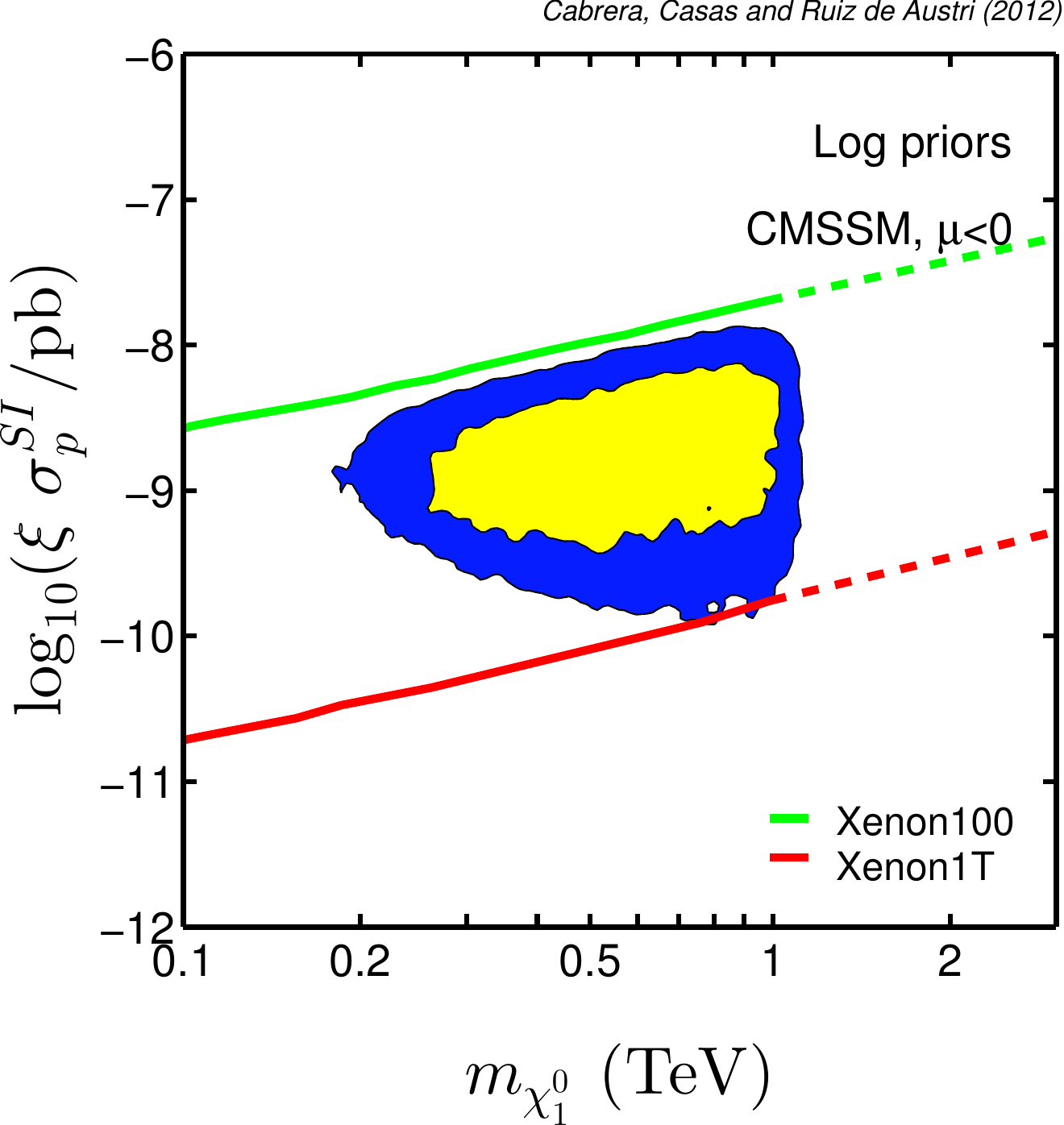}
\caption[test]{As in fig.~\ref{fig:cmssm_all_scdm_2d_2} but considering that the LSP is not the solely DM component.\label{fig:cmssm_all_mcdm_2d_2}}
\end{center}
\end{figure}

Finally, in fig.~\ref{fig:cmssm_all_mcdm_2d_2} we show the probability in the ($m_\neut, \xi \sigma_p^{SI}$) plane.  The shape of the distribution is easy to understand from the above discussion. Neutralino masses of $\mathcal{O}(100)$ GeV become now allowed at the 
95 \% Bayesian credibility level, though the upper bound remains at of $\sim$ 1 TeV. In addition, values of $\sigsip > 10^{-8}$ pb are now within the 95 \% 
level due to the suppression caused by the $\xi$ factor in predicted recoil events signal. The XENON1T expected 90 \% exclusion limit contour is also displayed, showing that the region favored by the data at the 95 \% level can be fully probed by 
future tone-scale direct detection experiments.

\subsection{g-2}

The magnetic anomaly of the muon, $a_\mu= \frac{1}{2}(g-2)_\mu$ has been a classical and powerful test for new physics. 
The theoretical determination of $a_\mu$ depends crucially on the evaluation of the contribution coming from the hadronic vacuum-polarization diagram. This can be expressed in terms of the total hadronic cross section $e^+ e^-\rightarrow$ had or, alternatively, in terms of the hadronic $\tau-$decay data, which are theoretically related to the $e^+e^-$ hadronic cross section. Both approaches lead to different results, although in the last times they have notably converged. On the other hand, in either case the results are in remarkable disagreement with the experimental determination. More precisely, defining $\delta a_\mu = \delta^{\rm exp}a_\mu- \delta^{\rm SM}a_\mu$, the recent results are $\delta a_\mu = (25.5\pm 8.0)\times 10^{-10}$ (using $e^+e^-$ data) and $\delta a_\mu = (15.7\pm 8.2)\times 10^{-10}$ (using $\tau-$decay data) \cite{Davier:2009zi,Davier:2009ag}. This corresponds to a discrepancy of 3.2$\sigma$ and 1.9$\sigma$ respectively.

This potential signal of new physics may be attributed to supersymmetric contributions to the muon anomaly, $\delta^{\rm MSSM}a_\mu$, which arise mainly from 1-loop diagrams with chargino or neutralino exchange \cite{Degrassi:1998es,Heinemeyer:2003dq,Heinemeyer:2004yq,Marchetti:2008hw,vonWeitershausen:2010zr}. This contribution i ncreases with increasing $\tan\beta$ and decreasing supersymmetric masses. The problem is that the large Higgs mass revealed by the experiment, well above $M_Z$, requires large supersymmetric masses, which is hardly conciliable with a sizeable $\delta^{\rm MSSM}a_\mu $. This is the reason why we have not used the $g-2$ result in the computation of the likelihood in our analysis. Besides, the $g-2$ computation has been rather controversial in the last years and probably the situation is not clear enough yet.

The Bayesian approach allows for a quantification of the discrepancy between the observed Higgs mass and the assumption that the supersymmetric contributions are responsible for  $\delta a_\mu$. We briefly describe the procedure (for more details see ref.\cite{Trotta:2005ar,Trotta:2008qt,Feroz:2009dv}), which will also be used in sect.6 in other context. The idea is to separate the complete set of data in two subsets:
\bea
\label{DDD}
\{{\rm data}\}  = \{\mathscr{D}, D \} .
\eea
Here $\mathscr{D}$ represents the subset of observations, whose compatibity with the rest of the observations, ${D}$, we want to test. In our case, $\mathscr{D}$ is the experimental value of $a_\mu$, whereas $D$ denotes the remaining experimental data, including the experimental Higgs mass. Then one constructs the quantity $p(\mathscr{D}| D)$, i.e. the probability of getting a certain result when measuring $\mathscr{D}\equiv a_\mu$, given the known values of the remaining observables,
\bea
\label{pDD}
p(\mathscr{D}| D)\ =\ \frac{p(\mathscr{D}, D)}{p(D)}.
\eea
Here $p(\mathscr{D}, D)$ is probability of measuring both sets of data at the same time, while $p(D)$ is the analogue, but just for the $D$ subset. These probabilities are given by the integral of the likelihood times the prior in the parameter space, and therefore coincide with the normalization factor appearing in denominator of eq.(\ref{Bayes}), sometimes called the {\em evidence}. E.g.
\bea
\label{BayesEvid}
p(\mathscr{D}, D)=p({\rm data})\ =\ \int d\theta_1\cdots\ d\theta_N \  p({\rm data}|\theta_i)\ p(\theta_i).
\eea
Now, the consistency of $\mathscr{D}^{\rm obs}$ (i.e. the actually measured muon anomaly) with the rest of data, $D$, can be tested by comparing the probability $p(\mathscr{D}^{\rm obs}|D)$ with the one obtained for the value of $\mathscr{D}$ that maximizes such probability, say  $\mathscr{D}^{\rm max}$ (assuming the same reported error at the new central value). $\mathscr{D}^{\rm max}$ is somehow the "prediction" of the model for the muon anomaly, given the rest of experimental data. The quantity
\bea
\label{Ltest}
\mathscr{L} (\mathscr{D}^{\rm obs}|D) \ \equiv\ 
\frac{p(\mathscr{D}^{\rm obs}|D)}{p(\mathscr{D}^{\rm max}|D)}\ =\ 
\frac{p(\mathscr{D}^{\rm obs},D)}{p(\mathscr{D}^{\rm max},D)}
\eea
is a global likelihood ratio (integrated to all possible values of the parameters of the model). Note that the $p(D)$ factor cancels out in the expression of $\mathscr{L} (\mathscr{D}^{\rm obs}|D)$, which is simply given by the ratio of the global evidences. This was called the $\mathscr{L}-$test in ref.\cite{Feroz:2009dv}. If this likelihood has a Gaussian-like profile, the quantity  $-2\ln \mathscr{L} (\mathscr{D}^{\rm obs}|D)$ can be interpreted as $\sigma-$deviation from the theoretical prediction.

Obviously, for very large $m_h$, the supersymmetric particles must be very heavy. In that limit SUSY decouples, so $\mathscr{D}^{\rm max}$ should approach the SM prediction, i.e. $-2\ln \mathscr{L} (\mathscr{D}^{\rm obs}|D)\rightarrow 3.2^2$ (asuming the theoretical calculation of $a_\mu$ based on $e^+ e^-$ data); which is indeed the case \cite{Cabrera:2010xx}. Now, one can compute $\mathscr{L}$ for any value of $m_h$, in particular for $m_h=125.9\pm 0.4$ GeV, as given by the experiment. Our calculation indicates that, essentially, for this value of the Higgs mass, $\mathscr{L}$ has already reached its maximum (decoupling limit) value. In other words, the tension between the experimental Higgs mass and the experimental $a_\mu$ is already as large as for the SM (and thus larger than 3$\sigma$). This means that in the regions of the CMSSM where the prediction for $a_\mu$ is closer to the experiment than the SM prediction are not favored since they present an onerous disagreement with t
 he observed Higgs mass. In some tiny regions this price can be somewhat alleviated, but they are statistically quite irrelevant, which reflects the cost of the underlying fine-tuning.

\section{NUHM}

Let us now (more briefly) describe the results when the MSSM model considered is the Non
Universal Higgs Masses one (NUHM). Most of the discussion (and many of the results) are in fact analogous and it is not necessary to repeat them here. Hence, we will focus on the new features and their origin.

\begin{figure}[t]
\begin{center}
\includegraphics[angle=0,width=0.35\linewidth]{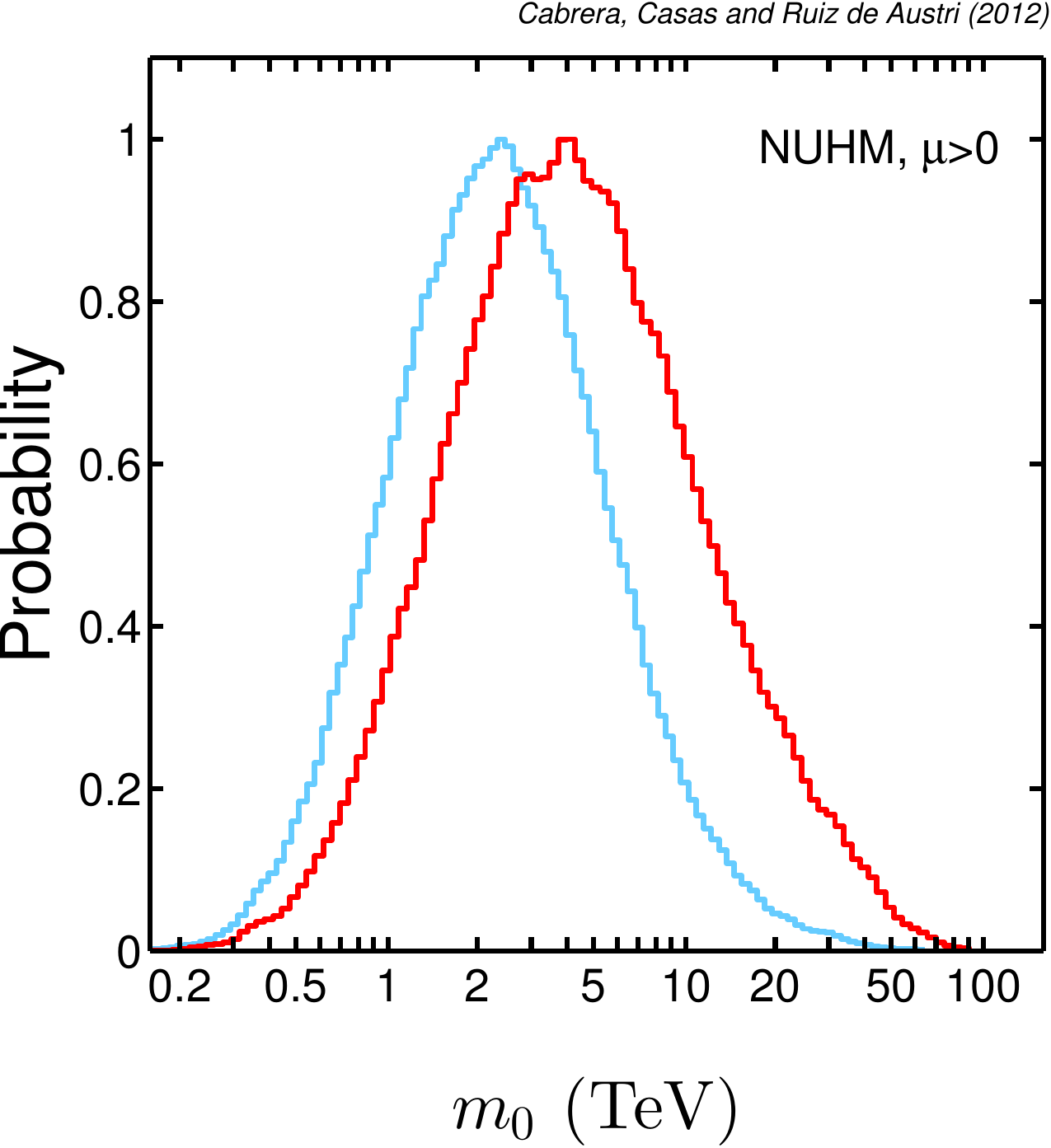} \hspace{1.2cm}
\includegraphics[angle=0,width=0.35\linewidth]{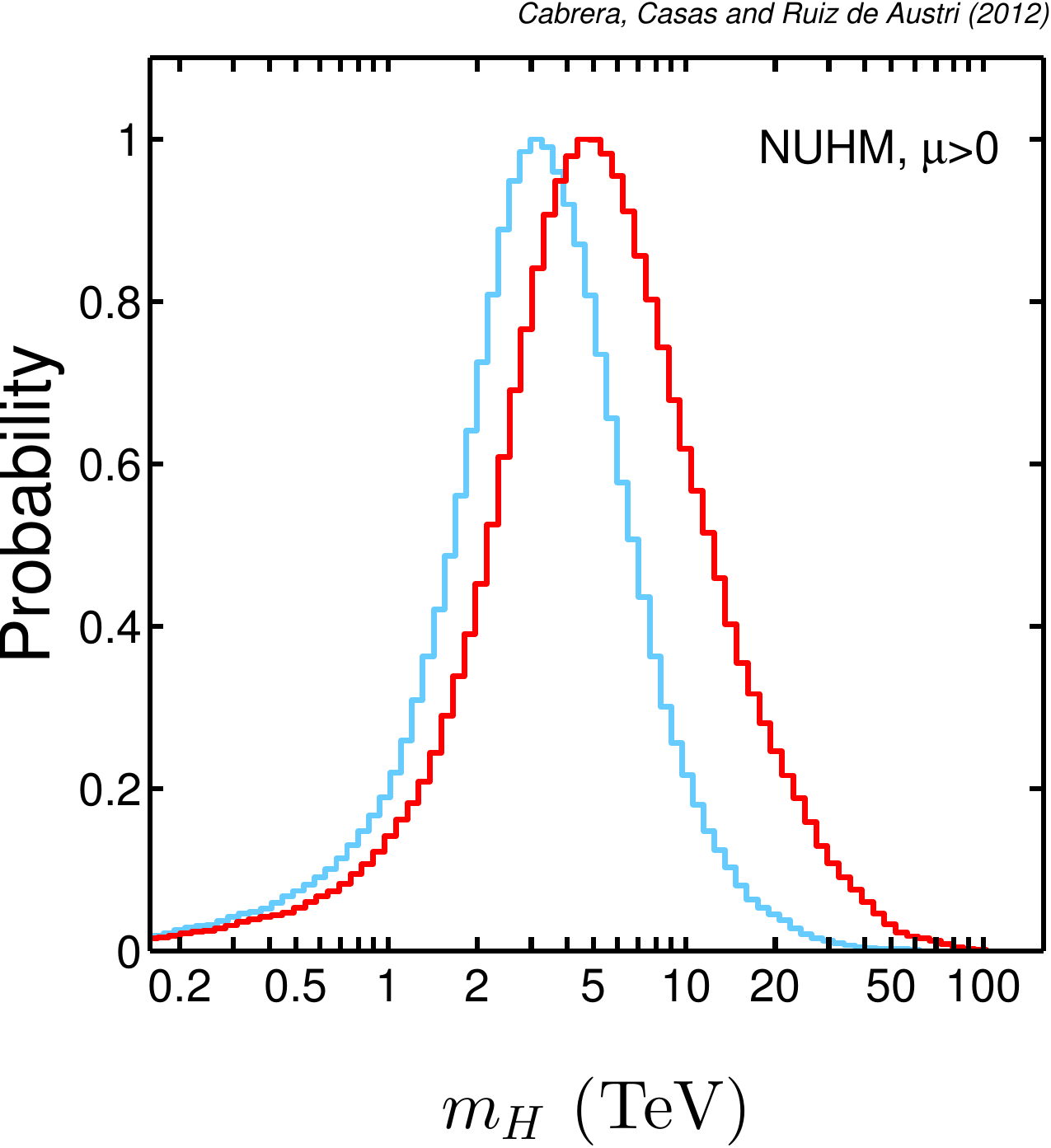}\\ \vspace{0.5cm}
\includegraphics[angle=0,width=0.35\linewidth]{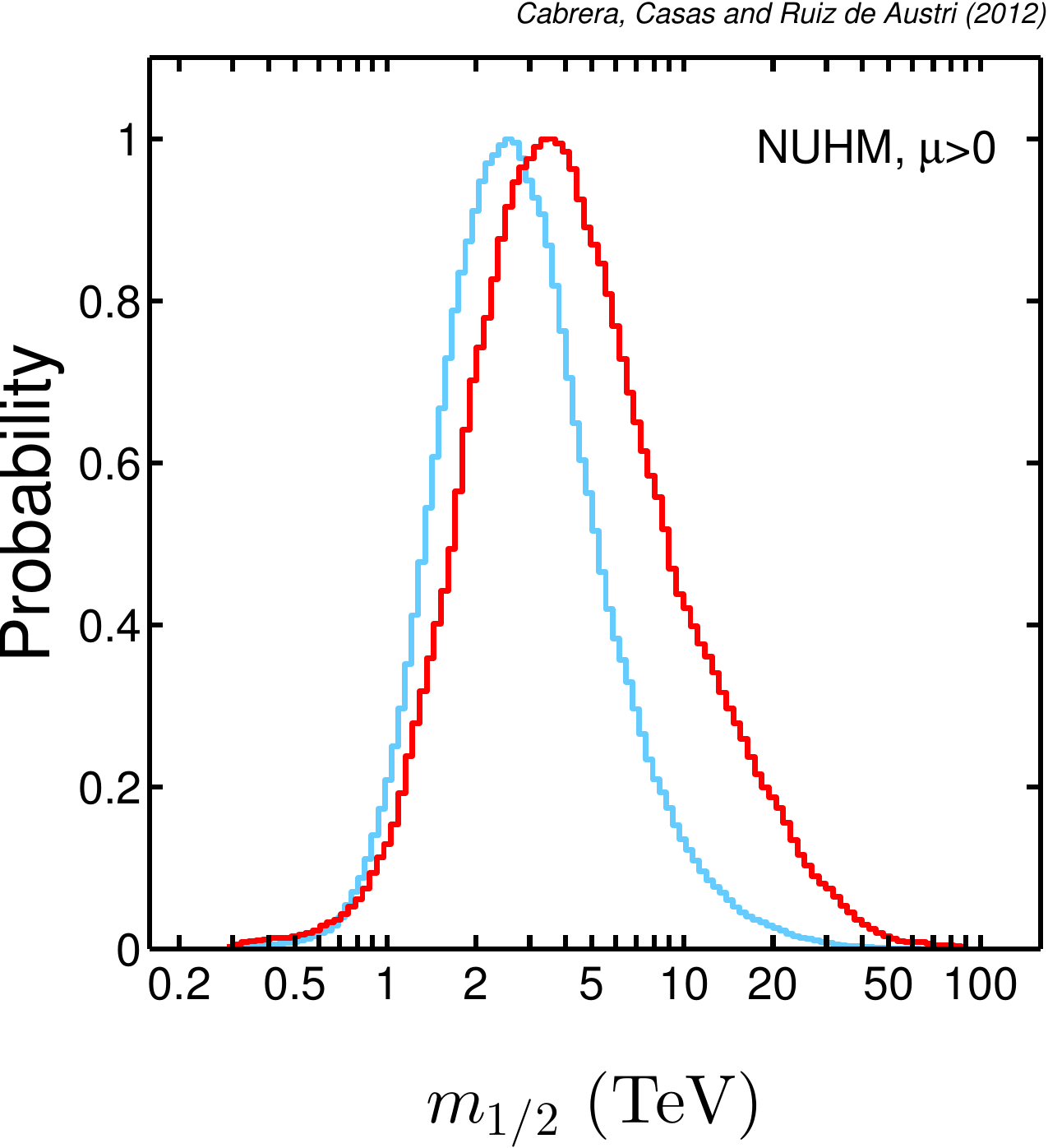} \hspace{1.2cm}
\includegraphics[angle=0,width=0.35\linewidth]{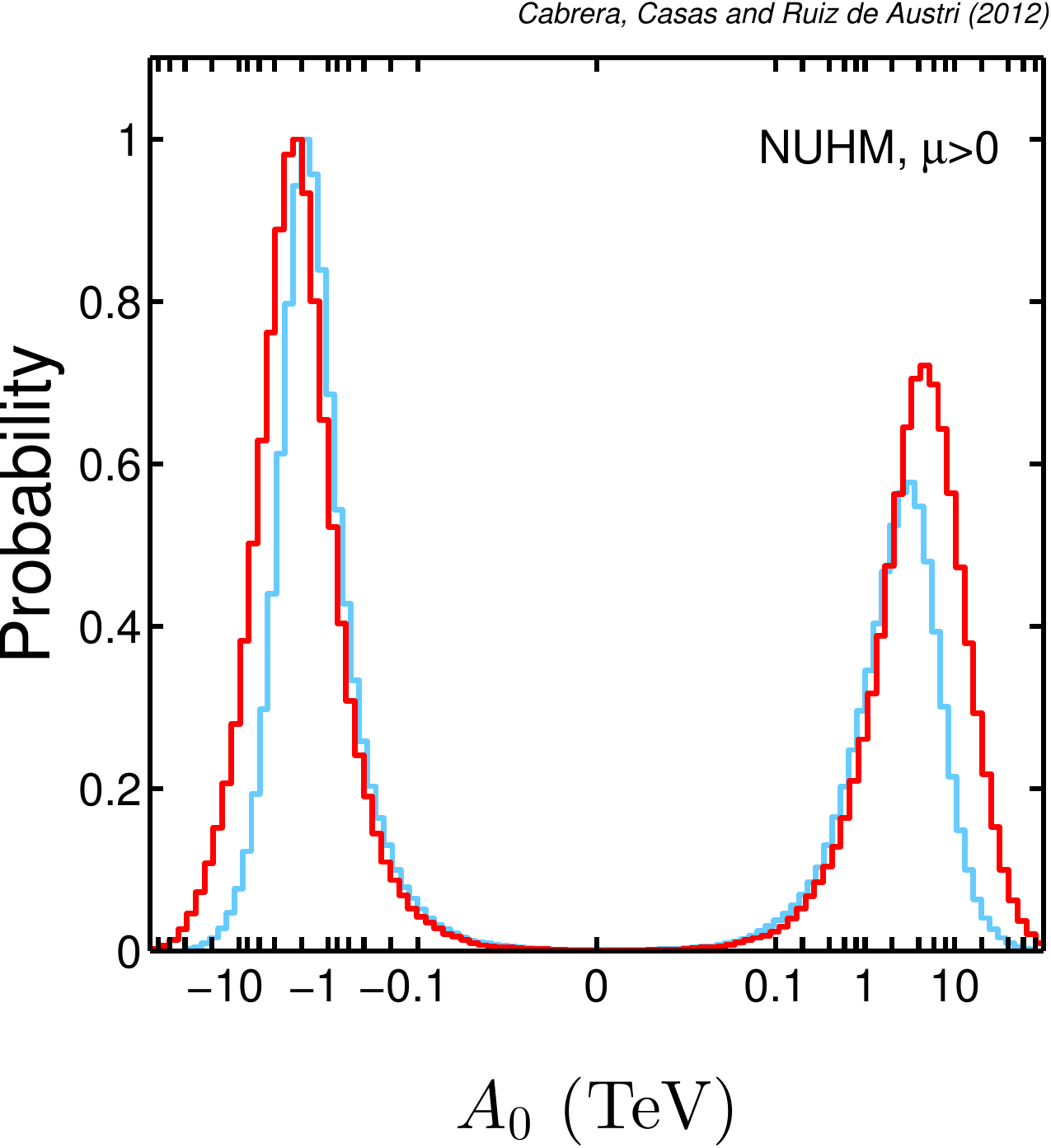} \\ \vspace{0.5cm}
\includegraphics[angle=0,width=0.35\linewidth]{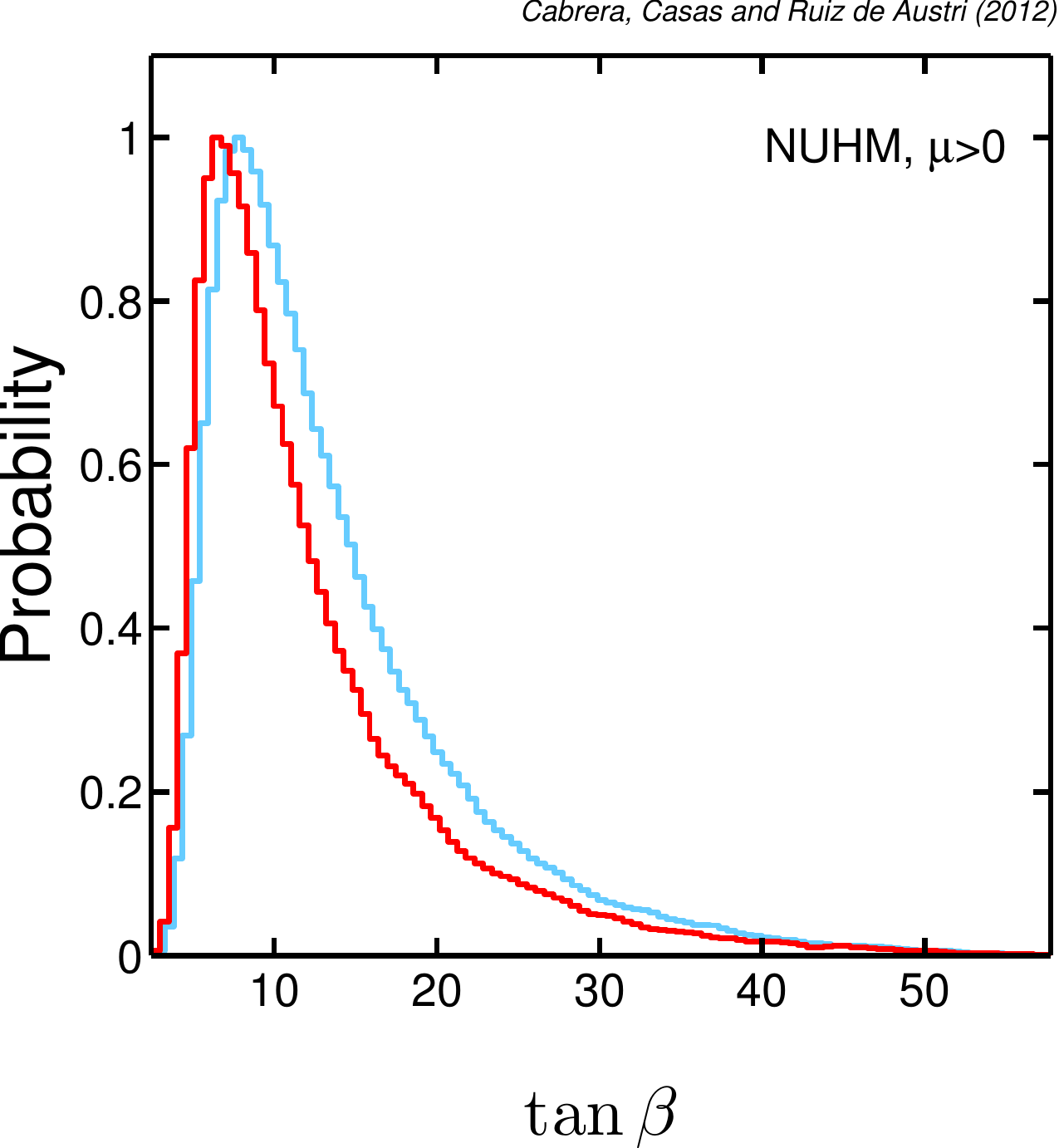}
\caption[test]{1D marginalized posterior probability distribution of the NUHM
  parameters: $m_0,m_H,m_{1/2}, A$ and $\tan \beta$. The red(cyan)
  corresponds to flat(logarithmic) priors. 
\label{fig:nuhm_all_nodm_1d}}
\end{center}
\end{figure}

A slight difference has to do with the choice of the priors. Actually, the priors we have used for the NUHM are the same as the ones described in eq.~(\ref{rangos_mMABmu}) - (\ref{MS_marg_2}), but now we have to include the common mass for the Higgses at $M_X$, i.e. $m_H$ as an additional independent parameter. Then, the log prior reads,

\begin{figure}[t]
\begin{center}
\includegraphics[angle=0,width=0.35\linewidth]{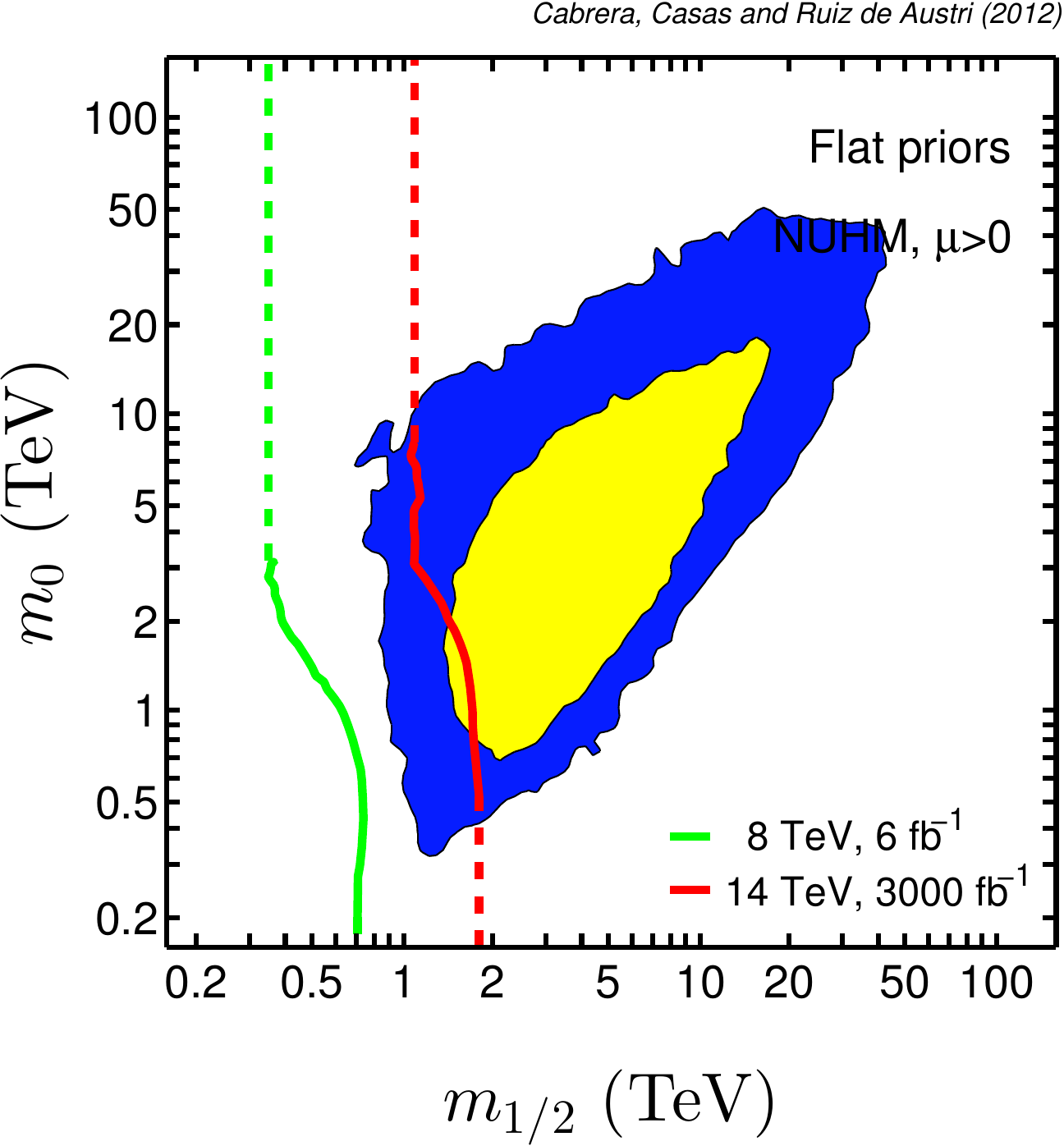} \hspace{1.2cm}
\includegraphics[angle=0,width=0.35\linewidth]{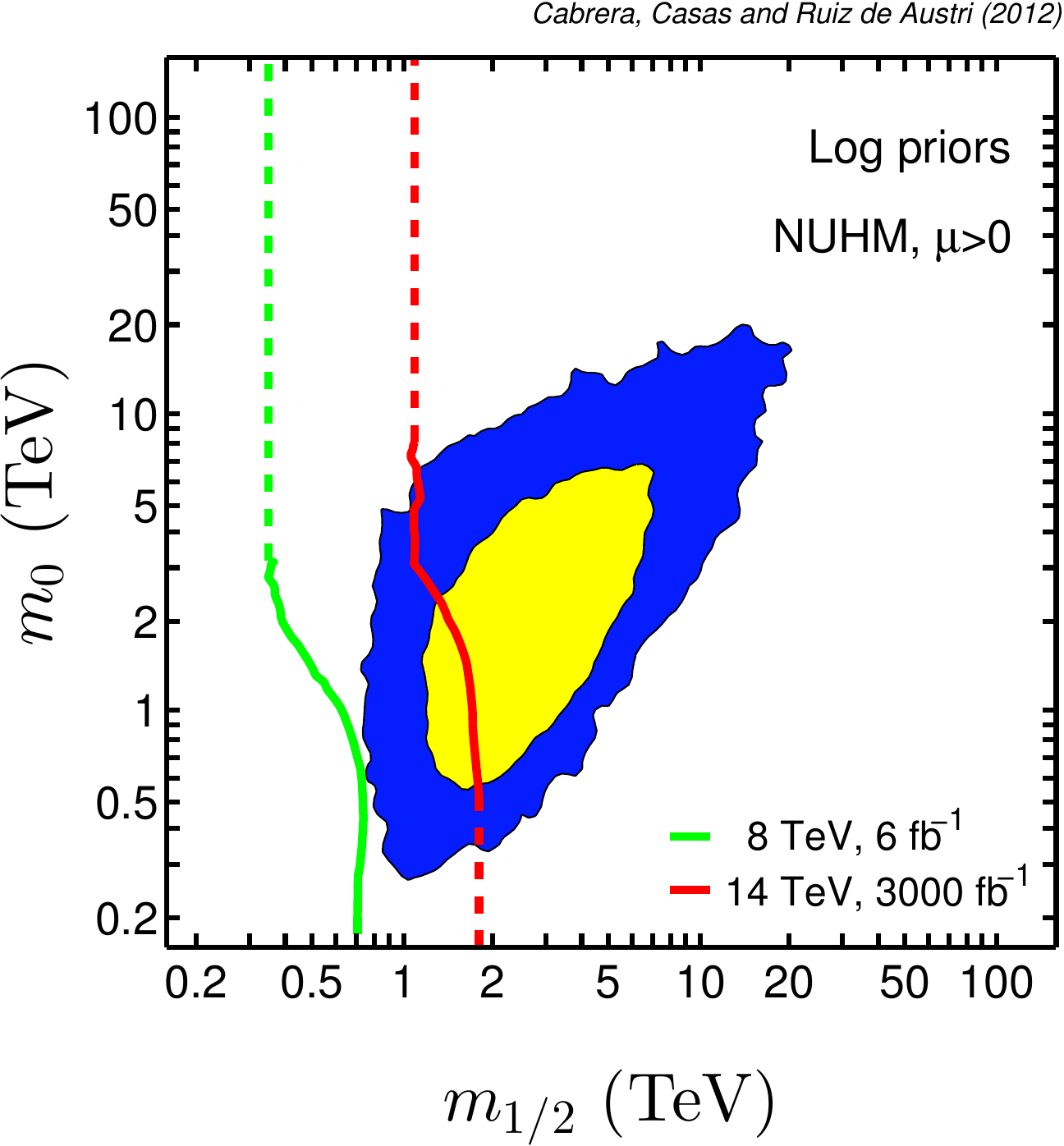}\\ \vspace{0.5cm} 
\includegraphics[angle=0,width=0.35\linewidth]{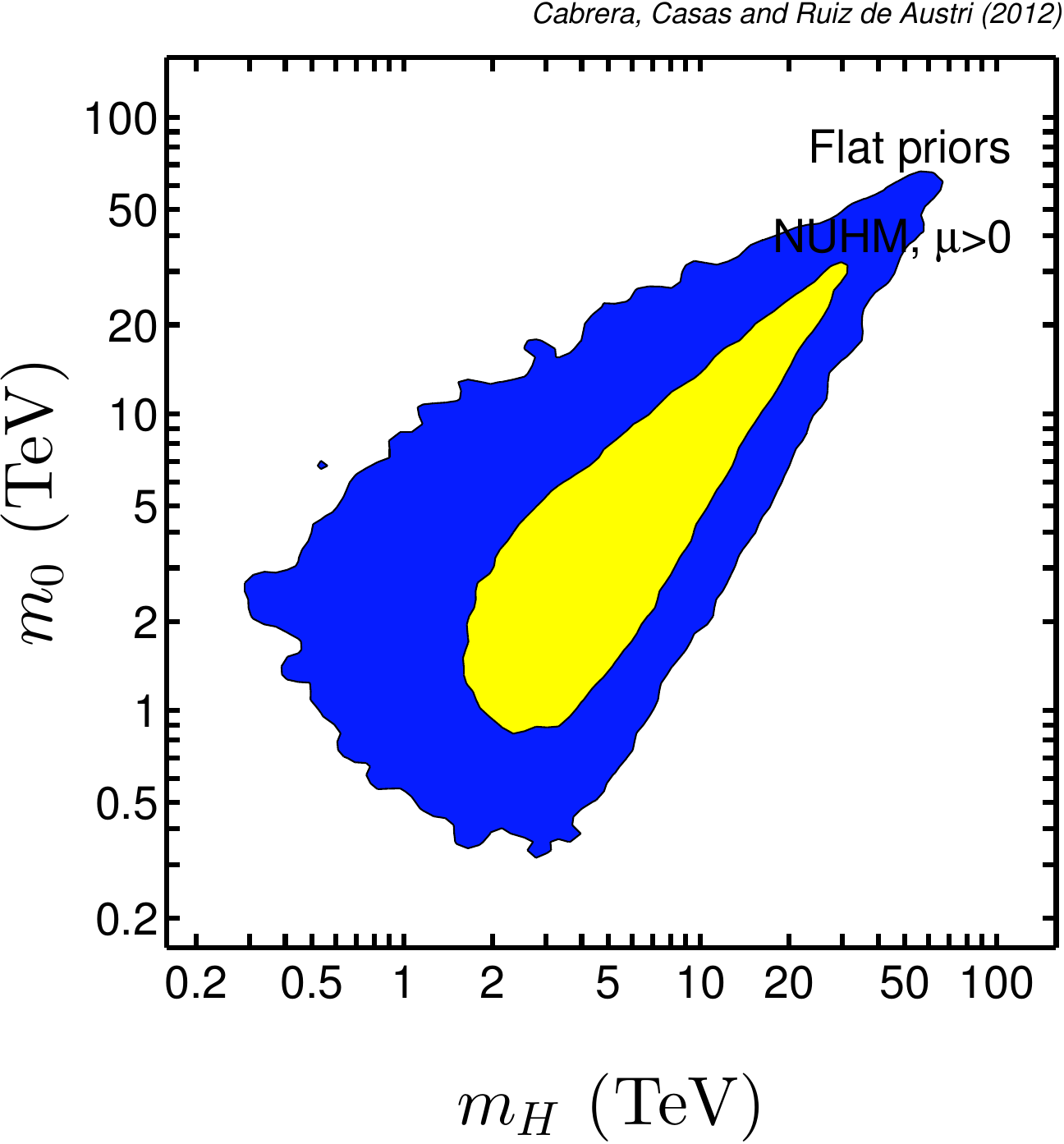} \hspace{1.2cm}
\includegraphics[angle=0,width=0.35\linewidth]{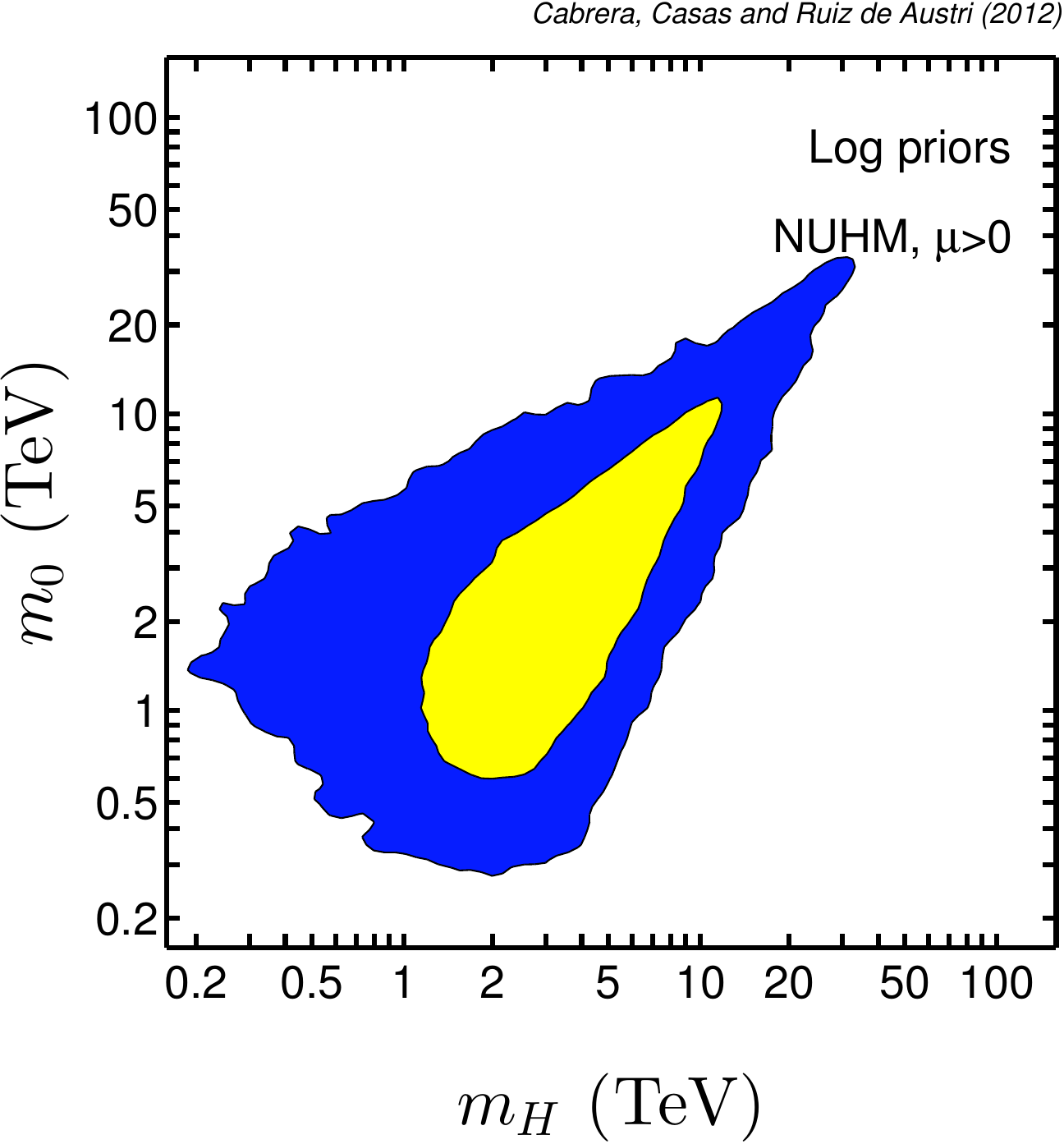}\\ \vspace{0.5cm} 
\includegraphics[angle=0,width=0.35\linewidth]{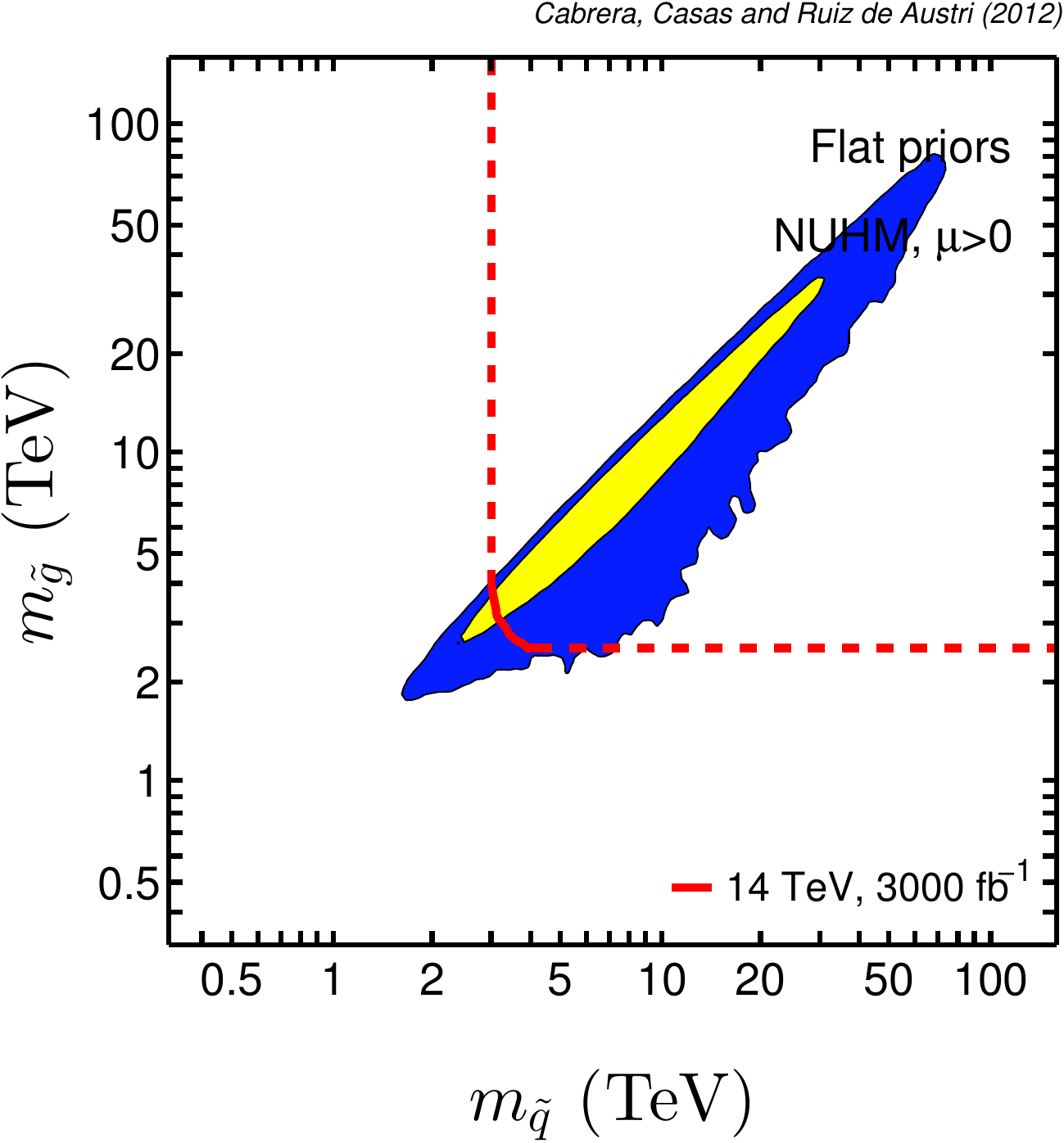} \hspace{1.2cm} 
\includegraphics[angle=0,width=0.35\linewidth]{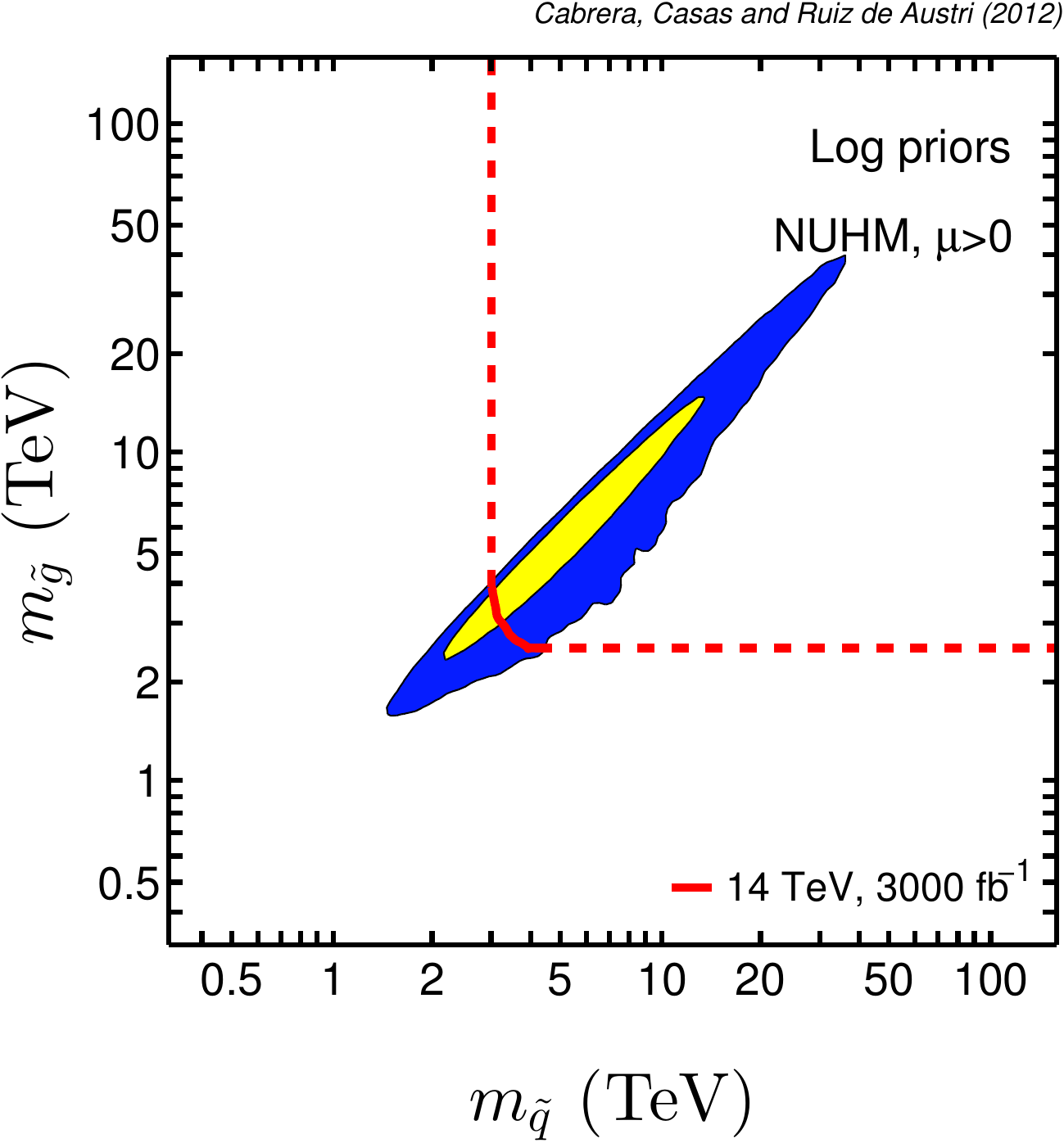}
\caption[test]{2D marginalized posterior probability distribution of the NUHM
  in $m_{1/2}-m_0$, $m_H-m_0$ and $m_{\tilde g}-m_{\tilde q}$ planes.
Left panels show the flat priors case and right panels the logarithmic one.
The inner and outer contours enclose respective 68\% and 95\% joint regions.
\label{fig:nuhm_all_nodm_2d}}
\end{center}
\end{figure}

\begin{figure}[t]
\begin{center}
\includegraphics[angle=0,width=0.35\linewidth]{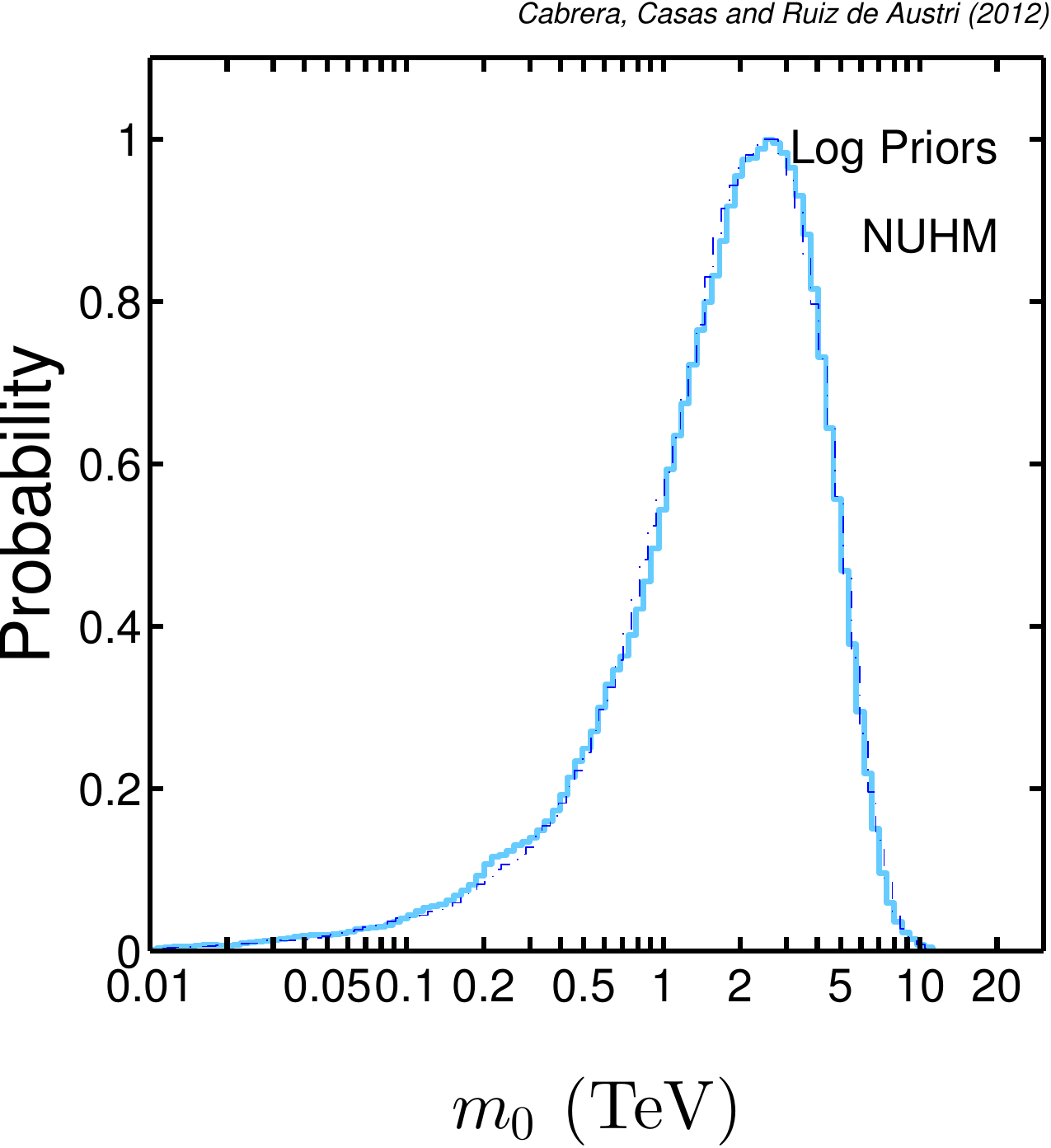} \hspace{1.2cm}
\includegraphics[angle=0,width=0.35\linewidth]{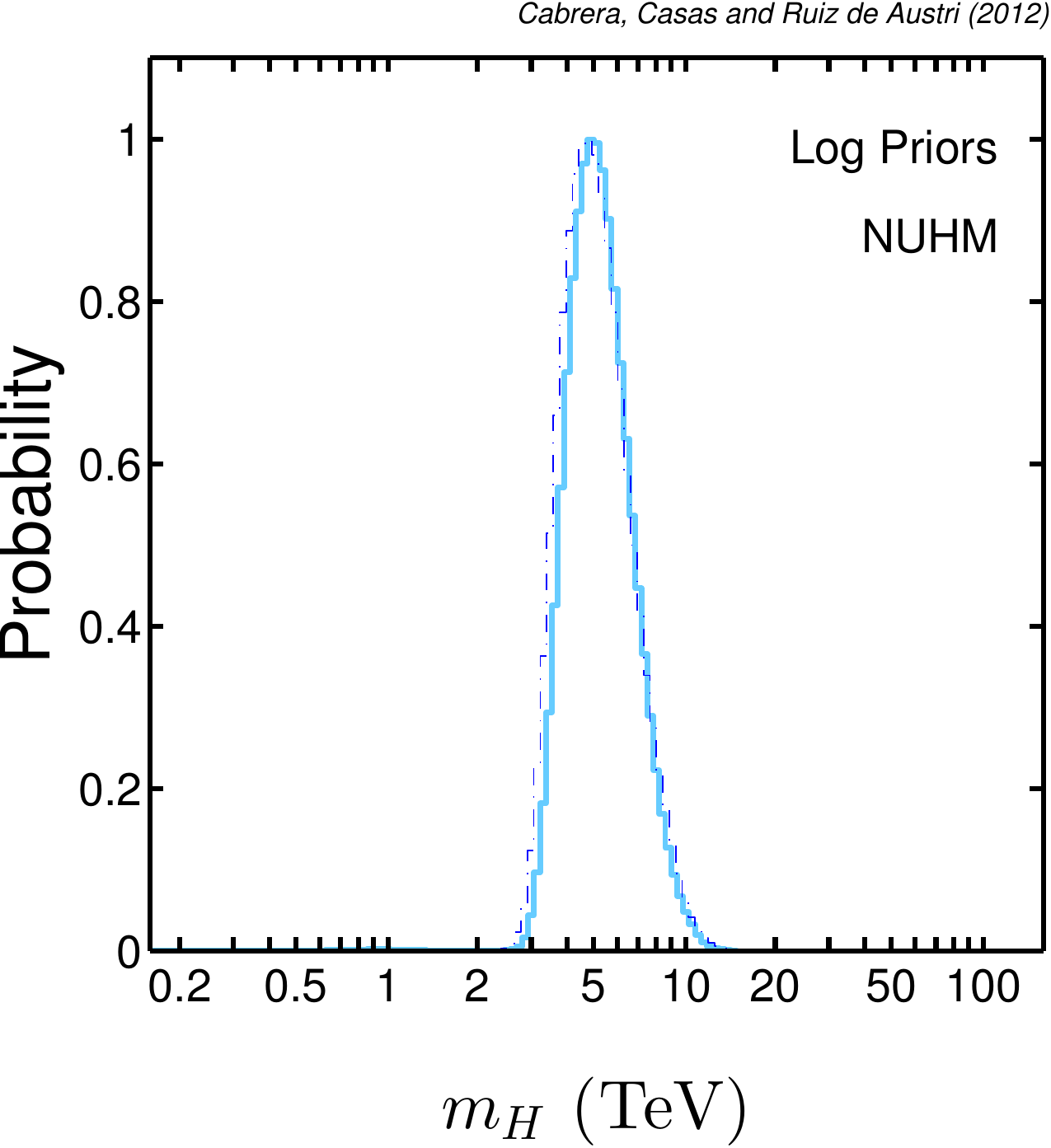} \\ \vspace{0.5cm}
\includegraphics[angle=0,width=0.35\linewidth]{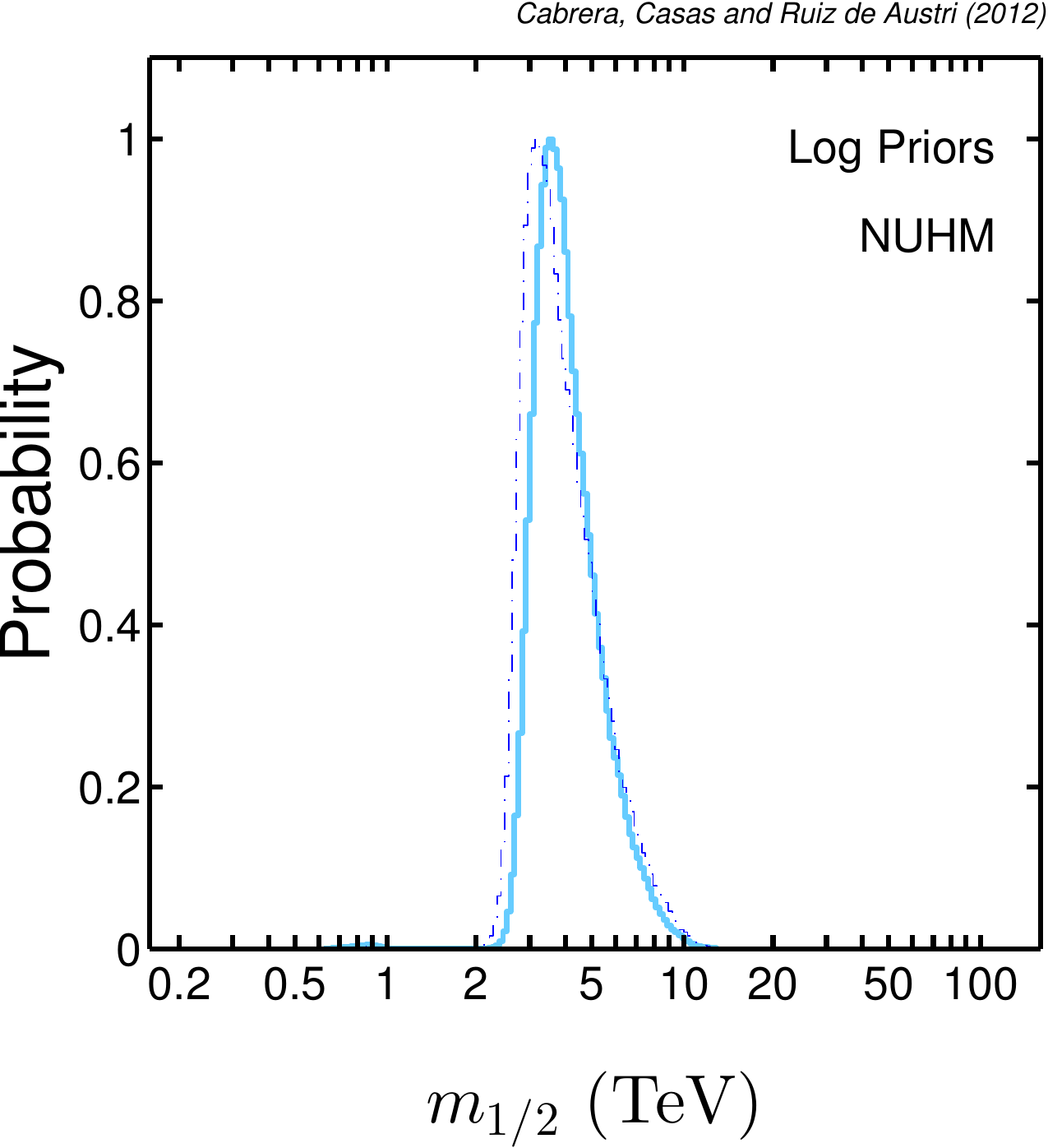} \hspace{1.2cm}
\includegraphics[angle=0,width=0.35\linewidth]{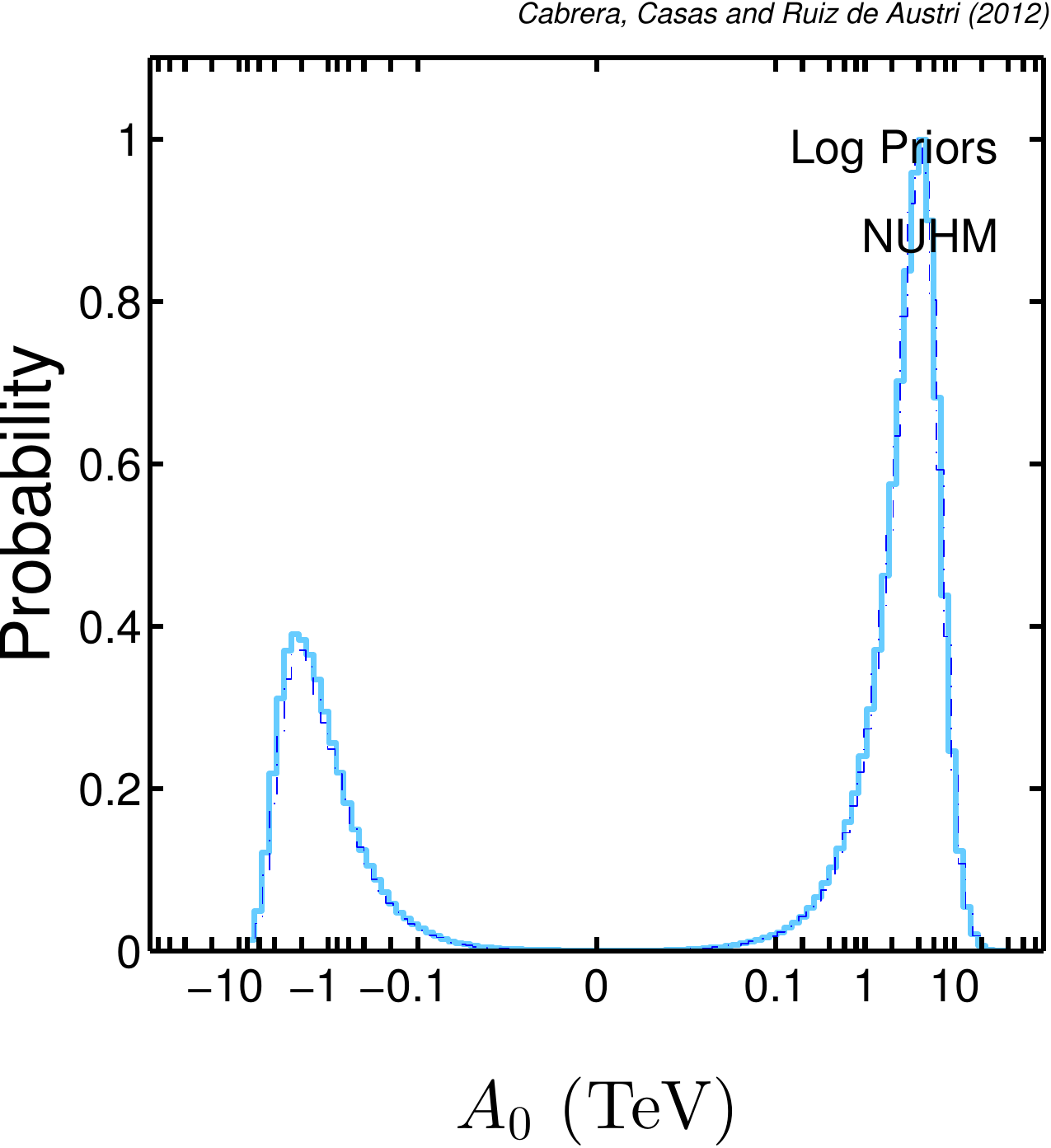} \\ \vspace{0.5cm}
\includegraphics[angle=0,width=0.35\linewidth]{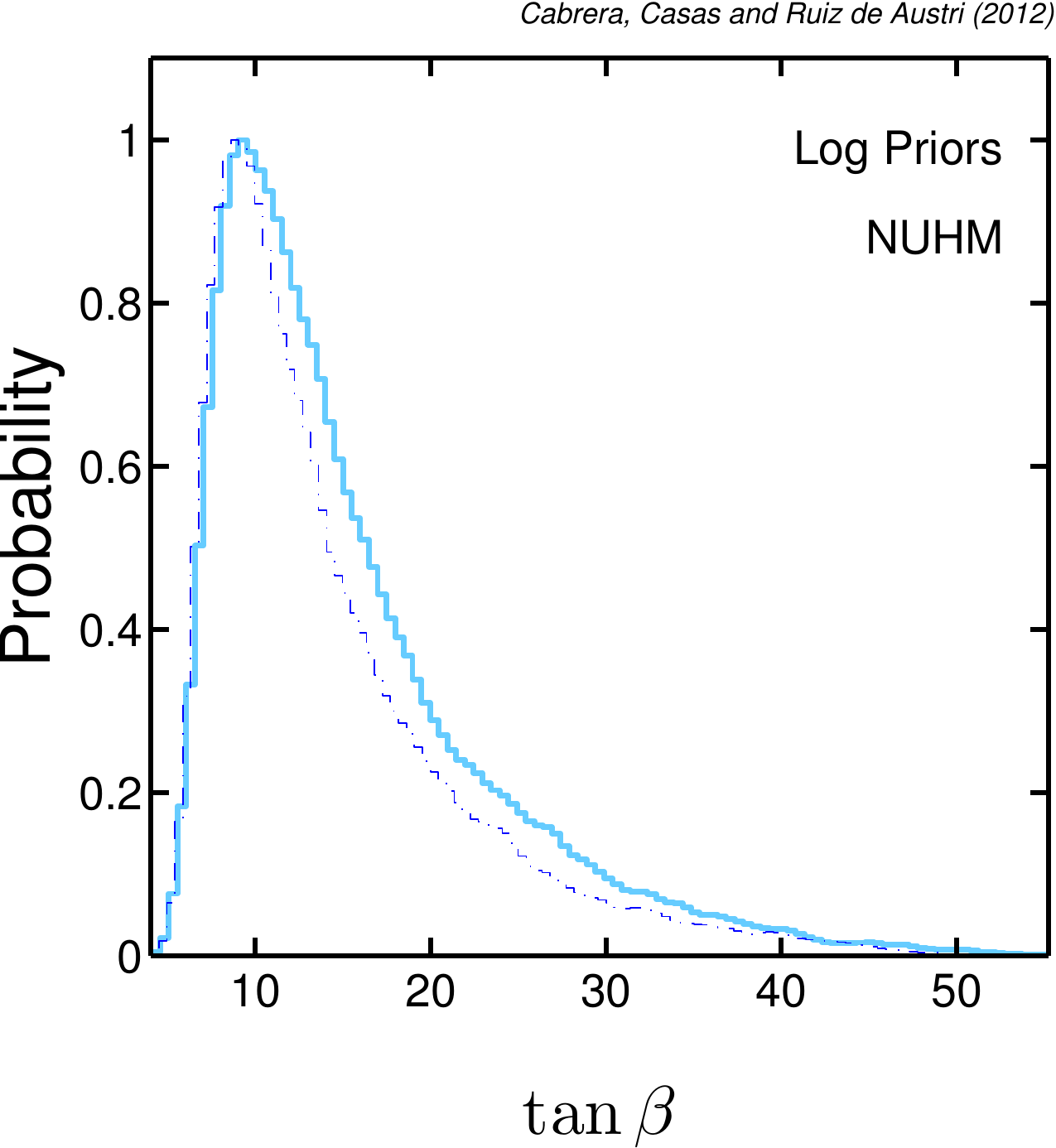}
\caption[test]{As in fig.~\ref{fig:nuhm_all_nodm_1d} but with an additional 
constraint from DM 
(WMAP 7-years and XENON100) considering that the LSP is the solely DM 
component. Here the cyan line corresponds to a positive $\mu$ whereas 
the  blue dashed to a negative $\mu$.\label{fig:nuhm_all_cdm_1d}}
\end{center}
\end{figure}

\begin{eqnarray}
  \label{eq:plog_nuhm}
  p(m_0, m_H, m_{1/2}, A, B,\mu) \propto \frac{1}{[\max\{m_0, m_H, m_{1/2}, |A|, |B|,\mu, M_S^0\}]^6},
\end{eqnarray}
and the flat prior,
\begin{eqnarray}
  \label{eq:pflat_nuhm}
  p(m_0, m_H, m_{1/2}, A, B,\mu) \propto \ \frac{1}{[\max\{m_0, m_H, m_{1/2}, |A|, |B|,\mu, M_S^0\}]^4}.
\end{eqnarray}

\begin{figure}[t]
\begin{center}
\includegraphics[angle=0,width=0.35\linewidth]{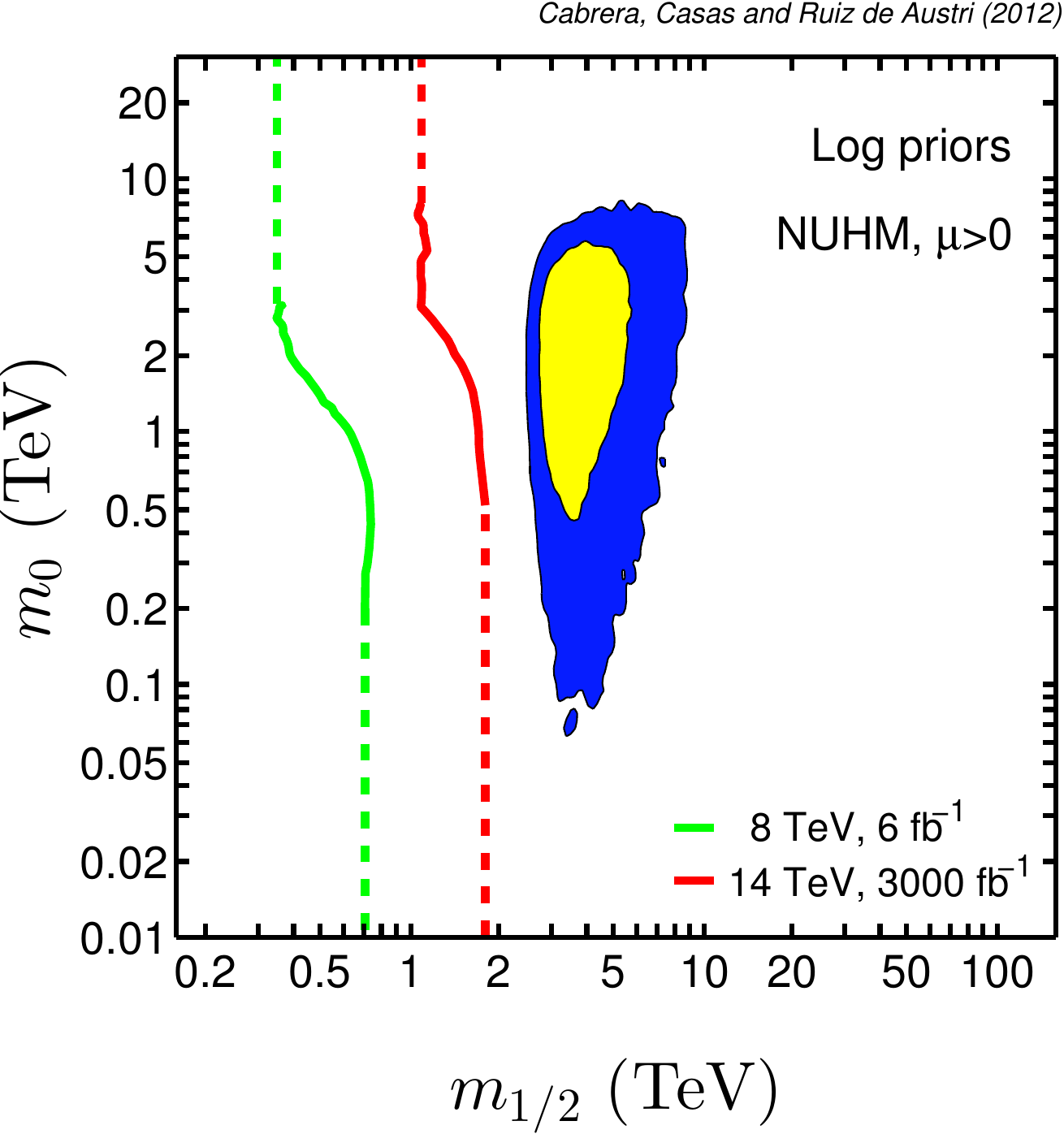} \hspace{1.2cm}
\includegraphics[angle=0,width=0.35\linewidth]{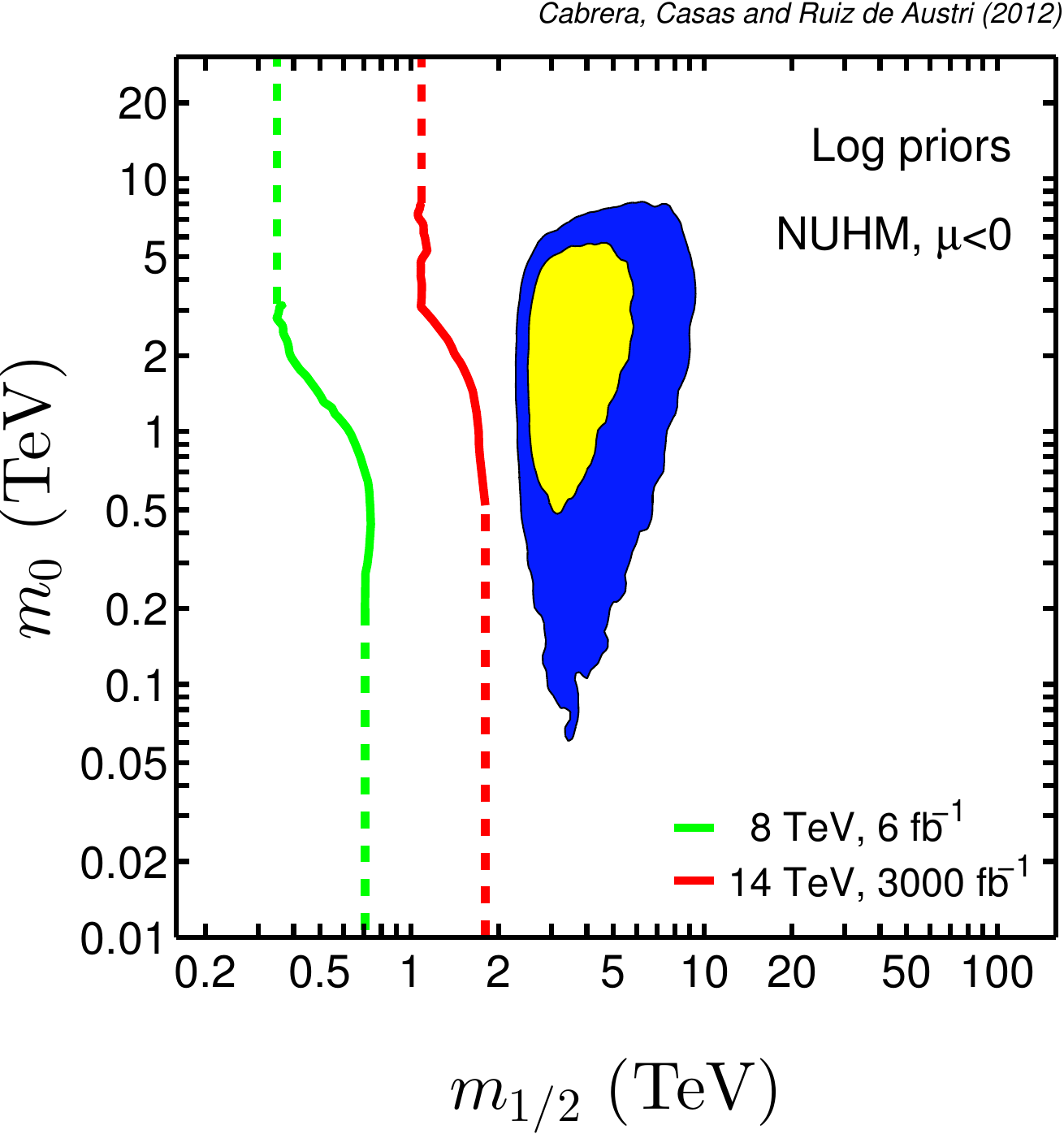}\\ \vspace{0.5cm}
\includegraphics[angle=0,width=0.35\linewidth]{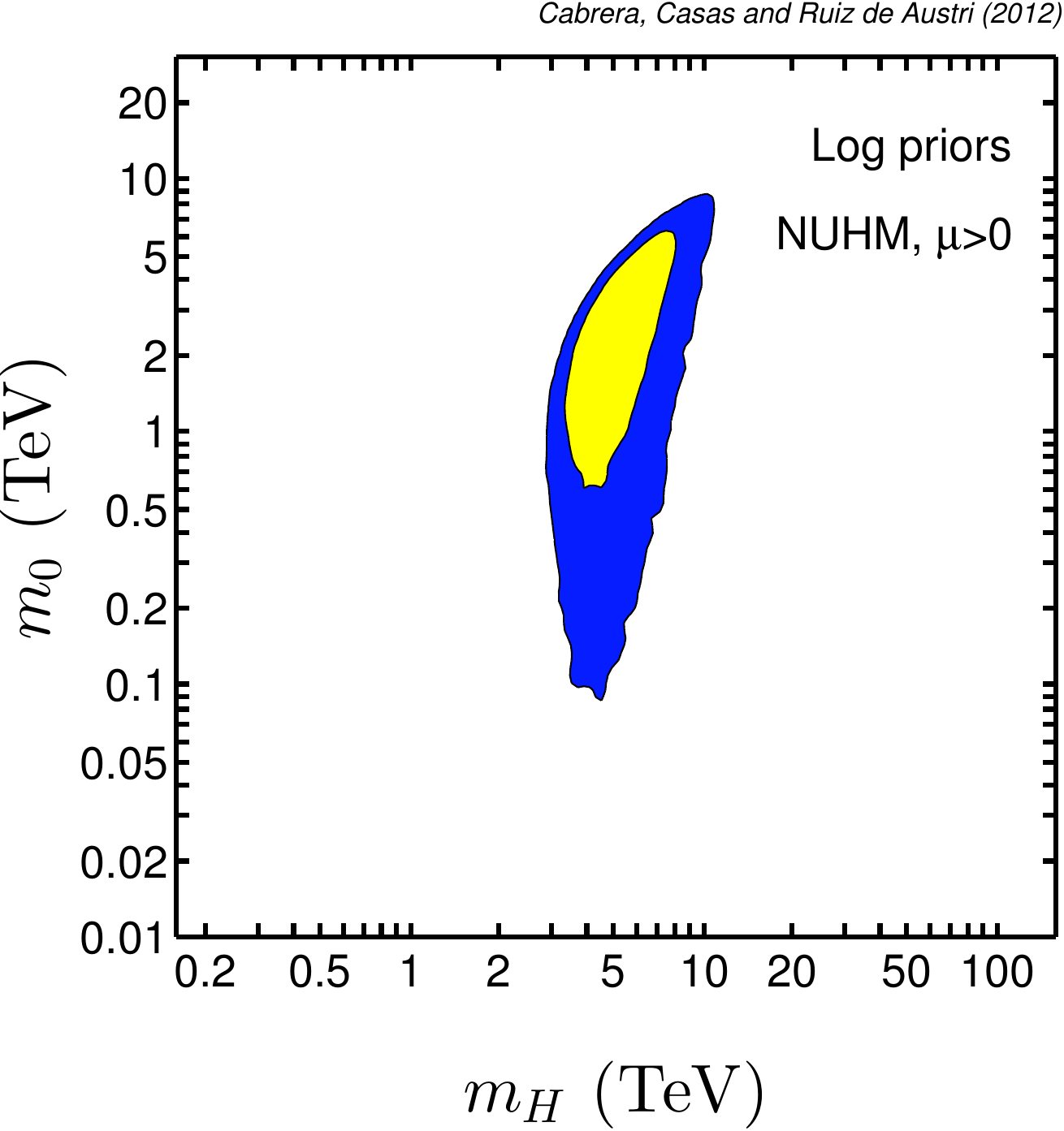} \hspace{1.2cm}
\includegraphics[angle=0,width=0.35\linewidth]{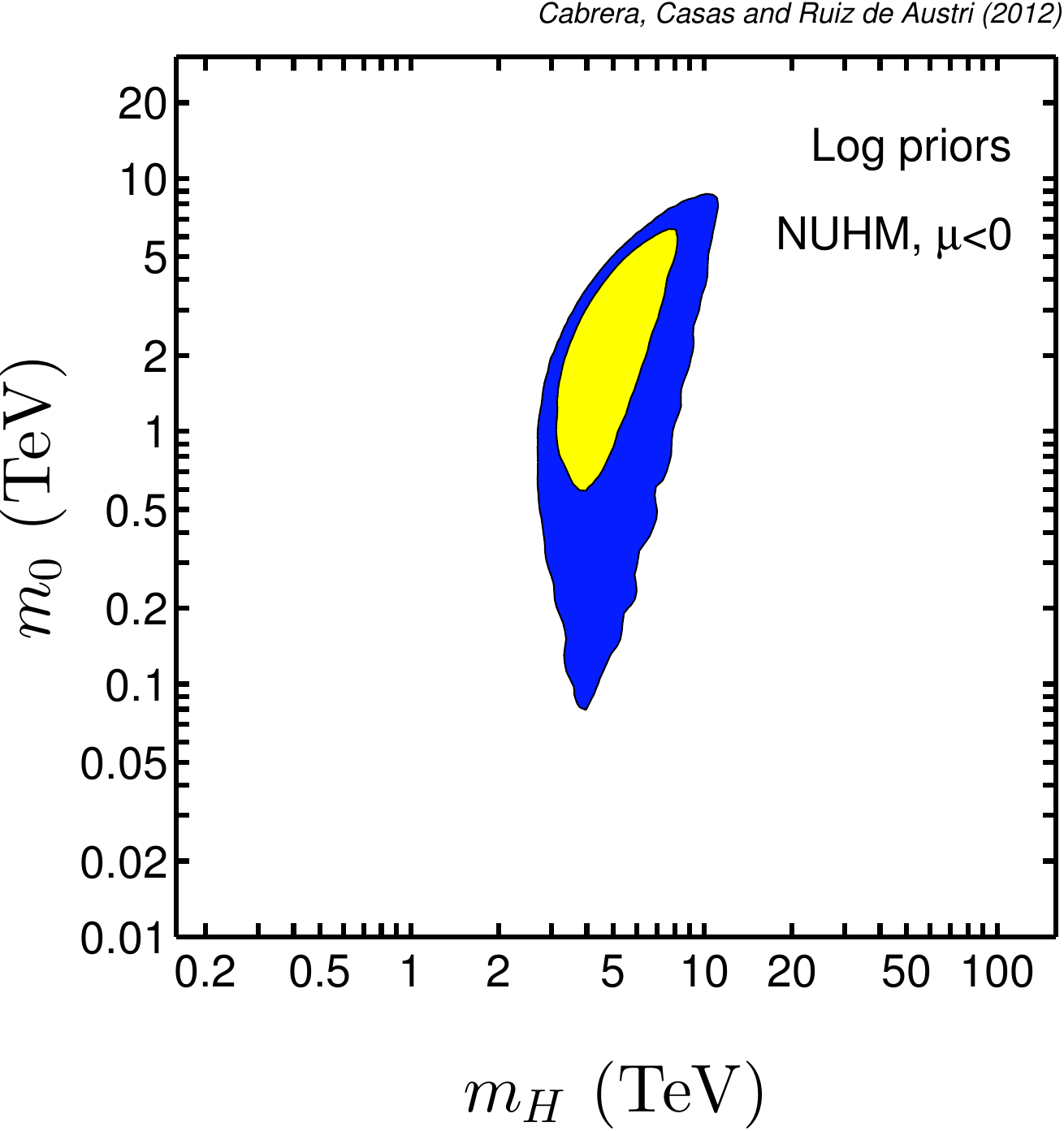} \\ \vspace{0.5cm}
\includegraphics[angle=0,width=0.35\linewidth]{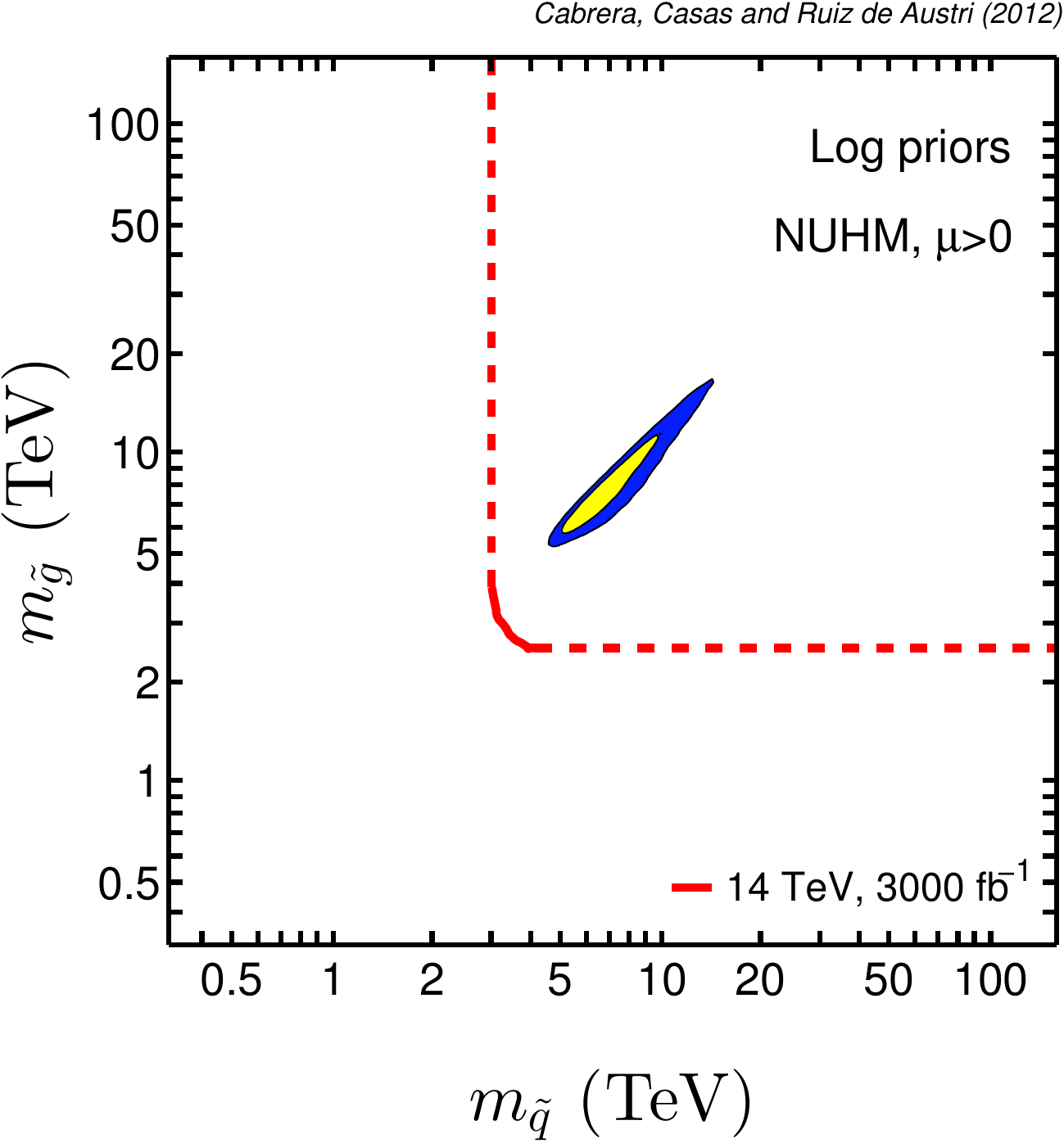} \hspace{1.2cm}
\includegraphics[angle=0,width=0.35\linewidth]{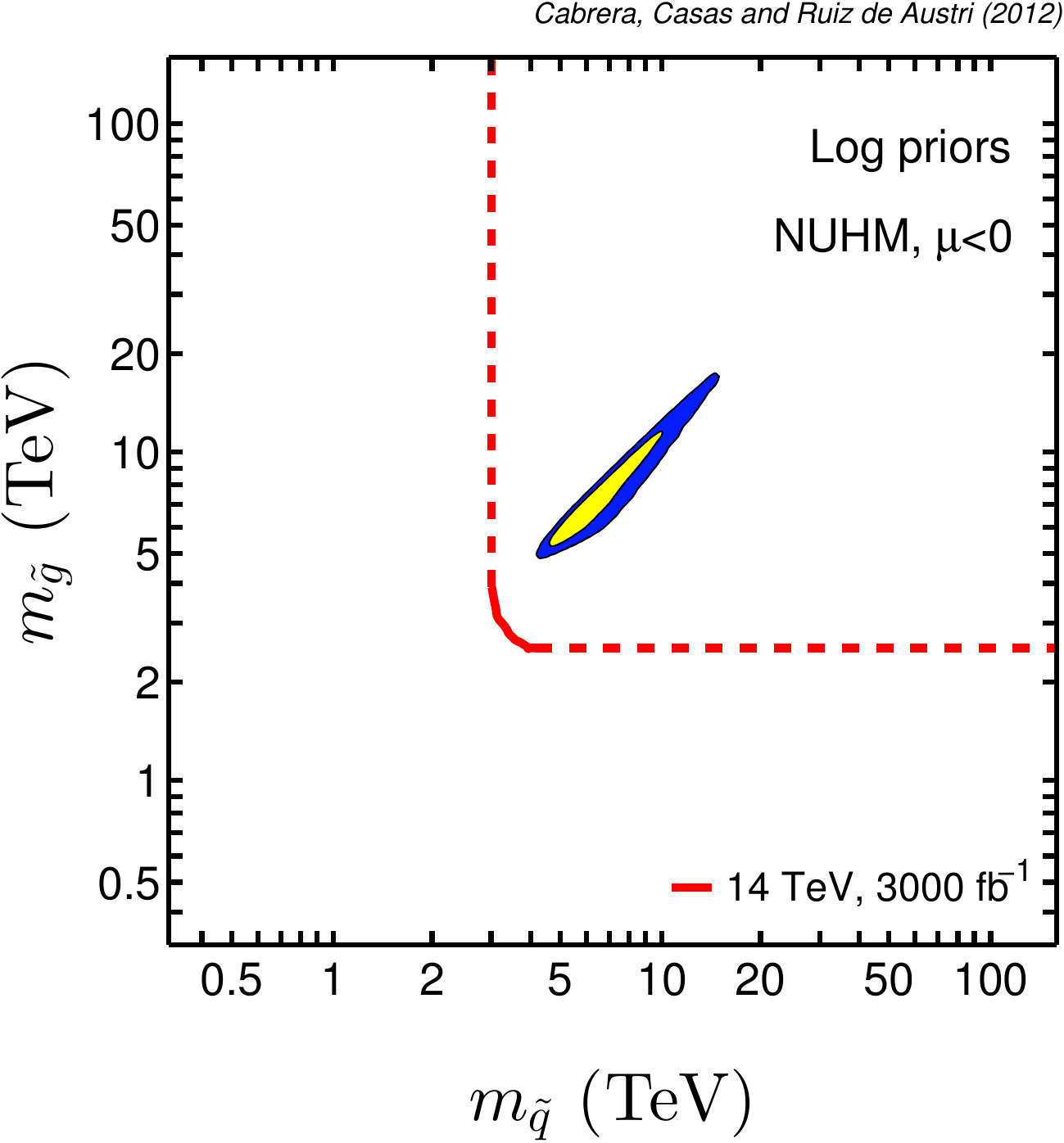}
\caption[test]{As in fig.~\ref{fig:nuhm_all_nodm_2d} but with an additional constraint from DM (WMAP 7-years and XENON100) considering that the LSP is the solely DM component.\label{fig:nuhm_all_scdm_2d_1}}
\end{center}
\end{figure}


\noindent
Figs.~\ref{fig:nuhm_all_nodm_1d}, \ref{fig:nuhm_all_nodm_2d} show the results when DM constraints are not included in the analysis. They are analogous to Figs. \ref{fig:cmssm_all_nodm_1d}, \ref{fig:cmssm_all_nodm_2d} for the CMSSM. Fig.~\ref{fig:nuhm_all_nodm_1d} shows the posterior probability
distribution function for $m_0$, $m_{H}$, $m_{1/2}$,
$A_0$ and $\tan\beta$, for log and flat priors and $\mu>0$. The results for $\mu<0$ are quite similar and are not shown.
As in the
CMSSM, the preferred region is in the high-energy region, mainly due to
the large value of the Higgs mass.

From the $m_0$ and $m_H$ pdfs in fig.~\ref{fig:nuhm_all_nodm_1d} or from the two-dimensional pdf in the $m_H-m_0$ plane in fig.~\ref{fig:nuhm_all_nodm_2d} it is clear that the NUHM naturally prefers $m_0<m_H$. The reason is that, as discussed in the Appendix, the separation of the $m_0$ and $m_H$ values can be used to decrease the value of $m_0$ at which $\mu^2$ becomes very small (and eventually negative). Recall that in the CMSSM this occurred for large $m_0$, and corresponded to the region of high $m_0$ in fig.~\ref{fig:cmssm_all_nodm_2d}. Actually, after including DM constraints, that ``Focus-Point" region was the only surviving one. Now, it is possible (and pays) to live with smaller $m_0$, which statistically is advantageous. This is the origin of the low$-m_0$ region in Figs.~\ref{fig:nuhm_all_nodm_1d}, \ref{fig:nuhm_all_nodm_2d}. 

Finally, Figs.~\ref{fig:nuhm_all_cdm_1d}, \ref{fig:nuhm_all_scdm_2d_1} show the results after including DM constraints (assuming that the DM is purely LSP). The results are quite similar to those for the CMSSM, but shifted to the low-$m_0$ region for the above-commented reasons. The same happens if we allow multi-component DM (we do not show the plots to avoid proliferation of figures).

In general, the results for the NUHM are similar to those of the CMSSM, but preferring lower values of $m_0$, which slightly improves the chances of detection at LHC.

\section{Comparison of evidence between models}

In order to compare the relative probability of two models, one has to evaluate the corresponding Bayesian evidences, i.e. the piece $p({\rm data})$ in the denominator of eq.(\ref{Bayes}), often called ${\cal Z}$. Note that, integrating both sides of eq.(\ref{Bayes}), and using the fact that $p(\theta_i|{\rm data})$ must be correctly normalized, one simply obtains
\bea
\label{BayesEvid}
{\cal Z}\ \equiv\ p({\rm data})\ =\ \int d\theta_1\cdots\ d\theta_N \  p({\rm data}|\theta_i)\ p(\theta_i),
\eea
i.e. the evidence is the integral of the likelihood times the prior, and therefore it is a measure of the global probability of the model. Denoting ${\cal M}_1$, ${\cal M}_2$ the two models to be compared, the relative posterior probability, given a set of data, is simply
\begin{equation}
\frac{p({\cal M}_{1}|data)}{p({\cal M}_{2}|data)}
=\frac{\mathcal{Z}_{1}\ p({\cal M}_{1})}{\mathcal{Z}_{2}\ p({\cal M}_{2})}
=B_{12}\frac{p({\cal M}_{1})}{p({\cal M}_{2})},
\label{bfactor}
\end{equation}
where $p({\cal M}_1)$, $p({\cal M}_2)$ are the prior probabilities of the two models (usually taken the same, so that $p({\cal M}_{1})/p({\cal M}_{2})$ is set to unity) and
$B_{12} \equiv {\cal Z}_1/{\cal Z}_2$ is called the Bayes factor.

The natural logarithm of the Bayes factor provides a useful indication of the different performance of two models. A conventional way to interpret the Bayes factor is the so-called Jeffrey's scale \cite{Jeffreys}, shown in Table~\ref{tab:Jeffreys}.

\begin{table}
\begin{center}
\begin{tabular}{|c|c|c|c|}
\hline
$|\ln B_{12}|$ & Odds & Probability & Strength of evidence \\
\hline\hline
$<1.0$ & $\lesssim 3:1$ & $<0.750$ & Inconclusive \\
$1.0$ & $\sim 3:1$ & $0.750$ & Weak Evidence \\
$2.5$ & $\sim 12:1$ & $0.923$ & Moderate Evidence \\
$5.0$ & $\sim 150:1$ & $0.993$ & Strong Evidence \\ \hline
\end{tabular}
\end{center}
\caption{Jeffrey's scale for the interpretation of model probabilities.}
\label{tab:Jeffreys}
\end{table}

The evaluation of the Bayesian evidence is in general a numerically challenging
task, as it involves a multidimensional integral over the whole parameter space. However, the MultiNest algorithm is able to perform the calculation in a reasonable computing time.

Along this paper we have considered two main theoretical models, CMSSM and NUHM. Both models contain in fact two sub-models, corresponding to the $\mu>0$ and $\mu<0$ branches in the parameter space. Besides, we have studied the models in a variety of instances, namely ignoring dark matter constraints, requiring that the dark matter is purely supersymmetric or allowing for multicomponent dark matter. Finally, the analyses have been performed using flat or logarithmic priors for the initial parameters. We have evaluated the global evidence in all these cases, to establish the comparison between models. The main conclusions are the following.

The first observation is that the evidences of the $\mu>0$ and $\mu<0$ branches are very similar in all the instances (i.e. for CMSSM or for NUHM, with or without dark matter constraints and for flat or logarithmic priors). The associated logarithm of the Bayes factor is always well below 1, i.e. completely inconclusive (although systematically showing a slight preference for the $\mu<0$ branch). This is shown in Table~\ref{tab:evidences_mu}. The reason is simply that, since the favored regions are at quite high values of the supersymmetric masses, SUSY is decoupled for most of the observables sensitive to the sign of $\mu$, e.g. $B\rightarrow X_S \gamma$.
Concerning the relative probability of the CMSSM and NUHM scenarios, we show in Table~\ref{tab:evidences} the log of the Bayes factor, $B = {\cal Z}_{CMSSM}/{\cal Z}_{NUHM}$ in the various cases. Since the NUHM contains one parameter more than the CMSSM, one could think that the associated evidence should be better, but this is not necessarily the case. The expression of the evidence, eq.~(\ref{BayesEvid}), contains a likelihood term which certainly rewards the better fitting that one can achieve by enlarging the set of independent parameters. But it also contains an integration in the parameter space, which disfavors an extended model where a large portion of the parameter space gives a bad likelihood. This is an ``Occam's razor" effect. The Bayes factor tells us to which extent the improvement in the likelihood compensates the Occam's razor effect. Actually, from Table~\ref{tab:evidences} the Bayes factor shows a weak evidence towards the simpler, CMSSM, scenario. However, the strength is not enough to say that the data show a clear preference for the CMSSM. In our opinion both models are still on approximately the same foot and are therefore equally interesting to interpret experimental data.

\begin{table}
\begin{center}
\begin{tabular}{|c|c|c|c|c|}
\hline
\multirow{2}{*}{Observables} & \multicolumn{2}{|c|}
{CMSSM} & \multicolumn{2}{|c|}{NUGM} \\\cline{2-5}
&  $\ln  B$; \footnotesize{flat} & $\ln  B$; \footnotesize{log} &  $\ln \, B$; \footnotesize{flat} & $\ln B$; \footnotesize{log}    \\ \hline \hline
No DM constraints    &  $-0.04\ \pm 0.04$ &  $-0.05\ \pm 0.04$ &  $-0.05\ \pm 0.05$ &  $-0.17\ \pm 0.05 $  \\
Single-component DM  &  $-0.31\ \pm 0.05$ &  $-0.26\ \pm 0.05$ &  $-0.05\ \pm 0.06$ &  $-0.17\ \pm 0.07 $    \\
Multi-component DM   &  $-0.36\ \pm 0.04$ &  $-0.43\ \pm 0.04$ &  $0.71\ \pm 0.06$ &  $-0.42\ \pm 0.04 $   \\
\hline
\end{tabular}
\caption{The natural log of the Bayes factor ($\ln  B$) for the comparison
between $\mu > 0$ and $\mu < 0$; for flat and log priors. Positive $\ln B$ favors $\mu>0$.}
\label{tab:evidences_mu}
\end{center}
\end{table}

\begin{table}
\begin{center}
\begin{tabular}{|c|c|c|c|c|}
\hline
Observables  &  $\ln B$; \footnotesize{$\mu>0$, flat} & $\ln B$; \footnotesize{$\mu>0$, log} &  $\ln B$; \footnotesize{$\mu<0$, flat} & $\ln B$; \footnotesize{$\mu<0$, log}    \\ \hline \hline
No DM constraints        &  $1.30\ \pm 0.05$ &   $1.06\ \pm 0.05$ & $1.18\ \pm 0.05$ & $1.02\ \pm 0.05 $  \\
Single-component DM    &  $1.00\ \pm 0.08$ &   $0.30\ \pm 0.05$ & $1.14\ \pm 0.06$ & $0.20\ \pm 0.05 $    \\
Multi-component DM   &  $0.66\ \pm 0.05$ &   $0.05\ \pm 0.06$ & $0.60\ \pm 0.06$ & $-0.12\ \pm 0.06 $   \\
\hline
\end{tabular}
\caption{The natural log of the Bayes factor ($\ln B$) for the comparison between the CMSSM and NUHM models; for $\mu > 0$
and $\mu < 0$, and for flat and log priors. Positive $\ln B$ favors CMSSM.}
\label{tab:evidences}
\end{center}
\end{table}

\section{Can still SUSY be natural? The impact of the Higgs discovery in the naturalness of SUSY}

The results of sets. 4, 5 show that the CMSSM and NUHM scenarios require large supersymmetric masses, which challenges the natural coexistence of SUSY with the electroweak scale (the problem is actually quite general for the MSSM, as discussed below). There are more-or-less sophisticated criteria in the literature to quantify the amount of fine-tuning of a given model \cite{deCarlos:1993yy,Anderson:1994dz,Ciafaloni:1996zh,Giusti:1998gz,Casas:2004gh,Casas:2005ev,Allanach:2006jc,Athron:2007ry,Fichet:2012sn,Ghilencea:2012qk}. However, the most rigorous and free-of-prejudice way to treat the fine-tuning is precisely through the Bayesian analysis, which automatically incorporates a fine-tuning penalization, encoded in the Jacobian factor of eqs.(\ref{J}, \ref{approx_eff_prior}). On the other hand, it is important to note that a Bayesian analysis starts with the assumption that the scenario under consideration is the true one, and then it gives the regions of higher probability in the associated parameter space. It does not directly give a hint about the plausibility of the scenario, unless one establish a comparison with an alternative one. 

In spite of this, it is very interesting the fact that we can use the Bayesian approach to study how the plausibility of the MSSM has changed after the Higgs discovery, as we are about to see in this section. For that purpose we will use a similar strategy to the one followed in subsect.4.3. to study the tension posed by the $g-2$ data. As in that case, we separate the experimental data in two subsets
\bea
\label{DDDmh}
\{{\rm data}\}  = \{\mathscr{D}, D \},
\eea
where ${\mathscr D}$ represents now the Higgs mass data and ${D}$ the rest of experimental information. As explained in subsect. 4.3 we construct now the $\mathscr{L}-$parameter, defined by
\bea
\label{Ltestmh}
\mathscr{L} (\mathscr{D}^{\rm obs}|D) \ \equiv\ 
\frac{p(\mathscr{D}^{\rm obs}|D)}{p(\mathscr{D}^{\rm max}|D)}\ =\ 
\frac{p(\mathscr{D}^{\rm obs},D)}{p(\mathscr{D}^{\rm max},D)}.
\eea
Here $\mathscr{D}^{\rm max}$ is the would-have-been most probable value of the Higgs mass in the MSSM context, given the values of the other experimental data. Our analysis shows that, restricting ourselves to the CMSSM scenario, this maximizing value is $m_h\simeq 117$ GeV. In other words this would have been the central value of the CMSSM `prediction" for $m_h$ just before the LHC discovery. This can also be seen from ref.~\cite{Cabrera:2009dm}. Then the $\mathscr{L}-$parameter defined above gives the relative global likelihood of the actually measured $m_h$ with respect to the maximal one. The results of our computation of $\mathscr{L}$ are
\bea
\label{Lmh}
\ln\ \mathscr{L}&=& -1.92 \hspace{1cm} {\rm (log\  priors)}
\nonumber\\
\ln\ \mathscr{L}&=& -0.70 \hspace{1cm} {\rm (flat\ priors)}
\eea
which correspond to $1.9\sigma$ and $1.18\sigma$ respectively. (For this computation we have not included dark matter constraints.) The previous results show that, although the experimental Higgs mass is not certainly the preferred value in the CMSSM context, it is not as improbable as it would seem. This is in part due to the fact that an important CMSSM region, from the point of view of statistical weight, is the Focus-Point one. This region corresponds to large scalar masses (but with low fine-tuning due to the small value of $\mu$) and thus to naturally large Higgs mass. The results for the NUHM are similar.

One can summarize this statement by saying that the measured Higgs mass introduces  an internal inconsistency of observables in the CMSSM/NUHM at the level of  $\sim 2\sigma$. Certainly, as past experiences have shown, a $2\sigma$ tension is not particularly dramatic and physicists are hardly impressed by it. Notice however that the meaning of this tension is consistent with the degree of fine-tuning that is often attributed to supersymmetry due to the experimental Higgs mass (a few percent). Namely, a Bayesian forecast performed before the discovery of the Higgs, under the assumption that the CMSSM/NUHM was true, would have showed that the probability of having $m_h\simeq 126$ GeV (or larger) was about $5\%$.

So the conclusion is that, although the CMSSM or the NUHM are quite unlikely to show up at the LHC (at least using the present strategies to find out SUSY), they are still interesting and plausible models after the Higgs discovery. Let us say they are at $\sim 2\ \sigma$.

Finally, we would like to address a question related to the naturalness of SUSY, namely:
To which extent the problems of the CMSSM and NUHM remain in general MSSMs? 
To address this question, let us recall that the original motivations for the CMSSM are 1) Minimal Flavor and CP violation, 2) Simplicity, 3) The fact that it arises in some theoretically motivated scenarios (like minimal SUGRA or Dilaton-dominated SUSY breaking). From these motivations only the first one is robust, but the experimental results do not require completely universal soft terms. E.g. the third generation of squarks and sleptons could have very different masses. The degeneracy of gaugino masses at $M_X$ is not experimentally justified either. Therefore, going beyond the CMSSM (or the NUHM) is very plausible. Does this solve the problems of the CMSSM?

This question can be explored by promoting the CMSSM to more general MSSMs compatible with the flavour and CP constraints. Playing with the initial parameters it is possible to have a lighter third generation, as invoked in "natural SUSY" models. It also allows certain types of supersymmetric spectrum that can evade the detection at LHC. In particular if the LSP is heavy or the masses of the supersymmetric states are not far from each other (``compressed spectrum" \cite{Martin:2007gf}), then the supersymmetric events have small $p_T$s and are triggered out in the experimental analyses. It is undoubtedly possible to arrange the MSSM spectrum in these ways in order to fool the LHC. But, in the absence of a solid theoretical reason for them, it sounds a rather artificial possibility.

In any case, in {\em any} MSSM the third generation cannot be too light, since it is needed to be rather heavy in order to reproduce (thanks to the radiative corrections) the experimental mass of the Higgs. This requirement is alleviated if $X_t$ is close to its maximizing value, but still the required stop masses imply that the electroweak breaking is quite fine-tuned, although perhaps not as much as it is often said, as we have discussed in this section. This situation could only improve substantially if one goes {\em beyond} the MSSM, which are out of the scope of this paper.

\section{Comparison with previous literature}

There are several works in the literature that have incorporated the impact of a Higgs at $\sim 125$ GeV and/or the new XENON100 data into Bayesian and frequentist analyses \cite{Balazs:2012qc,Fowlie:2012im, Akula:2012kk, Strege:2012bt, Buchmueller:2012hv}. Some of them are previous to the announcement of the Higgs discovery, and thus have a slightly different central value and uncertainty of $m_h$ from that reported in ref.\cite{:2012gk, :2012gu}. However, that departure is not dramatic, even though small shifts in the Higgs mass have an important effect in the SUSY studies, as is clear e.g. from the discussion in sect. 4. On the other hand, there are important differences between the approaches followed in the various analyses and the results obtained in them.
The aim of this section is to show the consistency of all the results, to explain the origin of the discrepancies and to argue what are the most reliable results and conclusions.

Let us start with other Bayesian studies of the CMSSM and/or the NUHM after the Higgs discovery, namely those of refs.\cite{Fowlie:2012im, Akula:2012kk, Strege:2012bt}. One can check that, in spite of being based on more or less the same data, the results do not look the same (compared with each other and with the present paper). To be more concrete, consider the probability distribution function in the $m_{1/2}-m_0$ plane in the CMSSM with $\mu>0$, using log priors and including the DM constraints. In ref.\cite{Akula:2012kk} they allow for multi-component DM, so the top-left plot of their figure 1 must be compared to the top-left plot of our fig.~\ref{fig:cmssm_all_mcdm_2d_1}. Actually, they are quite consistent (the preferred region is in the area of the Focus-Point), although the plot of ref.\cite{Akula:2012kk} is cut at 8 TeV and does not show most of the allowed region visible in our fig.~\ref{fig:cmssm_all_mcdm_2d_1}. This cut is not an artifact of the representation, but the actual upper limit considered for $m_0$ in ref.\cite{Akula:2012kk}. But otherwise, the results are in very good agreement. On the other hand, the left plot of figure 2 of ref.\cite{Fowlie:2012im} and the first plot in the middle row of figure 4 of ref.\cite{Strege:2012bt} correspond to the requirement that the DM is supersymmetric. Hence, they should be compared with the top-left plot of our fig.~\ref{fig:cmssm_all_scdm_2d_1}. Actually, the plots of those two references are quite consistent with each other, but they show a preferred region, with quite small soft masses, which does {\em not} show up at all, neither in our plot nor in the plot of ref.\cite{Akula:2012kk} (which being less restrictive should display it). What is going on?

\begin{figure}[t]
\begin{center}
\includegraphics[angle=0,width=0.35\linewidth]{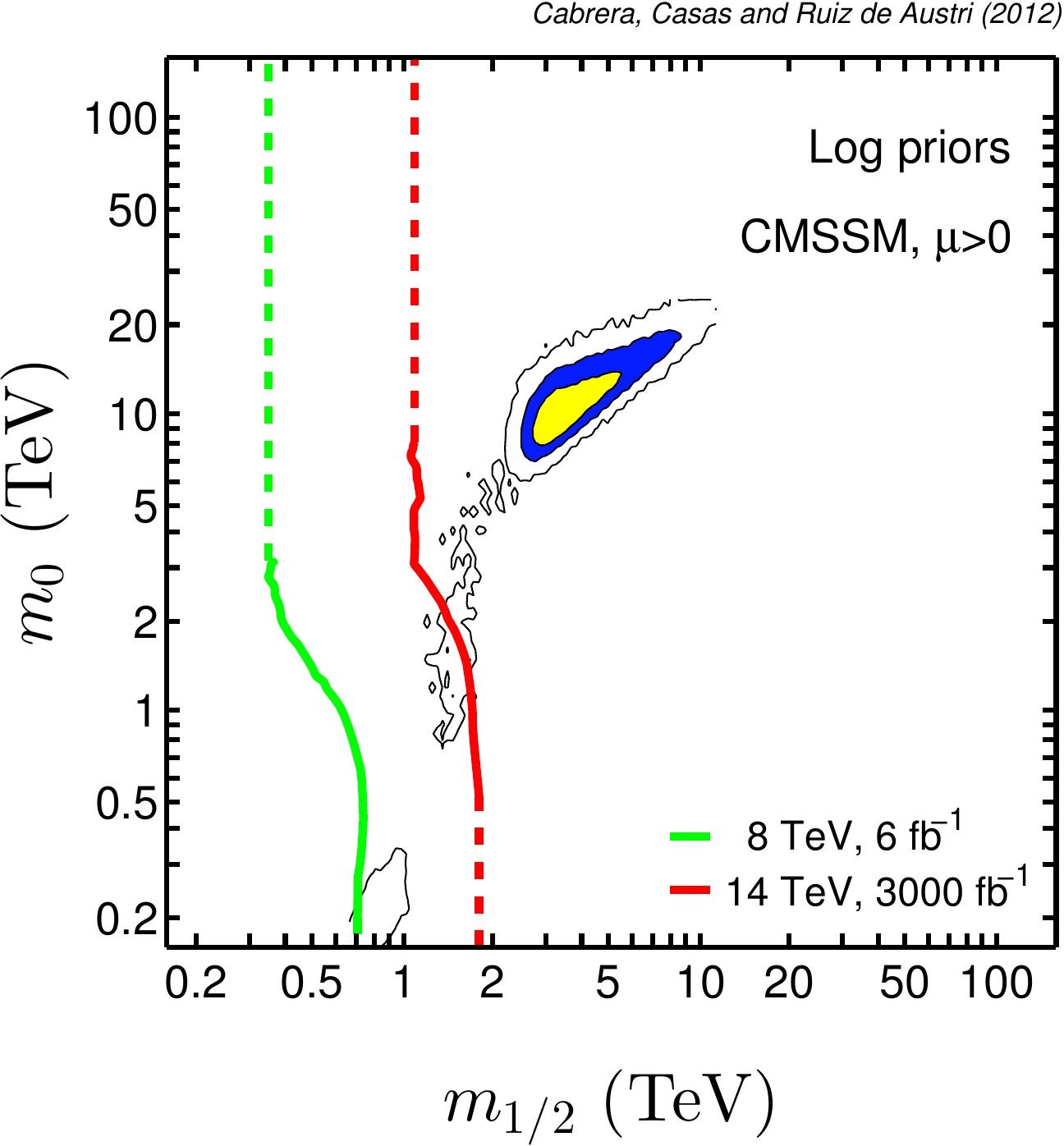}
\caption[test]{2D marginalized posterior probability distribution of the CMSSM in the $m_{1/2}$--$m_0$ plane, using log priors and DM constraints.
The yellow and blue areas enclose respective 68\% and 95\% joint regions. The white one corresponds to 99.9\% c.l.\label{fig:comparative}}
\end{center}
\end{figure}

The origin of these discrepancies lies on the different ranges used for the soft parameters. Not only one can leave aside a most interesting region if the ranges are too restrictive. In addition, as discussed after eq.(\ref{marg}), the results obtained by marginalizing the non-shown parameters may depend on the ranges of integration. The upper-limits of the ranges for ($m_0$, $m_{1/2}$) in refs. \cite{Akula:2012kk, Fowlie:2012im, Strege:2012bt} were $(8\ {\rm TeV}, 5\ {\rm TeV})$, $(4 {\rm TeV}, 2\ {\rm TeV})$ and $(4\ {\rm TeV}, 4\ {\rm TeV})$. This explains why the latter two references missed the region where the bulk of the probability is (the Focus-Point) and show only the less satisfactory stau-coannihilation and $A-$funnel regions. As discussed in subsect. 4.2 these two regions are disfavored, as they corresponds to rather small soft terms, difficult to reconcile with the Higgs mass (they are still consistent thanks to the theoretical uncertainty assumed in the supersymmetric calculation). This is why these regions do not belong to the $68\%$ and $95\%$ c. l. regions in the plots of ref. \cite{Akula:2012kk} and the present paper. In fact, one can force these two regions to appear by plotting the contours of higher confidence levels. We have checked that they become visible for the $99.9\%$ c.l. region, as fig.~\ref{fig:comparative} shows. The new contours are consistent with those shown in refs. \cite{Fowlie:2012im, Strege:2012bt}, but they are extremely disfavored.

The conclusion here is that it is very convenient to use an approach that makes the actual ranges for the parameters irrelevant. This is in fact possible, taking profit of the Jacobian factor discussed in sect. 2, which penalizes fine-tuned regions at high scale. In this sense, we honestly think that the results presented here are very reliable and represent an improvement with respect to previous analyses.

Let us finally comment briefly on frequentist approaches. These are based on the analysis of the likelihood function in the parameter space. The scan of the parameter space, evaluating the likelihood, leads to zones of estimated probability (inside contours of constant likelihood) around the best fit points. In doing so, the unplotted variables are optimized to obtain the best likelihood. Two recent frequentist analyses incorporating a Higgs mass at 125 GeV are those of ref. \cite{Buchmueller:2012hv} and ref. \cite{Strege:2012bt} (the first one was made before the Higgs discovery). They are quite consistent with each other, obtaining a best-fit point in the stau-coannihilation region.  The preferred areas are inside this region and in the $A-$funnel one.  Roughly speaking, they coincide with those obtained in the Bayesian analyses of refs.~\cite{Fowlie:2012im, Strege:2012bt}. The fact that the Focus-Point region with pure Higgsino LSP does not appear is probably a consequence of the limited ranges used in the scan of $m_{1/2}, m_0$ (though there could be also scanning issues, \cite{Akrami:2009hp, Feroz:2011bj}). In addition, a frequentist analysis is not sensitive to the fine-tuning. Fine-tuning has to do with statistical weight and a frequentist analysis is based entirely on likelihood, i.e. the ability to reproduce the experiment, and thus cannot ``see" the fine-tuning. As discussed in the introduction, it may happen that a point (o a region) in the parameter space can present an optimal likelihood, but only after an extreme tuning of the unplotted parameters, involving cancellations. Usually that point is considered very implausible or disfavored since, a priori, cancellations are not likely unless there exists some known theoretical reason for them. However, as long as the point is capable to reproduce the experimental data, the fine-tuning considerations do not affect its privileged condition in a frequentist analysis. 
This fact can favor points in the frequentist approach, e.g. in the low-energy regions, which are suppressed in the Bayesian one.

\section{Conclusions}

The most obvious conclusion from the results presented in this paper is that {\em if} the CMSSM or the NUHM are the model chosen by nature, the Bayesian analysis shows that they probably live in a region of the parameter space of large supersymmetric masses (squark and gluino masses above $\sim 2$ TeV), which makes very problematic their discovery at LHC. The quite large Higgs mass, $m_h\simeq 126$ GeV, is the main responsible for these negative results. Actually, given the experimental value of $m_h$, it would have been extremely improbable that the LHC had already detected a signal of new physics corresponding to the CMSSM or the NUHM. The Higgs mass is more powerful than the direct detection limits in pushing the MSSM parameter space into the large-energy region. Actually, we will be quite lucky if one of these models is correct {\em and} is still detectable in the next years, at least with the current techniques to look for physics beyond the SM at the LHC. We have quantified these probabilities with a rigorous and improved Bayesian analysis, whose results are remarkably independent of the choice of the priors and of the ranges assumed for the independent parameters. Besides, we have also performed a comparative study of the relative evidences of the CMSSM and the NUHM, showing they are quite similar, although with a slight preference for the CMSSM.

The previous conclusion is even more dramatic if one includes dark matter constraints in the analysis, in particular the last XENON100 data. These are able to discard a large portion of the Focus-Point region, while other regions (co-annihilation and $A-$funnel) are very disfavored from the experimental Higgs mass. There survives a fraction of the Focus-Point region with large scalar and gaugino masses, where the LSP neutralino is almost pure Higgsino. Its mass is constrained to be quite close to 1 TeV. This extreme conclusion is relaxed if one allows the dark matter to have other components (e.g. axions). Then a large part of the Focus-Point region is rescued. Interestingly, the future XENON1T data will be able to test these regions. Therefore, if the CMSSM or the NUHM are the true models of nature, it is more plausible to discover them in experiments of direct-detection of dark matter than in the LHC.

Our Bayesian analysis contains an automatic (non-imposed) penalization of fine-tuned regions. This penalization arises from the Bayesian analysis itself when the value of $M_Z$ is treated on the same foot as the rest of the experimental information (and not as a constraint of the model). This fine-tuning penalization is encoded in a model-independent Jacobian factor, which should never be ignored. This factor is also the responsible for the great stability against changes in the ranges of the parameters. We have shown that the different choices made for these ranges in previous analyses (which did not include the Jacobian penalization) have led to apparent discrepancies between the results obtained by different groups. We demonstrate that all of them are consistent once the freedom in the ranges is taken into account. This has motivated also that some previous analyses have missed important regions in the supersymmetric parameter space.

Although the 125 GeV Higgs pushes SUSY to regions where is hardly detectable at the LHC, this does not necessarily mean that SUSY is very fine-tuned. Actually, the Focus-Point region is not particularly fine-tuned, despite having large supersymmetric masses, since the value of $\mu$ is small. In this paper we have evaluated to which extent the Higgs discovery has diminished the naturalness of SUSY (focussing on the CMSSM/NUHM for this exercise). It is important to note that a Bayesian analysis starts with the assumption that the scenario under consideration is the true one, and then it gives the regions of higher probability in the associated parameter space. It does not directly give a hint about the plausibility of the scenario, unless one establish a comparison with an alternative one. Still, it is remarkable that we can use the Bayesian approach to study how the plausibility of the MSSM has changed after the Higgs discovery. The key is to perform the same Bayesian analysis with the same data except the Higgs mass, say as if we were in the year 2011. Then one obtains a preferred value for the Higgs mass (namely, around 117 GeV) with a probability profile; and one can compute how far apart from that ``prediction" the actual experimental value is. This can be done with the help of Bayesian tecnhiques (comparing the evidence for different values of $m_h$). The result is that the measured Higgs mass introduces  an internal inconsistency of observables in the CMSSM/NUHM at the level of  $\sim 2\sigma$. Admittedly, as past experiences have shown, a $2\sigma$ tension is not particularly dramatic and physicists are hardly impressed by it. Notice however that the meaning of this tension is consistent with the degree of fine-tuning that is often attributed to supersymmetry due to the experimental Higgs mass (a few percent). Namely, a Bayesian forecast performed before the discovery of the Higgs, under the assumption that the CMSSM/NUHM was true, would have showed that the probability of having $m_h\simeq 126$ GeV (or larger) was about $5\%$.

So the final conclusion is that, although the CMSSM or the NUHM are quite unlikely to show up at the LHC (at least using the present strategies to find out SUSY), they are still interesting and plausible models after the Higgs discovery; and the chances of discovering them in future experiments of direct detection of dark matter are quite high. 

\appendix

\section{Focus-Point behavior in the CMSSM and the NUHM}

Let us now discuss the impact of the more resent experimental data in the Non
Universal Higgs Masses (NUHM). We will briefly comment the similarities
between the CMSSM and NUHM and we will focus on new features that appear when
breaking the sfermions and higgs mass degeneracy at $M_X$.

As we have seen in previous sections, the impact of the Higgs mass
measurement pushes the preferred region to higher masses, in particular to the
focus point region. In order to simplify following explanations, let us
continue with a short review of the focus point.\\


As it was pointed by Feng et al \cite{Feng:1999mn}, the focus point is a scale
at which for a fixed $m_{1/2}$, $m_0$, $A_0$ and $\tan\beta$ in the CMSSM the
mass of the $m_{H_u}^2$ is constant independently of the value of $m_0$. The
value of this scale, that we are going to call $Q_{\rm FP}$, depends mainly of the
top yukawa and therefore of $\tan\beta$. Surprisingly $Q_{\rm FP}\simeq M_Z$.

From the minimization of the tree-level form of the scalar potential,
\begin{eqnarray}
\label{eqn:mu}
\mu^2 = \frac{m_{H_d}^2 - m_{H_{u}}^2 \tan^2{\beta}
}{\tan^2{\beta}-1} - \frac{1}{2}M_Z^2,
\end{eqnarray}
one can see that for large values of $\tan\beta$ ($\gtrsim 8$) a good
approximation is to consider
\begin{eqnarray}
  \label{eq:muLargetb1}
  \mu^2 \simeq - m_{H_{u,d}}^2.
\end{eqnarray}
Equation (\ref{eq:muLargetb1}) implies that arbitrarily large values of $m_0$
do not increase the value of $\mu^2(Q_{\rm FP})$. Then, if the EW breaking
scale ($Q_{\rm EW}$) is equal to $Q_{\rm FP}$, for small values of $m_{1/2}$
one can have large enough squark masses to get sizable radiative correction
for the Higgs mass and small $\mu$ to avoid fine-tuning. However, the
EW-breaking scale is chosen in such a way that the Next-to-Leading-Log
contribution to the Higgs potential are small, the optimal scale is given by
the geometrical average of the stop masses, $Q_{\rm EW}=\sqrt{\tilde{t}_1
  \tilde{t_2}}$ \cite{Casas:1994us,Carena:1995bx,Haber:1996fp}, therefore the EW-scale increases with $m_0$ making
impossible to have $Q_{\rm EW} = Q_{\rm FP}$ for different values of
$m_0$.

Rather than being interested in the focus point, we will be
interested in the behavior of $\mu^2$ around the focus point (what is commonly called the
focus point region). 
Now, let us write $\mu^2$ in terms of $m_{H_u}^2(Q_{\rm FP})$ and the running
for going from $Q_{\rm FP}$ to $Q_{\rm EW}$ ($\delta m_{H_{u,d}}^2$). In the
limit of large $\tan\beta$ one can rewrite eq.~(\ref{eq:muLargetb1}) as,
\begin{eqnarray}
\label{eqn:muLargetb2}
\mu(Q_{\rm EW})^2 \simeq - m_{H_{u}}^2(Q_{\rm EW}) = - (m_{H_{u}}^2(Q_{\rm FP}) + \delta m_{H_{u}}^2).
\end{eqnarray}
If $Q_{\rm EW}$ is greater than $Q_{\rm FP}$, which is the case for large
values of $m_0$, $\delta m_{H_{u}}^2$ will be positive and proportional to
$\delta m_0$. This means that for large enough $\delta m_0$ there will be a
cancellation between $\delta m_{H_{u}}^2$ and $ m_{H_u}^2(Q_{\rm FP})$ such
that $m_{H_u}^2$ and therefore $\mu^2$ will be very small\footnote{Notice that
  eq.(\ref{eqn:muLargetb2}) is not a good approximation any more.}. At this
point if one continue increasing $\delta m_0$, at some point $\mu^2$ will
becomes negative and it will be not possible anymore to have the correct
EW-breaking.

These explains why in the CMSSM it is not possible to get arbitrarily large
values of $m_0$ for a given $m_{1/2}$, $m_0$, $A_0$ and $\tan\beta$. Notice
also that for large values of $\tan\beta$ the maximal values of $m_0$ are
smaller since $m_{H_d}^2$ get a larger penalization respect to
$m_{H_u}^2$ in (\ref{eqn:mu}).\\

The focus point in NUHM1 shows one different feature, $Q_{\rm FP}$ will not
only depend on the Yukawas but also on the relation between $m_0^2$ and
$m_{H_{u,d}}^2$. Let us briefly describe the focus point behavior in NUHM, we
will take a CMSSM point and considerer different relations between $\delta
m_0$ and $\delta m_{H_u}^2$. If one leave $m_{H_u}^2(Q_{\rm GUT})$ fixed and
increases the value of $m_0$ then $\delta m_{H_{u}}^2$ will be always
negative. If instead one leave $m_0$ fixed and increases $m_{H_u}^2(Q_{\rm
  GUT})$ then $\delta m_{H_{u}}^2$ will be always positive. Increasing both
one can get that the $Q_{\rm FP}$ is shifted to higher values when $\delta m_0
/ \delta m_{H_u}^2(Q_{\rm GUT}) > 1 $ and is shifted to lower values when $ 2
> \delta m_{H_u}^2(Q_{\rm GUT})/\delta m_0 > 1$.  For more details see
\cite{Feng:1999mn}.

In summary NUHM1 allows to get the correct EW-breaking for larger values of
$m_0$ with respect to the CMSSM, it also allows to shift the value of the focus
point to larger of smaller energies depending on the relation between $\delta
m_0$ and $\delta m_{H_u}^2(Q_{\rm GUT})$.\\

{\em Acknowledgements.} 
We would like to thank Roberto Trotta and Charlotte Strege for providing 
us with the Xenon100 likelihood implementation. We acknowledge also John Ellis and Keith Olive for useful comments and 
observations for the final version of the paper.
R. RdA, is supported by the Ram\'on y Cajal program of the Spanish MICINN 
and also thanks the support of the 
Spanish MICINN's Consolider-Ingenio 2010 Programme under the grant MULTIDARK 
CSD2209-00064 and the Invisibles European ITN project (FP7-PEOPLE-2011-ITN, 
PITN-GA-2011-289442-INVISIBLES). 
The use of IFT-UAM High Performance Computing Service is gratefully 
acknowledged.

\bibliographystyle{JHEP-2}
\bibliography{references}

\providecommand{\href}[2]{#2}\begingroup\raggedright\begin{thebibliography}{100}

\bibitem{:2012gk}
{\bf ATLAS} Collaboration, G.~Aad {\em et.~al.}, {\it {Observation of a new
  particle in the search for the Standard Model Higgs boson with the ATLAS
  detector at the LHC}},  {\em Phys.Lett.} {\bf B716} (2012) 1--29
  [\href{http://arXiv.org/abs/1207.7214}{{\tt 1207.7214}}].

\bibitem{:2012gu}
{\bf CMS} Collaboration, S.~Chatrchyan {\em et.~al.}, {\it {Observation of a
  new boson at a mass of 125 GeV with the CMS experiment at the LHC}},  {\em
  Phys.Lett.} {\bf B716} (2012) 30--61
  [\href{http://arXiv.org/abs/1207.7235}{{\tt 1207.7235}}].

\bibitem{ATLAS-CONF-2012-162}
{\it Updated atlas results on the signal strength of the higgs-like boson for
  decays into ww and heavy fermion final states},  Tech. Rep.
  ATLAS-CONF-2012-162, CERN, Geneva, Nov, 2012.

\bibitem{CMS-PAS-HIG-12-045}
{\it Combination of standard model higgs boson searches and measurements of the
  properties of the new boson with a mass near 125 gev}, .

\bibitem{Farina:2011bh}
M.~Farina, M.~Kadastik, D.~Pappadopulo, J.~Pata, M.~Raidal {\em et.~al.}, {\it
  {Implications of XENON100 and LHC results for Dark Matter models}},  {\em
  Nucl.Phys.} {\bf B853} (2011) 607--624
  [\href{http://arXiv.org/abs/1104.3572}{{\tt 1104.3572}}].

\bibitem{Balazs:2012qc}
C.~Balazs, A.~Buckley, D.~Carter, B.~Farmer and M.~White, {\it {Should we still
  believe in constrained supersymmetry?}},
  \href{http://arXiv.org/abs/1205.1568}{{\tt 1205.1568}}.

\bibitem{Fowlie:2012im}
A.~Fowlie, M.~Kazana, K.~Kowalska, S.~Munir, L.~Roszkowski {\em et.~al.}, {\it
  {The CMSSM Favoring New Territories: The Impact of New LHC Limits and a 125
  GeV Higgs}},  {\em Phys.Rev.} {\bf D86} (2012) 075010
  [\href{http://arXiv.org/abs/1206.0264}{{\tt 1206.0264}}].

\bibitem{Akula:2012kk}
S.~Akula, P.~Nath and G.~Peim, {\it {Implications of the Higgs Boson Discovery
  for mSUGRA}},  {\em Phys.Lett.} {\bf B717} (2012) 188--192
  [\href{http://arXiv.org/abs/1207.1839}{{\tt 1207.1839}}].

\bibitem{Buchmueller:2012hv}
O.~Buchmueller, R.~Cavanaugh, M.~Citron, A.~De~Roeck, M.~Dolan {\em et.~al.},
  {\it {The CMSSM and NUHM1 in Light of 7 TeV LHC, $B_s to mu+mu-$ and XENON100
  Data}},  {\em Eur.Phys.J.} {\bf C72} (2012) 2243
  [\href{http://arXiv.org/abs/1207.7315}{{\tt 1207.7315}}].

\bibitem{Arbey:2012dq}
A.~Arbey, M.~Battaglia, A.~Djouadi and F.~Mahmoudi, {\it {The Higgs sector of
  the phenomenological MSSM in the light of the Higgs boson discovery}},  {\em
  JHEP} {\bf 1209} (2012) 107 [\href{http://arXiv.org/abs/1207.1348}{{\tt
  1207.1348}}].

\bibitem{Strege:2012bt}
C.~Strege, G.~Bertone, F.~Feroz, M.~Fornasa, R.~R. de~Austri {\em et.~al.},
  {\it {Global Fits of the cMSSM and NUHM including the LHC Higgs discovery and
  new XENON100 constraints}},  \href{http://arXiv.org/abs/1212.2636}{{\tt
  1212.2636}}.

\bibitem{Cabrera:2008tj}
M.~E. Cabrera, J.~A. Casas and R.~Ruiz~de Austri, {\it {Bayesian approach and
  Naturalness in MSSM analyses for the LHC}},  {\em JHEP} {\bf 03} (2009) 075
  [\href{http://arXiv.org/abs/0812.0536}{{\tt 0812.0536}}].

\bibitem{Strumia:1999fr}
A.~Strumia, {\it {Naturalness of supersymmetric models}},
  [\href{http://arXiv.org/abs/hep-ph/9904247}{{\tt hep-ph/9904247}}]. 
  
\bibitem{Ellis:1986yg}
J.~R. Ellis, K.~Enqvist, D.~V. Nanopoulos and F.~Zwirner, {\it {Observables in
  Low-Energy Superstring Models}},  {\em Mod.Phys.Lett.} {\bf A1} (1986) 57.

\bibitem{Barbieri:1987fn}
R.~Barbieri and G.~Giudice, {\it {Upper Bounds on Supersymmetric Particle
  Masses}},  {\em Nucl.Phys.} {\bf B306} (1988) 63.

\bibitem{softsusy}
B.~Allanach, {\it {SOFTSUSY: a program for calculating supersymmetric
  spectra}},  {\em Comput.Phys.Commun.} {\bf 143} (2002) 305--331
  [\href{http://arXiv.org/abs/hep-ph/0104145}{{\tt hep-ph/0104145}}].

\bibitem{Kastening:1991gv}
B.~M. Kastening, {\it {Renormalization group improvement of the effective
  potential in massive phi**4 theory}},  {\em Phys.Lett.} {\bf B283} (1992)
  287--292.

\bibitem{Ford:1992mv}
C.~Ford, D.~Jones, P.~Stephenson and M.~Einhorn, {\it {The Effective potential
  and the renormalization group}},  {\em Nucl.Phys.} {\bf B395} (1993) 17--34
  [\href{http://arXiv.org/abs/hep-lat/9210033}{{\tt hep-lat/9210033}}].

\bibitem{Bando:1992np}
M.~Bando, T.~Kugo, N.~Maekawa and H.~Nakano, {\it {Improving the effective
  potential}},  {\em Phys.Lett.} {\bf B301} (1993) 83--89
  [\href{http://arXiv.org/abs/hep-ph/9210228}{{\tt hep-ph/9210228}}].

\bibitem{Casas:1994us}
J.~A. Casas, J.~R. Espinosa, M.~Quiros and A.~Riotto, {\it {The Lightest Higgs
  boson mass in the minimal supersymmetric standard model}},  {\em Nucl. Phys.}
  {\bf B436} (1995) 3--29 [\href{http://arXiv.org/abs/hep-ph/9407389}{{\tt
  hep-ph/9407389}}].

\bibitem{Ibanez:1983di}
L.~E. Ibanez and C.~Lopez, {\it {N=1 Supergravity, the Weak Scale and the
  Low-Energy Particle Spectrum}},  {\em Nucl.Phys.} {\bf B233} (1984) 511.

\bibitem{Hall:1993gn}
L.~J. Hall, R.~Rattazzi and U.~Sarid, {\it {The Top quark mass in
  supersymmetric SO(10) unification}},  {\em Phys.Rev.} {\bf D50} (1994)
  7048--7065 [\href{http://arXiv.org/abs/hep-ph/9306309}{{\tt
  hep-ph/9306309}}].

\bibitem{Nelson:1993vc}
A.~E. Nelson and L.~Randall, {\it {Naturally large tan BETA}},  {\em
  Phys.Lett.} {\bf B316} (1993) 516--520
  [\href{http://arXiv.org/abs/hep-ph/9308277}{{\tt hep-ph/9308277}}].

\bibitem{Kaplunovsky:1993rd}
V.~S. Kaplunovsky and J.~Louis, {\it {Model independent analysis of soft terms
  in effective supergravity and in string theory}},  {\em Phys.Lett.} {\bf
  B306} (1993) 269--275 [\href{http://arXiv.org/abs/hep-th/9303040}{{\tt
  hep-th/9303040}}].

\bibitem{Giudice:1988yz}
G.~Giudice and A.~Masiero, {\it {A Natural Solution to the mu Problem in
  Supergravity Theories}},  {\em Phys.Lett.} {\bf B206} (1988) 480--484.

\bibitem{Allanach:2007qk}
B.~C. Allanach, K.~Cranmer, C.~G. Lester and A.~M. Weber, {\it {Natural priors,
  CMSSM fits and LHC weather forecasts}},  {\em JHEP} {\bf 0708} (2007) 023
  [\href{http://arXiv.org/abs/0705.0487}{{\tt 0705.0487}}].

\bibitem{Cabrera:2009dm}
M.~E. Cabrera, J.~A. Casas and R.~Ruiz~d Austri, {\it {MSSM Forecast for the
  LHC}},  {\em JHEP} {\bf 05} (2010) 043
  [\href{http://arXiv.org/abs/0911.4686}{{\tt 0911.4686}}].

\bibitem{topmass:1}
{\bf Tevatron Electroweak Working Group, CDF Collaboration, D0 Collaboration}
  Collaboration, {\it {Combination of CDF and D0 results on the mass of the top
  quark using up to 5.8~fb-1 of data}},
  \href{http://arXiv.org/abs/1107.5255}{{\tt 1107.5255}}.

\bibitem{pdg07}
{\bf Particle Data Group} Collaboration, W.~Yao {\em et.~al.}, {\it {Review of
  Particle Physics}},  {\em J.Phys.} {\bf G33} (2006) 1--1232.

\bibitem{Hagiwara:2006jt}
K.~Hagiwara, A.~Martin, D.~Nomura and T.~Teubner, {\it {Improved predictions
  for g-2 of the muon and alpha(QED) (M**2(Z))}},  {\em Phys.Lett.} {\bf B649}
  (2007) 173--179 [\href{http://arXiv.org/abs/hep-ph/0611102}{{\tt
  hep-ph/0611102}}].

\bibitem{deAustri:2006pe}
R.~R. de~Austri, R.~Trotta and L.~Roszkowski, {\it {A Markov chain Monte Carlo
  analysis of the CMSSM}},  {\em JHEP} {\bf 0605} (2006) 002
  [\href{http://arXiv.org/abs/hep-ph/0602028}{{\tt hep-ph/0602028}}].

\bibitem{lepwwg}
{\em {http://lepewwg.web.cern.ch/LEPEWWG}}.

\bibitem{hfag}
{\bf Heavy Flavor Averaging Group} Collaboration, Y.~Amhis {\em et.~al.}, {\it
  {Averages of b-hadron, c-hadron, and tau-lepton properties as of early
  2012}},  \href{http://arXiv.org/abs/1207.1158}{{\tt 1207.1158}}.

\bibitem{Aaij:2011qx}
{\bf LHCb} Collaboration, R.~Aaij {\em et.~al.}, {\it {Measurement of the
  $B^0_s - \bar{B}^0_s$ oscillation frequency $\Delta m_s$ in $B^0_s \to
  D_s^-(3) \pi$ decays}},  {\em Phys.Lett.} {\bf B709} (2012) 177--184
  [\href{http://arXiv.org/abs/1112.4311}{{\tt 1112.4311}}].

\bibitem{Aubert:2008af}
{\bf BABAR} Collaboration, B.~Aubert {\em et.~al.}, {\it {Measurement of
  Branching Fractions and CP and Isospin Asymmetries in $B \to K^{*} \gamma$}},
   \href{http://arXiv.org/abs/0808.1915}{{\tt 0808.1915}}.

\bibitem{Aubert:2007dsa}
{\bf BABAR} Collaboration, B.~Aubert {\em et.~al.}, {\it {Observation of the
  semileptonic decays $B \to D^{*} \tau^{-} \bar{\nu}$( $\tau^{)}$ and evidence
  for $B \to D \tau^{-} \bar{\nu}$( $\tau^{)}$}},  {\em Phys.Rev.Lett.} {\bf
  100} (2008) 021801 [\href{http://arXiv.org/abs/0709.1698}{{\tt 0709.1698}}].

\bibitem{Antonelli:2008jg}
{\bf FlaviaNet Working Group on Kaon Decays} Collaboration, M.~Antonelli {\em
  et.~al.}, {\it {Precision tests of the Standard Model with leptonic and
  semileptonic kaon decays}},  \href{http://arXiv.org/abs/0801.1817}{{\tt
  0801.1817}}.

\bibitem{:2012ct}
{\bf LHCb} Collaboration, R.~Aaij {\em et.~al.}, {\it {First evidence for the
  decay $B_s \to \mu^+ \mu^-$}},  \href{http://arXiv.org/abs/1211.2674}{{\tt
  1211.2674}}.

\bibitem{wmap}
N.~Jarosik, C.~Bennett, J.~Dunkley, B.~Gold, M.~Greason {\em et.~al.}, {\it
  {Seven-Year Wilkinson Microwave Anisotropy Probe (WMAP) Observations: Sky
  Maps, Systematic Errors, and Basic Results}},  {\em Astrophys.J.Suppl.} {\bf
  192} (2011) 14 [\href{http://arXiv.org/abs/1001.4744}{{\tt 1001.4744}}].

\bibitem{CMS}
{\em {https://twiki.cern.ch/twiki/bin/view/CMSPublic/PhysicsResults}}.

\bibitem{LHCSUSY}
{\it Search for squarks and gluinos with the atlas detector using final states
  with jets and missing transverse momentum and 5.8 fb$^{-1}$ of $\sqrt{s}$=8
  tev proton-proton collision data},  Tech. Rep. ATLAS-CONF-2012-109, CERN,
  Geneva, Aug, 2012.

\bibitem{LHCSUSYNUHM}
{\bf CMS} Collaboration, {\it Search for neutral higgs bosons decaying into tau
  leptons in the dimuon channel with cms in pp collisions at 7 tev}, .

\bibitem{Aprile:2012nq}
{\bf XENON100} Collaboration, E.~Aprile {\em et.~al.}, {\it {Dark Matter
  Results from 225 Live Days of XENON100 Data}},  {\em Phys.Rev.Lett.} {\bf
  109} (2012) 181301 [\href{http://arXiv.org/abs/1207.5988}{{\tt 1207.5988}}].

\bibitem{Dedes:2003km}
A.~Dedes, G.~Degrassi and P.~Slavich, {\it {On the two loop Yukawa corrections
  to the MSSM Higgs boson masses at large tan beta}},  {\em Nucl.Phys.} {\bf
  B672} (2003) 144--162 [\href{http://arXiv.org/abs/hep-ph/0305127}{{\tt
  hep-ph/0305127}}].

\bibitem{Awramik:2003rn}
M.~Awramik, M.~Czakon, A.~Freitas and G.~Weiglein, {\it {Precise prediction for
  the W boson mass in the standard model}},  {\em Phys.Rev.} {\bf D69} (2004)
  053006 [\href{http://arXiv.org/abs/hep-ph/0311148}{{\tt hep-ph/0311148}}].

\bibitem{Awramik:2004ge}
M.~Awramik, M.~Czakon, A.~Freitas and G.~Weiglein, {\it {Complete two-loop
  electroweak fermionic corrections to $\sin^{2} \theta^{\rm lept}_{\rm eff}$
  and indirect determination of the Higgs boson mass}},  {\em Phys.Rev.Lett.}
  {\bf 93} (2004) 201805 [\href{http://arXiv.org/abs/hep-ph/0407317}{{\tt
  hep-ph/0407317}}].

\bibitem{dghhjw97}
A.~Djouadi, P.~Gambino, S.~Heinemeyer, W.~Hollik, C.~Junger {\em et.~al.}, {\it
  {Leading QCD corrections to scalar quark contributions to electroweak
  precision observables}},  {\em Phys.Rev.} {\bf D57} (1998) 4179--4196
  [\href{http://arXiv.org/abs/hep-ph/9710438}{{\tt hep-ph/9710438}}].

\bibitem{Degrassi:2007kj}
G.~Degrassi, P.~Gambino and P.~Slavich, {\it {SusyBSG: A Fortran code for BR[$B
  \rightarrow X(s) \gamma$] in the MSSM with Minimal Flavor Violation}},  {\em
  Comput.Phys.Commun.} {\bf 179} (2008) 759--771
  [\href{http://arXiv.org/abs/0712.3265}{{\tt 0712.3265}}].

\bibitem{Degrassi:2006eh}
G.~Degrassi, P.~Gambino and P.~Slavich, {\it {QCD corrections to radiative B
  decays in the MSSM with minimal flavor violation}},  {\em Phys.Lett.} {\bf
  B635} (2006) 335--342 [\href{http://arXiv.org/abs/hep-ph/0601135}{{\tt
  hep-ph/0601135}}].

\bibitem{D'Ambrosio:2002ex}
G.~D'Ambrosio, G.~Giudice, G.~Isidori and A.~Strumia, {\it {Minimal flavor
  violation: An Effective field theory approach}},  {\em Nucl.Phys.} {\bf B645}
  (2002) 155--187 [\href{http://arXiv.org/abs/hep-ph/0207036}{{\tt
  hep-ph/0207036}}].

\bibitem{Mahmoudi:2008tp}
F.~Mahmoudi, {\it {SuperIso v2.3: A Program for calculating flavor physics
  observables in Supersymmetry}},  {\em CPHCB,180,1579-1613.2009} {\bf 180}
  (2009) 1579--1613 [\href{http://arXiv.org/abs/0808.3144}{{\tt 0808.3144}}].

\bibitem{superbayes}
{\em {http://superbayes.org}}.

\bibitem{Aad:2011hh}
{\bf ATLAS} Collaboration, G.~Aad {\em et.~al.}, {\it {Search for supersymmetry
  using final states with one lepton, jets, and missing transverse momentum
  with the ATLAS detector in $\sqrt{s}=7$ TeV $pp$}},  {\em Phys.Rev.Lett.}
  {\bf 106} (2011) 131802 [\href{http://arXiv.org/abs/1102.2357}{{\tt
  1102.2357}}].

\bibitem{Allanach:2011ut}
B.~Allanach, {\it {Impact of CMS Multi-jets and Missing Energy Search on CMSSM
  Fits}},  {\em Phys.Rev.} {\bf D83} (2011) 095019
  [\href{http://arXiv.org/abs/1102.3149}{{\tt 1102.3149}}].

\bibitem{Feroz:2007kg}
F.~Feroz and M.~P. Hobson, {\it {Multimodal nested sampling: an efficient and
  robust alternative to MCMC methods for astronomical data analysis}},  {\em
  Mon. Not. Roy. Astron. Soc.} {\bf 384} (2008) 449--463
  [\href{http://arXiv.org/abs/0704.3704}{{\tt 0704.3704}}].

\bibitem{SkillingNS}
J.~Skilling, {\it Nested sampling},  in {\em {Bayesian Inference and Maximum
  Entropy Methods in Science and Engineering}} (R.~P. R.~Fischer and U.~von
  Toussaint, eds.), vol.~735, pp.~395--405, 2004.

\bibitem{Skilling:2006}
J.~Skilling, {\it Nested sampling for general bayesian computation},  {\em
  Bayesian Analysis} {\bf 1} (2006), no.~4 833--860.

\bibitem{Cabrera:2011bi}
M.~Cabrera, J.~Casas and A.~Delgado, {\it {Upper Bounds on Superpartner Masses
  from Upper Bounds on the Higgs Boson Mass}},  {\em Phys.Rev.Lett.} {\bf 108}
  (2012) 021802 [\href{http://arXiv.org/abs/1108.3867}{{\tt 1108.3867}}].

\bibitem{Giudice:2011cg}
G.~F. Giudice and A.~Strumia, {\it {Probing High-Scale and Split Supersymmetry
  with Higgs Mass Measurements}},  {\em Nucl.Phys.} {\bf B858} (2012) 63--83
  [\href{http://arXiv.org/abs/1108.6077}{{\tt 1108.6077}}].

\bibitem{HL_LHC}
A.~Collaboration, {\it Physics at a high-luminosity lhc with atlas},  Tech.
  Rep. ATL-PHYS-PUB-2012-001, CERN, Geneva, Aug, 2012.

\bibitem{Ellis:1990nz}
J.~R. Ellis, G.~Ridolfi and F.~Zwirner, {\it {Radiative corrections to the
  masses of supersymmetric Higgs bosons}},  {\em Phys. Lett.} {\bf B257} (1991)
  83--91.

\bibitem{Ellis:1991zd}
J.~R. Ellis, G.~Ridolfi and F.~Zwirner, {\it {On radiative corrections to
  supersymmetric Higgs boson masses and their implications for LEP searches}},
  {\em Phys. Lett.} {\bf B262} (1991) 477--484.

\bibitem{Okada:1990vk}
Y.~Okada, M.~Yamaguchi and T.~Yanagida, {\it {Upper bound of the lightest Higgs
  boson mass in the minimal supersymmetric standard model}},  {\em Prog. Theor.
  Phys.} {\bf 85} (1991) 1--6.

\bibitem{Okada:1990gg}
Y.~Okada, M.~Yamaguchi and T.~Yanagida, {\it {Renormalization group analysis on
  the Higgs mass in the softly broken supersymmetric standard model}},  {\em
  Phys. Lett.} {\bf B262} (1991) 54--58.

\bibitem{Haber:1990aw}
H.~E. Haber and R.~Hempfling, {\it {Can the mass of the lightest Higgs boson of
  the minimal supersymmetric model be larger than m(Z)?}},  {\em Phys. Rev.
  Lett.} {\bf 66} (1991) 1815--1818.

\bibitem{Barbieri:1990ja}
R.~Barbieri, M.~Frigeni and F.~Caravaglios, {\it {The Supersymmetric Higgs for
  heavy superpartners}},  {\em Phys. Lett.} {\bf B258} (1991) 167--170.

\bibitem{Carena:1995bx}
M.~S. Carena, J.~Espinosa, M.~Quiros and C.~Wagner, {\it {Analytical
  expressions for radiatively corrected Higgs masses and couplings in the
  MSSM}},  {\em Phys.Lett.} {\bf B355} (1995) 209--221
  [\href{http://arXiv.org/abs/hep-ph/9504316}{{\tt hep-ph/9504316}}].

\bibitem{Aparicio:2012iw}
L.~Aparicio, D.~Cerdeno and L.~Ibanez, {\it {A 119-125 GeV Higgs from a string
  derived slice of the CMSSM}},  {\em JHEP} {\bf 1204} (2012) 126
  [\href{http://arXiv.org/abs/1202.0822}{{\tt 1202.0822}}].

\bibitem{Jungman:1995df}
G.~Jungman, M.~Kamionkowski and K.~Griest, {\it {Supersymmetric dark matter}},
  {\em Phys.Rept.} {\bf 267} (1996) 195--373
  [\href{http://arXiv.org/abs/hep-ph/9506380}{{\tt hep-ph/9506380}}].

\bibitem{Knox:1992iy}
L.~Knox and M.~S. Turner, {\it {Inflation at the electroweak scale}},  {\em
  Phys.Rev.Lett.} {\bf 70} (1993) 371--374
  [\href{http://arXiv.org/abs/astro-ph/9209006}{{\tt astro-ph/9209006}}].

\bibitem{Ibarra:2008jk}
A.~Ibarra and D.~Tran, {\it {Decaying Dark Matter and the PAMELA Anomaly}},
  {\em JCAP} {\bf 0902} (2009) 021 [\href{http://arXiv.org/abs/0811.1555}{{\tt
  0811.1555}}].

\bibitem{Pato:2010zk}
M.~Pato, L.~Baudis, G.~Bertone, R.~Ruiz~de Austri, L.~E. Strigari {\em
  et.~al.}, {\it {Complementarity of Dark Matter Direct Detection Targets}},
  {\em Phys.Rev.} {\bf D83} (2011) 083505
  [\href{http://arXiv.org/abs/1012.3458}{{\tt 1012.3458}}].

\bibitem{Ellis:2008hf}
J.~R. Ellis, K.~A. Olive and C.~Savage, {\it {Hadronic Uncertainties in the
  Elastic Scattering of Supersymmetric Dark Matter}},  {\em Phys.Rev.} {\bf
  D77} (2008) 065026 [\href{http://arXiv.org/abs/0801.3656}{{\tt 0801.3656}}].

\bibitem{Bertone:2011nj}
G.~Bertone, D.~G. Cerdeno, M.~Fornasa, R.~Ruiz~de Austri, C.~Strege {\em
  et.~al.}, {\it {Global fits of the cMSSM including the first LHC and XENON100
  data}},  {\em JCAP} {\bf 1201} (2012) 015
  [\href{http://arXiv.org/abs/1107.1715}{{\tt 1107.1715}}].

\bibitem{ArkaniHamed:2006mb}
N.~Arkani-Hamed, A.~Delgado and G.~Giudice, {\it {The Well-tempered
  neutralino}},  {\em Nucl.Phys.} {\bf B741} (2006) 108--130
  [\href{http://arXiv.org/abs/hep-ph/0601041}{{\tt hep-ph/0601041}}].

\bibitem{Bertone:2010ww}
G.~Bertone, K.~Kong, R.~R. de~Austri and R.~Trotta, {\it {Global fits of the
  Minimal Universal Extra Dimensions scenario}},  {\em Phys.Rev.} {\bf D83}
  (2011) 036008 [\href{http://arXiv.org/abs/1010.2023}{{\tt 1010.2023}}].

\bibitem{Bertone:2010rv}
G.~Bertone, D.~G. Cerdeno, M.~Fornasa, R.~R. de~Austri and R.~Trotta, {\it
  {Identification of Dark Matter particles with LHC and direct detection
  data}},  {\em Phys.Rev.} {\bf D82} (2010) 055008
  [\href{http://arXiv.org/abs/1005.4280}{{\tt 1005.4280}}].

\bibitem{Davier:2009zi}
M.~Davier, A.~Hoecker, B.~Malaescu, C.~Z. Yuan and Z.~Zhang, {\it {Reevaluation
  of the hadronic contribution to the muon magnetic anomaly using new e+e-
  $\rightarrow$ pi+pi- cross section data from BABAR}},  {\em Eur. Phys. J.}
  {\bf C66} (2010) 1--9 [\href{http://arXiv.org/abs/0908.4300}{{\tt
  0908.4300}}].

\bibitem{Davier:2009ag}
M.~Davier {\em et.~al.}, {\it {The Discrepancy Between tau and e+e- Spectral
  Functions Revisited and the Consequences for the Muon Magnetic Anomaly}},
  {\em Eur. Phys. J.} {\bf C66} (2010) 127--136
  [\href{http://arXiv.org/abs/0906.5443}{{\tt 0906.5443}}].

\bibitem{Degrassi:1998es}
G.~Degrassi and G.~Giudice, {\it {QED logarithms in the electroweak corrections
  to the muon anomalous magnetic moment}},  {\em Phys.Rev.} {\bf D58} (1998)
  053007 [\href{http://arXiv.org/abs/hep-ph/9803384}{{\tt hep-ph/9803384}}].

\bibitem{Heinemeyer:2003dq}
S.~Heinemeyer, D.~Stockinger and G.~Weiglein, {\it {Two loop SUSY corrections
  to the anomalous magnetic moment of the muon}},  {\em Nucl.Phys.} {\bf B690}
  (2004) 62--80 [\href{http://arXiv.org/abs/hep-ph/0312264}{{\tt
  hep-ph/0312264}}].

\bibitem{Heinemeyer:2004yq}
S.~Heinemeyer, D.~Stockinger and G.~Weiglein, {\it {Electroweak and
  supersymmetric two-loop corrections to (g-2)(mu)}},  {\em Nucl.Phys.} {\bf
  B699} (2004) 103--123 [\href{http://arXiv.org/abs/hep-ph/0405255}{{\tt
  hep-ph/0405255}}].

\bibitem{Marchetti:2008hw}
S.~Marchetti, S.~Mertens, U.~Nierste and D.~Stockinger, {\it
  {Tan(beta)-enhanced supersymmetric corrections to the anomalous magnetic
  moment of the muon}},  {\em Phys.Rev.} {\bf D79} (2009) 013010
  [\href{http://arXiv.org/abs/0808.1530}{{\tt 0808.1530}}].

\bibitem{vonWeitershausen:2010zr}
P.~von Weitershausen, M.~Schafer, H.~Stockinger-Kim and D.~Stockinger, {\it
  {Photonic SUSY Two-Loop Corrections to the Muon Magnetic Moment}},  {\em
  Phys.Rev.} {\bf D81} (2010) 093004
  [\href{http://arXiv.org/abs/1003.5820}{{\tt 1003.5820}}].

\bibitem{Trotta:2005ar}
R.~Trotta, {\it {Applications of Bayesian model selection to cosmological
  parameters}},  {\em Mon. Not. Roy. Astron. Soc.} {\bf 378} (2007) 72--82
  [\href{http://arXiv.org/abs/astro-ph/0504022}{{\tt astro-ph/0504022}}].

\bibitem{Trotta:2008qt}
R.~Trotta, {\it {Bayes in the sky: Bayesian inference and model selection in
  cosmology}},  {\em Contemp. Phys.} {\bf 49} (2008) 71--104
  [\href{http://arXiv.org/abs/0803.4089}{{\tt 0803.4089}}].

\bibitem{Feroz:2009dv}
F.~Feroz, M.~P. Hobson, L.~Roszkowski, R.~Ruiz~de Austri and R.~Trotta, {\it
  {Are $BR(b\rightarrow s \gamma)$ and $(g-2)_\mu$ consistent within the
  Constrained MSSM?}},  \href{http://arXiv.org/abs/0903.2487}{{\tt 0903.2487}}.

\bibitem{Cabrera:2010xx}
M.~E. Cabrera, J.~Casas, R.~R. de~Austri and R.~Trotta, {\it {Quantifying the
  tension between the Higgs mass and $(g-2)_\mu$ in the CMSSM}},
  \href{http://arXiv.org/abs/1011.5935}{{\tt 1011.5935}}.
  
\bibitem{Jeffreys}
H. Jeffreys, Theory of probability, 3rd edn, Oxford Classics 
series (reprinted 1998) (Oxford University Press, Oxford, UK, 1961).

\bibitem{deCarlos:1993yy}
B.~de~Carlos and J.~Casas, {\it {One loop analysis of the electroweak breaking
  in supersymmetric models and the fine tuning problem}},  {\em Phys.Lett.}
  {\bf B309} (1993) 320--328 [\href{http://arXiv.org/abs/hep-ph/9303291}{{\tt
  hep-ph/9303291}}].

\bibitem{Anderson:1994dz}
G.~W. Anderson and D.~J. Castano, {\it {Measures of fine tuning}},  {\em
  Phys.Lett.} {\bf B347} (1995) 300--308
  [\href{http://arXiv.org/abs/hep-ph/9409419}{{\tt hep-ph/9409419}}].

\bibitem{Ciafaloni:1996zh}
P.~Ciafaloni and A.~Strumia, {\it {Naturalness upper bounds on gauge mediated
  soft terms}},  {\em Nucl.Phys.} {\bf B494} (1997) 41--53
  [\href{http://arXiv.org/abs/hep-ph/9611204}{{\tt hep-ph/9611204}}].

\bibitem{Giusti:1998gz}
L.~Giusti, A.~Romanino and A.~Strumia, {\it {Natural ranges of supersymmetric
  signals}},  {\em Nucl.Phys.} {\bf B550} (1999) 3--31
  [\href{http://arXiv.org/abs/hep-ph/9811386}{{\tt hep-ph/9811386}}].

\bibitem{Casas:2004gh}
J.~Casas, J.~Espinosa and I.~Hidalgo, {\it {Implications for new physics from
  fine-tuning arguments. 1. Application to SUSY and seesaw cases}},  {\em JHEP}
  {\bf 0411} (2004) 057 [\href{http://arXiv.org/abs/hep-ph/0410298}{{\tt
  hep-ph/0410298}}].

\bibitem{Casas:2005ev}
J.~Casas, J.~R. Espinosa and I.~Hidalgo, {\it {Implications for new physics
  from fine-tuning arguments. II. Little Higgs models}},  {\em JHEP} {\bf 0503}
  (2005) 038 [\href{http://arXiv.org/abs/hep-ph/0502066}{{\tt
  hep-ph/0502066}}].

\bibitem{Allanach:2006jc}
B.~Allanach, {\it {Naturalness priors and fits to the constrained minimal
  supersymmetric standard model}},  {\em Phys.Lett.} {\bf B635} (2006) 123--130
  [\href{http://arXiv.org/abs/hep-ph/0601089}{{\tt hep-ph/0601089}}].

\bibitem{Athron:2007ry}
P.~Athron and .~Miller, D.J., {\it {A New Measure of Fine Tuning}},  {\em
  Phys.Rev.} {\bf D76} (2007) 075010
  [\href{http://arXiv.org/abs/0705.2241}{{\tt 0705.2241}}].

\bibitem{Fichet:2012sn}
S.~Fichet, {\it {Quantified naturalness from Bayesian statistics}},
  \href{http://arXiv.org/abs/1204.4940}{{\tt 1204.4940}}.

\bibitem{Ghilencea:2012qk}
D.~Ghilencea and G.~Ross, {\it {The fine-tuning cost of the likelihood in SUSY
  models}},  {\em Nucl.Phys.} {\bf B868} (2013) 65--74
  [\href{http://arXiv.org/abs/1208.0837}{{\tt 1208.0837}}].

\bibitem{Martin:2007gf}
S.~P. Martin, {\it {Compressed supersymmetry and natural neutralino dark matter
  from top squark-mediated annihilation to top quarks}},  {\em Phys.Rev.} {\bf
  D75} (2007) 115005 [\href{http://arXiv.org/abs/hep-ph/0703097}{{\tt
  hep-ph/0703097}}].

\bibitem{Akrami:2009hp}
  Y.~Akrami, P.~Scott, J.~Edsjo, J.~Conrad and L.~Bergstrom,
  JHEP {\bf 1004} (2010) 057
  [arXiv:0910.3950 [hep-ph]].
  
 \bibitem{Feroz:2011bj}
  F.~Feroz, K.~Cranmer, M.~Hobson, R.~Ruiz de Austri and R.~Trotta,
  JHEP {\bf 1106} (2011) 042
  [arXiv:1101.3296 [hep-ph]].
  
\bibitem{Feng:1999mn}
J.~L. Feng, K.~T. Matchev and T.~Moroi, {\it {Multi - TeV scalars are natural
  in minimal supergravity}},  {\em Phys.Rev.Lett.} {\bf 84} (2000) 2322--2325
  [\href{http://arXiv.org/abs/hep-ph/9908309}{{\tt hep-ph/9908309}}].

\bibitem{Haber:1996fp}
  H.~E.~Haber, R.~Hempfling and A.~H.~Hoang,
  Z.\ Phys.\ C {\bf 75} (1997) 539
  [hep-ph/9609331].
  
  
\end{thebibliography}\endgroup

\end{document}